\RequirePackage[l2tabu, orthodox]{nag}
\newcommand*{\ATLASLATEXPATH}{}
\documentclass[USenglish,texlive=2011,PAPER,cernpreprint]{\ATLASLATEXPATH atlasdoc}
\pdfoutput=1
\usepackage[full]{\ATLASLATEXPATH atlaspackage}
\usepackage{comment}
\usepackage{\ATLASLATEXPATH atlasbiblatex}
\usepackage{\ATLASLATEXPATH atlascontribute}
\usepackage{\ATLASLATEXPATH atlasphysics}
\graphicspath{{}}
\usepackage{multirow}
\hypersetup{pdftitle={ATLAS draft},pdfauthor={The ATLAS Collaboration}}
\graphicspath{{}}

\usepackage{booktabs,siunitx}
\AtlasTitle{Measurement of jet charge in dijet events from $\sqrt{s}=8$ \TeV~$pp$ collisions with the ATLAS detector}
\AtlasJournal{Physical Review D}

\AtlasAbstract{%
The momentum-weighted sum of the charges of tracks associated to a jet is sensitive to the charge of the initiating quark or gluon.  This paper presents a measurement of the distribution of momentum--weighted sums, called {\it jet charge}, in dijet events using 20.3 fb$^{-1}$ of data recorded with the ATLAS detector at $\sqrt{s}=8$ \TeV~in $pp$ collisions at the LHC.  The jet charge distribution is unfolded to remove distortions from detector effects and the resulting particle-level distribution is compared with several models.  The $p_T$-dependence of the jet charge distribution average and standard deviation are compared to predictions obtained with several LO and NLO parton distribution functions.  The data are also compared to different Monte Carlo simulations of QCD dijet production using various settings of the free parameters within these models.  The chosen value of the strong coupling constant used to calculate gluon radiation is found to have a significant impact on the predicted jet charge.  There is evidence for a $p_\text{T}$-dependence of the jet charge distribution for a given jet flavor.~In agreement with perturbative QCD predictions, the data show that the average jet charge of quark-initiated jets decreases in magnitude as the energy of the jet increases.  
}

\PreprintIdNumber{CERN-PH-EP-2015-207}

\begin{document}
\maketitle

\section{Introduction}
\label{sec:intro}

Quarks and gluons produced in high-energy particle collisions hadronize before their electric charge can be directly measured.  However, information about the electric charge is embedded in the resulting collimated sprays of hadrons known as jets.  One jet observable sensitive to the electric charge of
quarks and gluons is the momentum-weighted charge sum constructed from charged-particle tracks in a jet~\cite{Feynman1978}.   
Called the {\it jet charge}, this observable was first used experimentally in deep inelastic scattering studies~\cite{Berge1980,Berge1981,Allen1981,Allen1982,Albanese1984,Barlag1982,Erickson1979} to establish a relationship between the quark model and hadrons.  Since then, jet charge observables have been used in a variety of applications, including tagging the charge of $b$-quark jets~\cite{SLD1995,Tasso1990,Delphi1991,Aleph1991,Opal1992,Opal1994,Delphi1996,CDF1999,Abazov2007,CDF2011,ATLAS2011} and hadronically decaying $W$ bosons~\cite{Barate1998,Abreu:2001rpa,Acciarri:1999kn,Abbiendi:2000ej,ATLAS-CONF-2013-086,CMS-PAS-JME-14-002} as well as distinguishing hadronically decaying $W$ bosons from jets produced in generic quantum chromodynamic (QCD) processes~\cite{CMS-PAS-JME-14-002} and quark jets from gluon jets~\cite{Aad:2014gea,ATLAS-CONF-2013-086,Acton:1992uu,Arnison:1986ja}.

The study presented in this paper is a measurement of the jet charge distribution in inclusive dijet events from $pp$ collisions at the LHC.  Inclusive dijet events provide a useful environment for measuring the jet charge as they are an abundant source of gluon-initiated and quark-initiated jets.  There are fewer theoretical ambiguities from close-by jets and large-angle radiation associated with assigning the jet flavor in events with two jets than in events with higher jet multiplicities.  Furthermore, the transverse momentum ($p_\text{T}$) range accessible in dijet events is broad, $\mathcal{O}(10)$ \GeV~up to $\mathcal{O}(1000)$ \GeV.  Since the initial state at the LHC has a net positive charge, the probability for positively charged quarks to be produced in $pp$ collisions is higher than that for negatively charged quarks.  The probability for collisions to involve a positively charged
valence up-quark in the proton increases with the parton center-of-mass
energy $\sqrt{\hat{s}}$. Thus the average jet charge in inclusive dijet events is expected to
increase with $\sqrt{\hat{s}}$ if it is correlated with the quark
charge. The parton distribution functions (PDFs) encode the
probabilities to find gluons and certain flavors of quarks at given
momentum fractions $x$ of the proton. The momentum fractions of the
two initial partons $x_1, x_2$, and the proton and parton center-of-mass
energies $\sqrt{s}$ and $\sqrt{\hat{s}}$ are related by
$\sqrt{\hat{s}}=\sqrt{x_1x_2s}$. The PDFs are fairly well constrained~\cite{Watt:2012tq,Harland-Lang:2014zoa,Aaron:2009aa,Abramowicz:2015mha,Abramowicz:2015mha,Ball:2014uwa} in the
$x$ range relevant for this study, $0.005$---$0.5$.  However, if the jet charge is directly sensitive to the parton flavor, its $p_\text{T}$ dependence can provide a consistency check using new information beyond the jet $p_\text{T}$, which is currently used in PDF fits.  The PDFs are not the only nonperturbative input needed to model the jet charge distribution and its evolution with $\sqrt{\hat{s}}$.  As a momentum-weighted sum over jet constituents, the jet charge is sensitive to the modeling of fragmentation.  Previous studies have shown that there are qualitative differences between the charged-particle track multiplicities of jets in data and as predicted by the leading models of hadron production~\cite{Aad:2014gea}.  Thus, a measurement of the jet charge distribution with a range of quark/gluon compositions can provide a constraint on models of jet formation.  

While the change in jet parton flavor due to the PDF $x$-dependence predicts most of the variation in the jet charge distribution as a function of $\sqrt{\hat{s}}$, there is a second contribution due to the energy-dependence of the fragmentation functions.  Ratios of the charge distribution moments at different values of $\sqrt{\hat{s}}$ can be calculated perturbatively.  Recent calculations~\cite{Waalewijn2012,Krohn2012} in the context of Soft Collinear Effective Theory~\cite{Bauer:2000ew,Bauer:2000yr,Bauer:2001ct,Bauer:2001yt} show a significant reduction in the magnitude of the average jet charge for a
given jet flavor as a function of jet energy.  Information from PDFs can be used to extract the energy-dependence of the average jet charge in the data for direct comparisons to the predictions.

This paper presents a measurement of the $p_\text{T}$-dependence of the jet charge distribution's mean and standard deviation in dijet events in $pp$ collisions at $\sqrt{s}=8$ \TeV~with the ATLAS detector.  The jet $p_\text{T}$ is a measurable quantity that is strongly related to $\sqrt{\hat{s}}$.  The average jet charge is extracted for both the leading and subleading jet and they are distinguished based on their relative orientation in rapidity\footnote{ATLAS uses a right-handed coordinate system with its origin at the nominal interaction point (IP) in the center of the detector and the $z$-axis along the beam pipe. The $x$-axis points from the IP to the center of the LHC ring, and the $y$-axis points upward. Cylindrical coordinates $(r,\phi)$ are used in the transverse plane, $\phi$ being the azimuthal angle around the beam pipe. The pseudorapidity is defined in terms of the polar angle $\theta$ as $\eta=-\ln\tan(\theta/2)$.  The variable $\Delta R=\sqrt{(\Delta\phi)^2+(\Delta\eta)^2}$ is a measure of how close two objects are in the $(\eta,\phi)$ plane.  The rapidity of a four-vector is defined as $y=0.5\text{ln}\left(\frac{E+p_z}{E-p_z}\right)$, where $E$ is the energy and $p_z$ the component of the momentum parallel to the beam axis.}.   After a description of the ATLAS detector (Sec.~\ref{sec:atlas}), the data and simulated samples (Sec.~\ref{sec:samples}) and the detector- and particle-level objects and selections used in the analysis (Sec.~\ref{sec:objects}), Sec.~\ref{sec:construcing} details the construction of the jet charge and some of its properties.   In order for the measured jet charge distribution to be compared with particle-level models, the data are unfolded to remove distortions from detector effects, as described in Sec.~\ref{sec:unfolding}.  Systematic uncertainties in the measured jet charge spectra are discussed in Sec.~\ref{sec:uncerts} and the results are presented in Sec.~\ref{sec:results}.  

\section{The ATLAS detector}
\label{sec:atlas}

ATLAS~\cite{Aad:2008zzm} is a general-purpose detector designed to measure the properties of particles produced in high-energy $pp$ collisions with nearly a full $4\pi$ coverage in solid angle.  The innermost subsystem of the detector is a series of tracking devices used to measure charged-particle trajectories bent in a 2~T axial field provided by a solenoid whose axis is parallel with the beam direction.   This inner detector (ID) consists of a silicon pixel detector surrounded by a semiconductor microstrip detector (SCT) and a straw-tube tracker.  It has full coverage in $\phi$ and can detect particles with $|\eta|<2.5$.   Charged-particle tracks are reconstructed from all three ID components, providing measurements of the transverse momentum of tracks with a resolution $\sigma_{p_\text{T}}/p_\text{T} = 0.05\%\times p_\text{T}/\text{ \GeV} \oplus 1\%$.  The track reconstruction algorithm fits five track parameters: $d_0$, $z_0$, $\phi$, $\theta,$ and $q/p$, where $d_0$ and $z_0$ are the transverse and longitudinal impact parameters, respectively, $q$ is the track charge and $p$ is the track momentum.  Excellent spatial precision is required to maintain a well-performing track reconstruction out to and exceeding charged-particle $p_\text{T}$ of 1 \TeV, where track sagittas are $\lesssim 0.2$~mm.

Surrounding the ID and solenoid are electromagnetic and hadronic calorimeters to measure showers from charged and neutral particles.   The high-granularity liquid-argon (LAr) sampling electromagnetic calorimeter is located just beyond the solenoid and spans the range $|\eta| < 3.2$.  Beyond the electromagnetic calorimeter is the two-component hadronic calorimeter that uses scintillator-tile sampling technology in the range $|\eta| < 1.7$ and LAr sampling technology for $1.5<|\eta| < 3.2$.  Additional calorimeters are located in the forward region. Surrounding the calorimeters is a muon spectrometer, with trigger and precision chambers, and incorporating three large toroid magnets composed of eight coils each.

Due to the large event rate, not every collision can be recorded for processing offline.  Events are selected using a three-level trigger system~\cite{Aad:2012xs} that is hardware-based at the first level and software-based for the two following levels.  An event must satisfy all three trigger levels to be recorded for further processing.  At each stage of the trigger, energy thresholds are placed on jet-like objects, with the similarity between online and offline jets increasing with each level.  The first level makes decisions based on low-granularity calorimeter towers with thresholds that are typically less than half of the energy required by jets at the second level.  A simple jet reconstruction algorithm is used at the second level in regions around the jets identified by the first level.  The third level (known as the Event Filter) clusters jets with the same algorithm as used offline over the entire detector with thresholds that are typically 20--30 \GeV~higher than at level two.   The single-jet trigger thresholds increase at each level due not only to differences in the jet reconstruction and calibrations, but also to meet the different bandwidth requirements at each trigger level.  For low-$p_\text{T}$ dijets, the event rate is far too large to save every event that passes the trigger selection and so most of the jet triggers are {\it prescaled} to artificially lower their recording rate. 

\section{Data and simulated samples}
\label{sec:samples}

This measurement uses the full dataset of $pp$ collisions recorded by the ATLAS detector in 2012, corresponding to an integrated luminosity of 20.3 fb${}^{-1}$ at a center-of-mass energy of $\sqrt{s}=8$ \TeV.  Events are only considered if they are collected during stable beam conditions and satisfy all data-quality requirements~\cite{ATLAS-CONF-2010-038}.  To reject noncollision events, there must be a primary vertex reconstructed from at least two tracks each with $p_\text{T}>400$ \MeV~\cite{ATLAS-CONF-2010-069}.  Due to the high instantaneous luminosity and the large total inelastic proton-proton
cross section, on average there are about $21$ simultaneous ({\it pileup}) collisions in each bunch crossing.

A set of single-jet triggers is used to collect dijet events with high efficiency.   Table~\ref{tab:triggermenu} shows the collected luminosity for each trigger as well as the offline jet $p_\text{T}$ ranges used, chosen such that the trigger is fully efficient.  The highest-$p_\text{T}$ trigger is not prescaled.

\begin{table}
\centering
\begin{tabular}{ccc}
Trigger threshold [\GeV] & Offline Selection [\GeV] & Luminosity [fb${}^{-1}$]  \\
\hline 
\hline
25 & [50,100] & 7.84$\times 10^{-5}$ \\
55 & [100, 136] & 4.42$\times 10^{-4}$\\
80 & [136, 190] & 2.32$\times 10^{-3}$ \\
110 & [190, 200] & 9.81$\times 10^{-3}$ \\
145 & [200, 225] & 3.63$\times 10^{-2}$ \\
180 & [225, 250] & 7.88$\times 10^{-2}$ \\
220 & [250, 300] & 2.61$\times 10^{-1}$ \\
280 & [300, 400] & 1.16 \\
360 & $\geq 400$ & 20.3 \\
\hline
\hline
\end{tabular}
\caption{The single-jet trigger menu used to collect dijet events with the 2012 dataset.  The first column is the level-three (Event Filter) jet $p_\text{T}$ threshold and the second column is the offline leading-jet $p_\text{T}$ range corresponding to the given trigger.  The luminosity collected with each trigger is in the last column.  The total 2012 dataset was 20.3 fb${}^{-1}$; the highest-$p_\text{T}$ trigger is not prescaled.}
\label{tab:triggermenu}
\end{table}

Monte Carlo (MC) simulated events are generated in $p_\text{T}$ slices in order to ensure a large number of events over a broad range of reconstructed jet $p_\text{T}$, given constraints on the available computing resources.   The $p_\text{T}$ slices span the interval $0$ to $5$ \TeV~in ranges that approximately double with each increasing slice, starting with a range of size $8$ \GeV~and ending with a range of size $2240$ \GeV.  The baseline sample used for the measurement is generated with {\sc Pythia} 8.175~\cite{Sjostrand:2007gs} with the AU2~\cite{ATL-PHYS-PUB-2012-003} set of tuned parameters (tune) and the next-to-leading-order (NLO) PDF set\footnote{A discussion on the use of NLO PDF sets with LO matrix elements is given in Refs.~\cite{Campbell:2006wx,Sherstnev:2007nd}.} CT10~\cite{Lai:2010vv,Gao:2013xoa}.  Another large sample of events is generated with {\sc Herwig++} 2.63~\cite{Bahr:2008pv,Arnold:2012fq} with tune EE3~\cite{Gieseke:2012ft} and leading-order (LO) PDF set CTEQ6L1~\cite{Pumplin:2002vw} (particle-level samples with CT10 and EE4 are also used for comparisons).  Both {\sc Pythia} and {\sc Herwig++} are LO in perturbative QCD for the ($2\rightarrow 2$) matrix element and resum the leading logarithms (LL) in the parton shower.  However, the ordering of emissions in the MC resummation in the shower differs between these two generators: {\sc Pythia} implements $p_\text{T}$-ordered showers~\cite{Sjostrand:2004ef} whereas {\sc Herwig++} uses angular ordering~\cite{Gieseke:2003rz}.  The phenomenological modeling of the non-pertubative physics also differs between {\sc Pythia} and {\sc Herwig++}.  In addition to different underlying-event models (Ref.~\cite{Sjostrand:2004pf} for {\sc Pythia} and an eikonal model~\cite{Bahr:2008dy} for {\sc Herwig++}) the hadronization models differ between {\sc Pythia} (Lund string model~\cite{string}) and {\sc Herwig++} (cluster model~\cite{Webber:1983if}).  These two schemes are known~\cite{Aad:2014gea} to predict different numbers of charged particles within jets and different distributions of the charged-particle energies within jets, both of which are important for the jet charge.   All tunes of the underlying event that are used with {\sc Pythia} and {\sc Herwig++} in this analysis use LHC data as input.  As discussed in Sec.~\ref{sec:intro}, the corrected data are compared to models with various PDF sets; for consistency, each set has a dedicated underlying-event tune constructed in the same way from a fixed set of data inputs (AU2) described in detail in Ref.~\cite{ATL-PHYS-PUB-2012-003}.  The PDF sets include LO sets CTEQ6L1~\cite{Pumplin:2002vw} and MSTW08LO~\cite{Watt:2012tq} as well as NLO sets CT10~\cite{Lai:2010vv,Gao:2013xoa}, NNPDF21 NLO~\cite{Ball:2010de}, and MSTW2008NLO~\cite{Watt:2012tq}.   A sample generated with a NLO matrix element from {\sc Powheg-Box} {\sc r2262}~\cite{Nason:2004rx,Frixione:2007vw,Alioli:2010xd,Frixione:2007nw} (henceforth referred to as {\sc Powheg}) with PDF set CT10 interfaced with {\sc Pythia} 8.175 and the AU2 tune is also used for comparisons.  

All MC samples are processed using the full ATLAS detector simulation~\cite{Aad:2010ah} based on GEANT4~\cite{Agostinelli:2002hh}. 

\section{Object reconstruction and event selection}
\label{sec:objects}

The reconstructed objects used for the jet charge as well as for the event selection are described in Sec.~\ref{sec:detectorlevel}.   The fiducial definition of the measurement, unfolded to particle level, is given in Sec.~\ref{sec:particlelevel}.

\subsection{Object reconstruction at detector level}
\label{sec:detectorlevel}

Jets are clustered using the anti-$k_t$ jet algorithm~\cite{Cacciari:2008gp} with radius parameter $R=0.4$ implemented in FastJet~\cite{Cacciari:2011ma} from topological calorimeter-cell clusters~\cite{TopoClusters}, calibrated using the local cluster weighting (LCW) algorithm \cite{EndcapTBelectronPion2002,Barillari:2009zza}.  An overall  jet energy calibration accounts for residual detector effects as well as contributions from pileup~\cite{areasATLAS} in order to make the reconstructed jet energy an unbiased measurement of the particle-level jet energy.  Jets are required to be central $(|\eta| < 2.1)$ so that their charged particles are within the $|\eta|<2.5$ coverage of the ID.   

When more than one primary vertex is reconstructed, the one with the highest $\sum p_\text{T}^2$ of tracks is selected as the hard-scatter vertex.  Events are further required to have at least two jets with $p_\text{T}>50$ \GeV~and only the leading two jets are considered for the jet charge measurement.  To select dijet topologies, the two leading jets must have $p_\text{T}^\text{lead}/p_\text{T}^\text{sublead} < 1.5$, where $p_\text{T}^\text{lead}$ and $p_\text{T}^\text{sublead} $ are the transverse momenta of the jets with the highest and second-highest $p_\text{T}$, respectively.  The jet with the smaller (larger) absolute pseudorapidity $|\eta|$ is classified as the more central (more forward) jet.  A measurement of the more forward and more central jet charge distributions can exploit the rapidity-dependence of the jet flavor to extract information about the jet charge for a particular flavor.  This is discussed in more detail in Sec.~\ref{sec:particlelevel}.

\begin{figure}
\begin{center}
\includegraphics[width=0.5\textwidth]{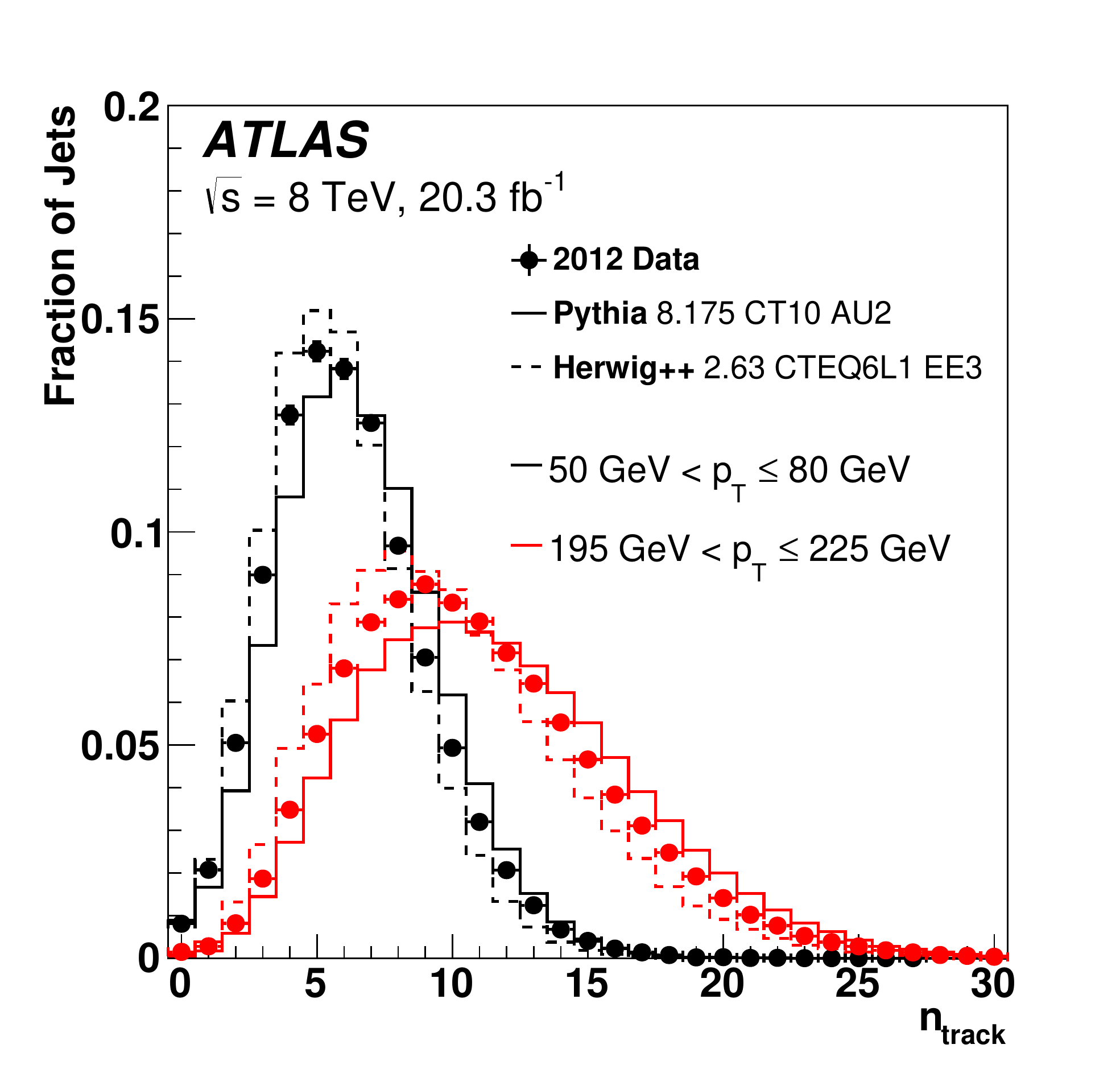}
\end{center}	
\caption{The distribution of the number of tracks associated with a jet in two example jet $p_\text{T}$ ranges.}
\label{fig:tracks}
\end{figure}

Tracks used to calculate the jet charge are required to have $p_\text{T} \geq$ 500 \MeV, $|\eta| < 2.5$,~and a $\chi^2$ per degree of freedom (resulting from the track fit) less than 3.0.    Additional quality criteria are applied to select tracks originating from the collision vertex and reject fake tracks reconstructed from random hits in the detector.  In particular, tracks must be well-matched to the hard-scatter vertex with $|z_0\sin(\theta)|<1.5$ mm and $|d_0|< 1$ mm, where $z_0$ and $d_0$ are calculated with respect to the primary vertex.  Tracks must furthermore have at least one hit in the pixel detector and at least six hits in the SCT.  The matching of tracks with the calorimeter-based jets is performed via the ghost-association technique~\cite{area}: the jet clustering process is repeated with the addition of {\it ghost} versions of measured tracks that have the same direction but infinitesimally small $p_\text{T}$, so that they do not change the properties of the calorimeter jets.  A track is associated with a jet if its ghost version is contained in the jet after reclustering.  The distribution of the number of tracks in two representative jet $p_\text{T}$ ranges is shown in Fig.~\ref{fig:tracks}.  The number of tracks increases with jet $p_\text{T}$ and the data fall between the predicted distributions of {\sc Pythia} and {\sc Herwig++}.

\subsection{Object definitions at particle level}
\label{sec:particlelevel}

\begin{figure}
\begin{center}
\subfloat[]{\includegraphics[width=0.5\textwidth]{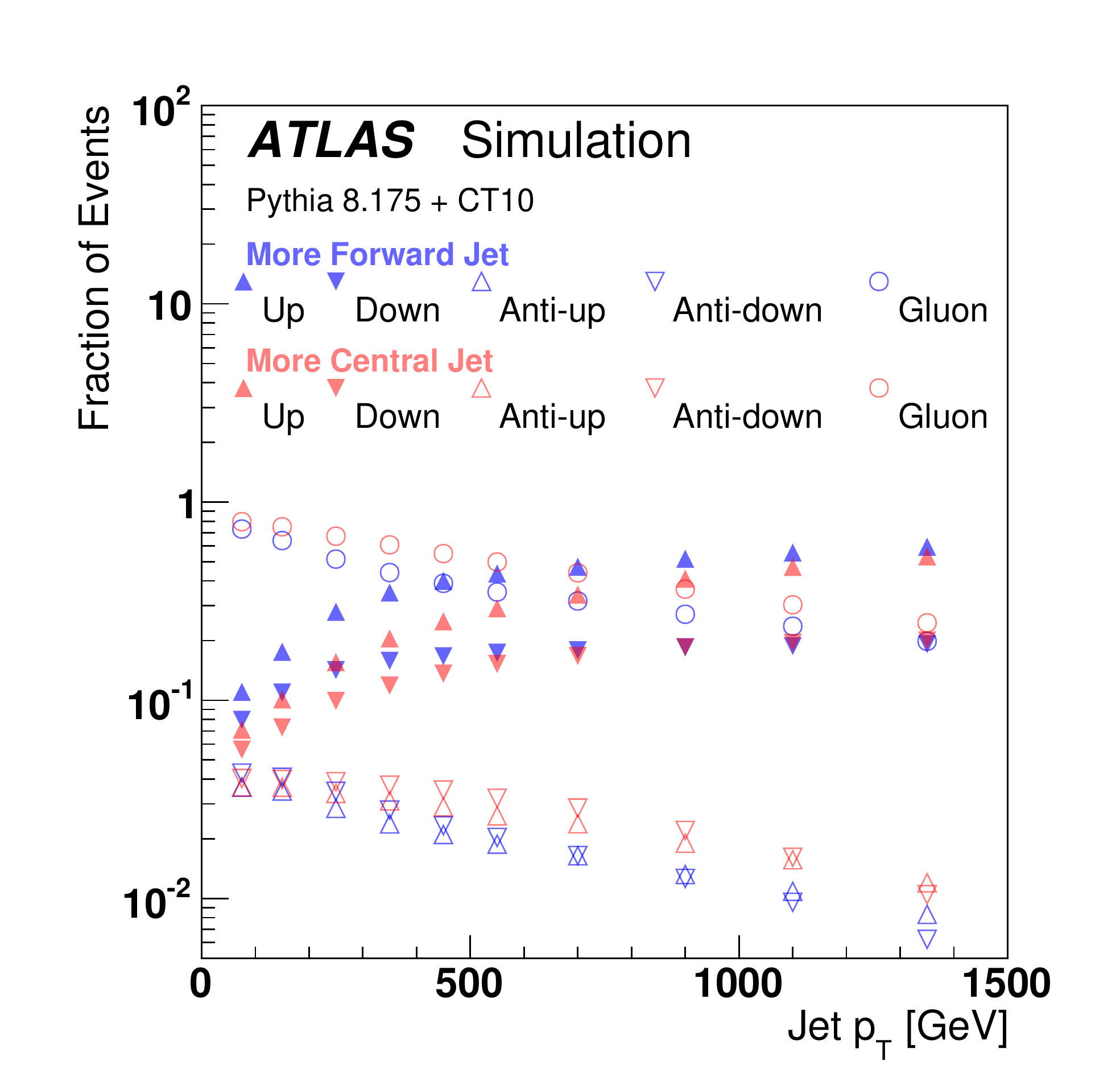}}\subfloat[]{\includegraphics[width=0.5\textwidth]{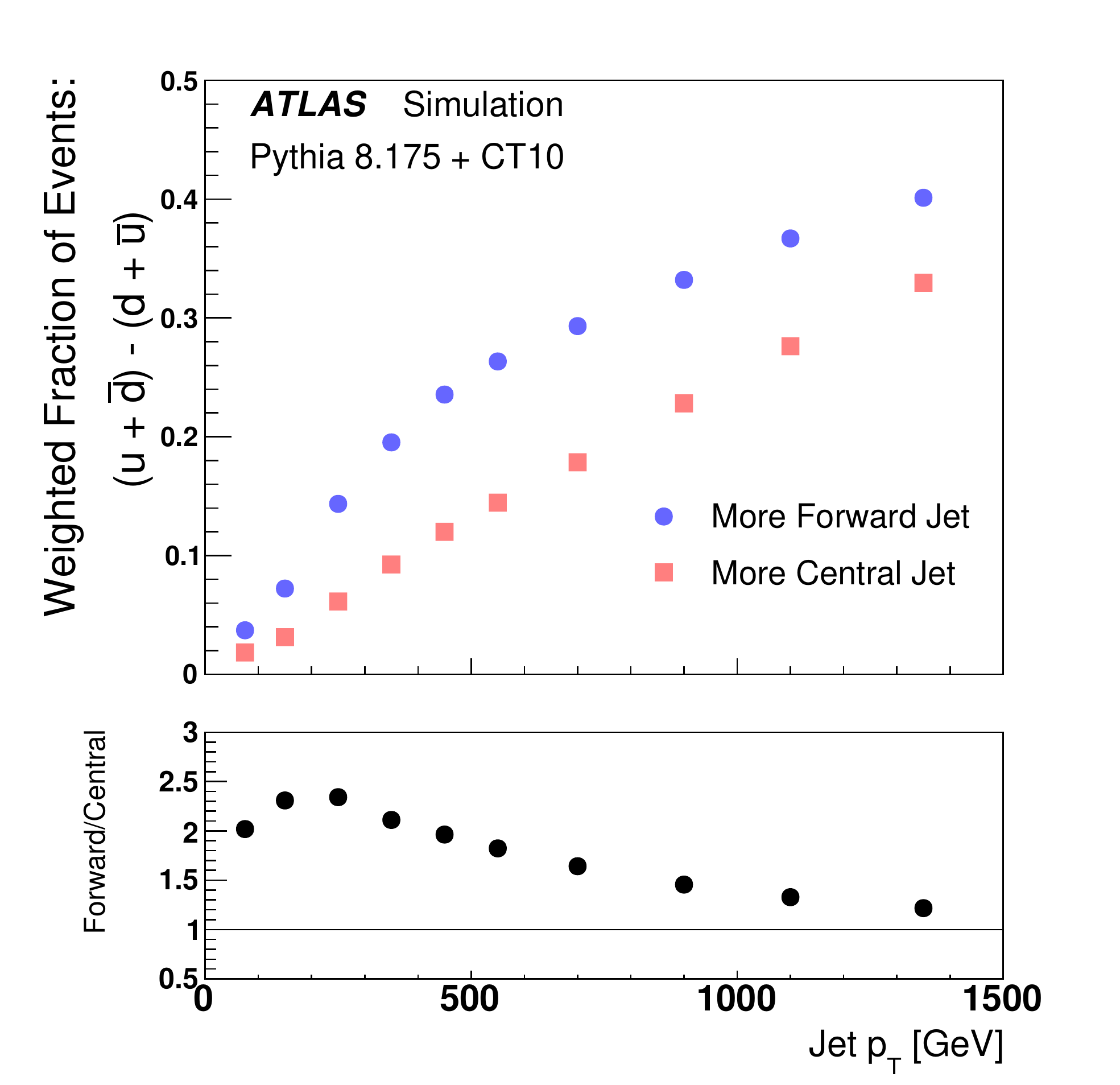}}
\end{center}	
\caption{For a given jet flavor, (a) shows the fraction $f$ of jets with that flavor in events passing the particle-level event selection and (b) shows the $p_\text{T}$ evolution of the flavor fractions weighted by charge-sign: $f_\text{up}+f_\text{anti-down}-f_\text{anti-up}-f_\text{down}$.  The forward-central differences between the flavor fractions are largest at low $p_\text{T}$, but the highest quark-jet purity occurs at high jet $p_\text{T}$.  The markers for the more forward and central jets are distinguished by their blue and red colors, respectively.}
\label{fig:flavorfrac}
\end{figure}

The measurement is carried out within a fiducial volume matching the experimental selection to avoid extrapolation into unmeasured kinematic regions that have additional model-dependence and related uncertainties.  Particle-level definitions of the reconstructed objects are chosen to be as close as possible to those described in Sec.~\ref{sec:detectorlevel}.  Particle-level jets are clustered from  generated stable particles with a mean lifetime $\tau>30$~ps,  excluding muons and neutrinos.  As with the detector-level jets, particle-level jets are clustered with the anti-$k_t$ $R=0.4$ algorithm.  In analogy to the ghost-association of tracks to jets performed at detector level, any charged particle clustered in a particle-level jet is considered for the jet charge calculation.\footnote{There is no $p_\text{T}>500$ MeV threshold applied to charged particles.  The impact of applying such a threshold is negligible for all $p_\text{T}$ bins except the first two where effects of up to 1\% are observed in the mean and standard deviation of the jet charge.}  There must be at least two jets with $|\eta|<2.1$ and $p_\text{T}>50$ \GeV.  The two highest-$p_\text{T}$ jets must satisfy the same $p_\text{T}$-balance requirement between the leading and subleading jet as at detector level ($p_\text{T}^\text{lead}/p_\text{T}^\text{sublead} < 1.5$).   Due to the high-energy and well-separated nature of the selected jets, the hard-scatter quarks and gluons can be cleanly matched to the outgoing jets.   While it is possible to classify jets as quark- or gluon-initiated beyond leading order in $m_\text{jet}/E_\text{jet}$~\cite{Banfi:2006hf}, the classification is algorithm-dependent and unnecessary for the present considerations.  In this analysis, the flavor of a jet is defined as that of the highest energy parton in simulation within a $\Delta R<0.4$ cone around the particle-jet axis.  The jet flavor depends on rapidity and so the two selected jets are classified as either more forward or more central; the more forward jet tends to be correlated to the higher-$x$ parton and is less likely to be a gluon jet.  Figure~\ref{fig:flavorfrac} shows the flavor fraction for the more forward and more central particle-level jets passing the event selection.  The $p_\text{T}$ evolution of the sum of the flavor fractions weighted by the sign of the parton charge is shown in Fig.~\ref{fig:flavorfrac}(b).  The forward-central differences between the flavor fractions are largest at low $p_\text{T}$, but the highest quark-jet purity occurs at high jet $p_\text{T}$.  

\clearpage
\newpage

\section{Constructing the jet charge}
\label{sec:construcing}

\begin{figure}
\begin{center}
\subfloat[]{\includegraphics[width=0.33\textwidth]{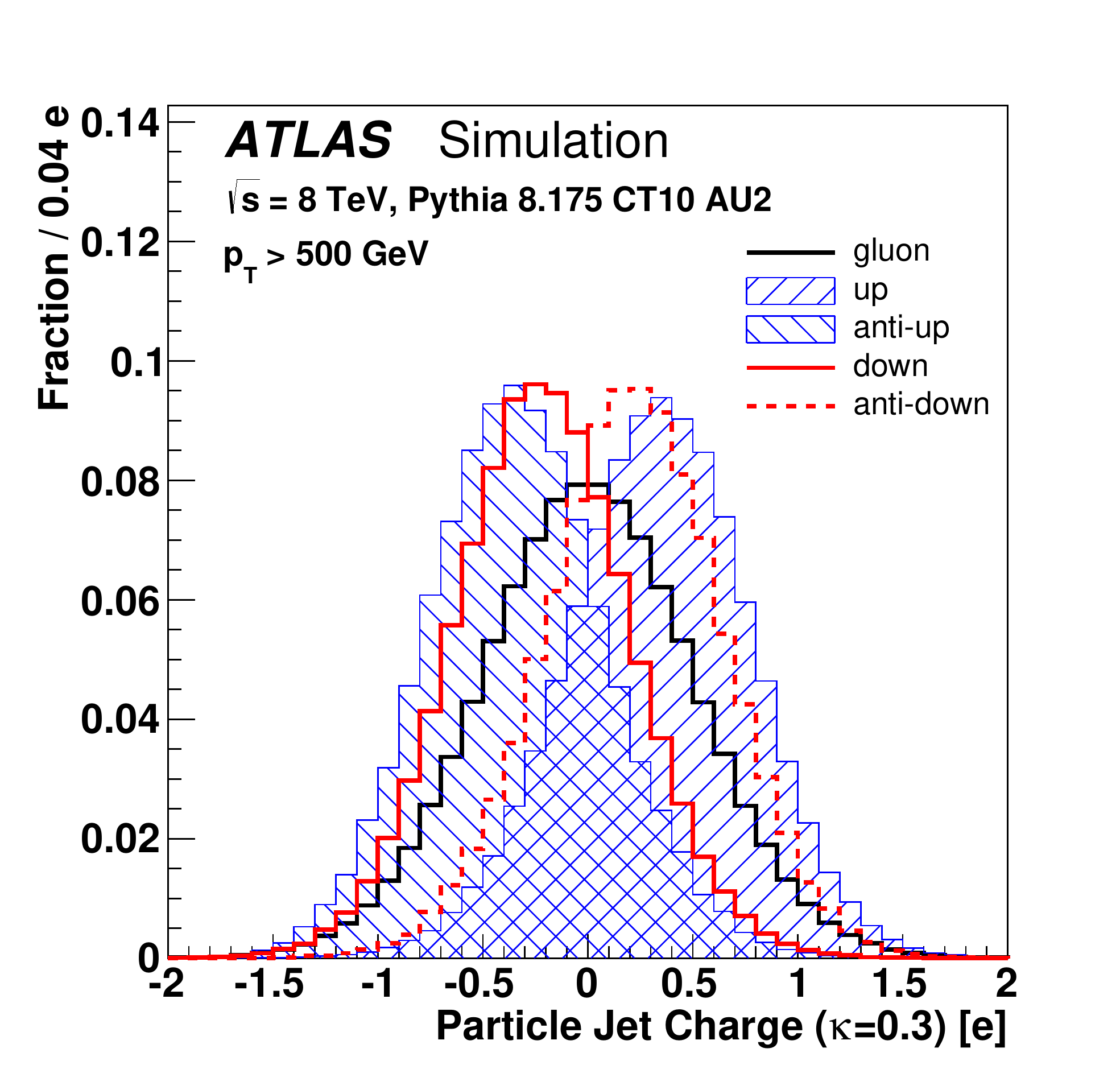}}\subfloat[]{\includegraphics[width=0.33\textwidth]{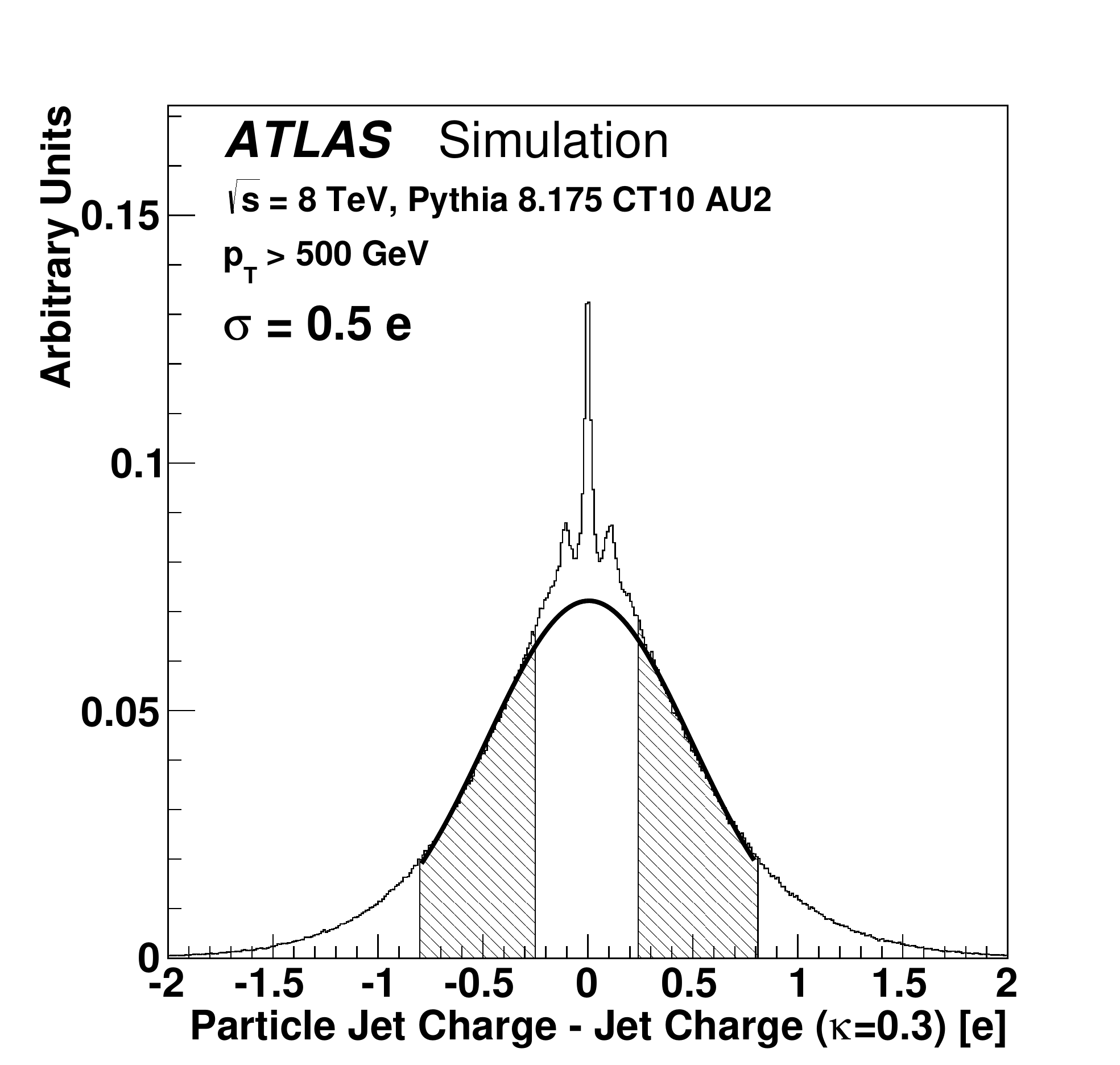}}\subfloat[]{\includegraphics[width=0.33\textwidth]{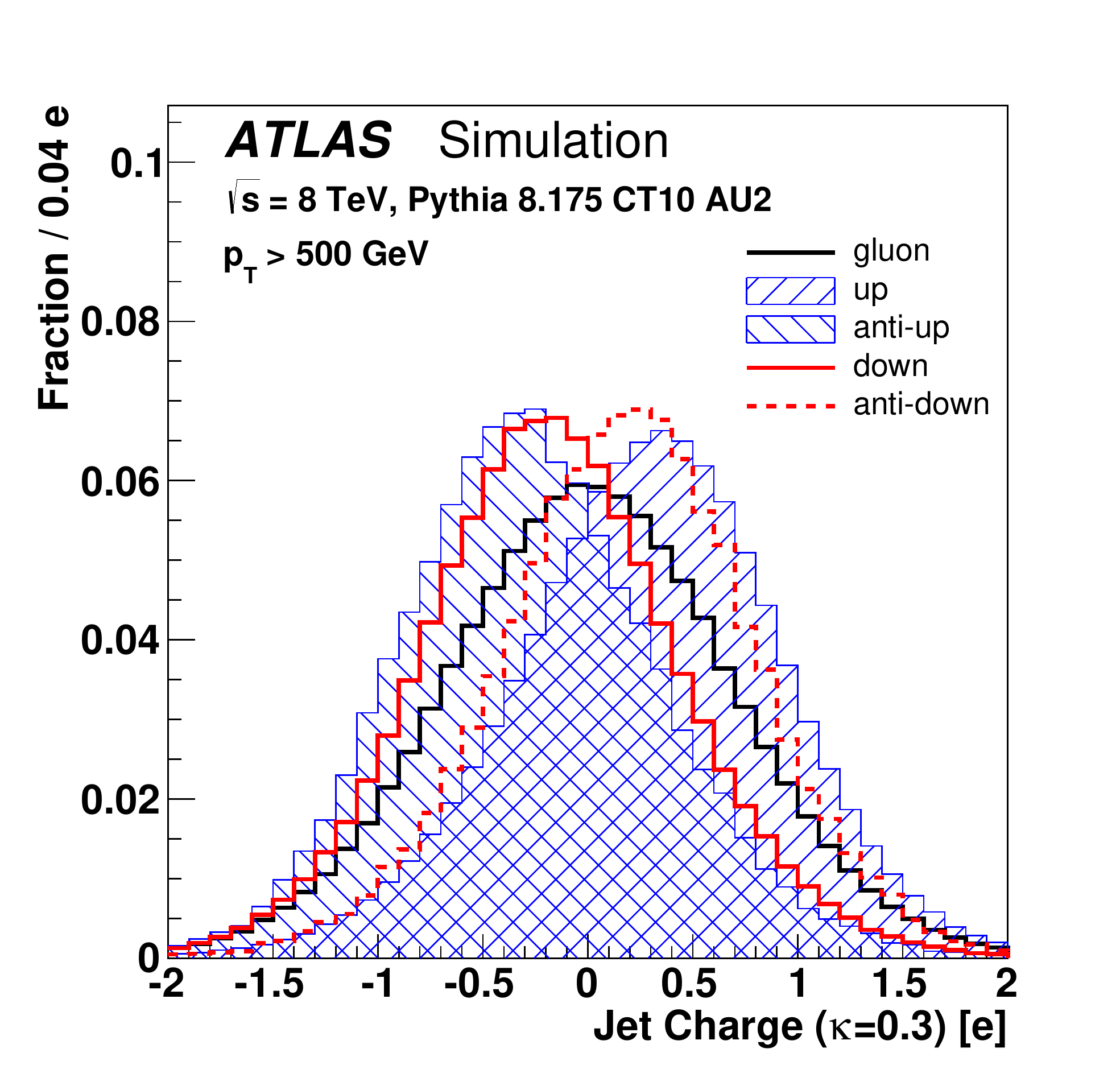}}
\end{center}	
\caption{The (a) the particle-level and (c) detector-level jet charge distribution for various jet flavors in a sample of jets with $p_\text{T}>500$ \GeV~for $\kappa=0.3$. (b): the distribution of the jet-by-jet difference between the particle-level and detector-level jet charge distributions.  The shaded region is used to fit a Gaussian function to extract the bulk response resolution, which is $\sigma\sim 0.5$ $e$, where $e$ is the positron charge.}
\label{fig:sortedbypartons}
\end{figure}

There is no unique way to define the jet charge.  The most na\"{i}ve construction is to add up the charge of all tracks associated with a jet.   However, this scheme is very sensitive to lost or extraneous soft radiation.  Therefore, a weighting scheme is introduced to suppress fluctuations.  Using the tracks assigned to a jet by ghost association, the jet charge $Q_J$ of a jet $J$ is calculated using a transverse-momentum-weighting scheme~\cite{Feynman1978}:

\begin{align}
  \label{chargedef}
  Q_J = \frac{1}{({p_\text{T}}_J)^\kappa}\sum_{i\in \text{\it Tracks}} q_i\times (p_\text{T,i})^\kappa,  
  \end{align} 

\noindent where $\text{\it Tracks}$ is the set of tracks associated with jet $J$, $q_i$ is the charge (in units of the positron charge) of track $i$ with associated transverse momentum $p_{\text{T},i}$, $\kappa$ is a free regularization parameter, and ${p_\text{T}}_J$ is the transverse momentum of the calorimeter jet.  The distributions of $Q_J$ for various jet flavors are shown in Fig.~\ref{fig:sortedbypartons} for $\kappa=0.3$.  In the simulation, there is a clear relationship between the jet charge and the initiating parton's charge, as up-quark jets tend to have a larger jet charge than gluon jets.   Furthermore, gluon jets tend to have a larger jet charge than down-quark jets.  However, the jet charge distribution is already broad at particle level and the jet charge response ($Q_\text{particle-level}-Q_\text{detector-level}$) resolution is comparable to the differences in the means of the distributions for different flavors, so one can expect only small changes in the inclusive jet charge distribution for changes in the jet flavor composition.   The three narrow distributions on top of the bulk response distribution in Fig.~\ref{fig:sortedbypartons}(b) are due to cases in which only one or two charged particles dominate the jet charge calculation at particle level.  The two off-center peaks are due to cases in which one of the two high-$p_\text{T}$-fraction tracks is not reconstructed and the widths of the two off-center and central peaks are due to the (single) track and jet $p_\text{T}$ resolutions.  The bulk response is fit to a Gaussian function with standard deviation $\sigma\sim 0.5$ $e$ (units of the positron charge).

\begin{figure}
\begin{center}
\subfloat[]{\includegraphics[width=0.45\textwidth]{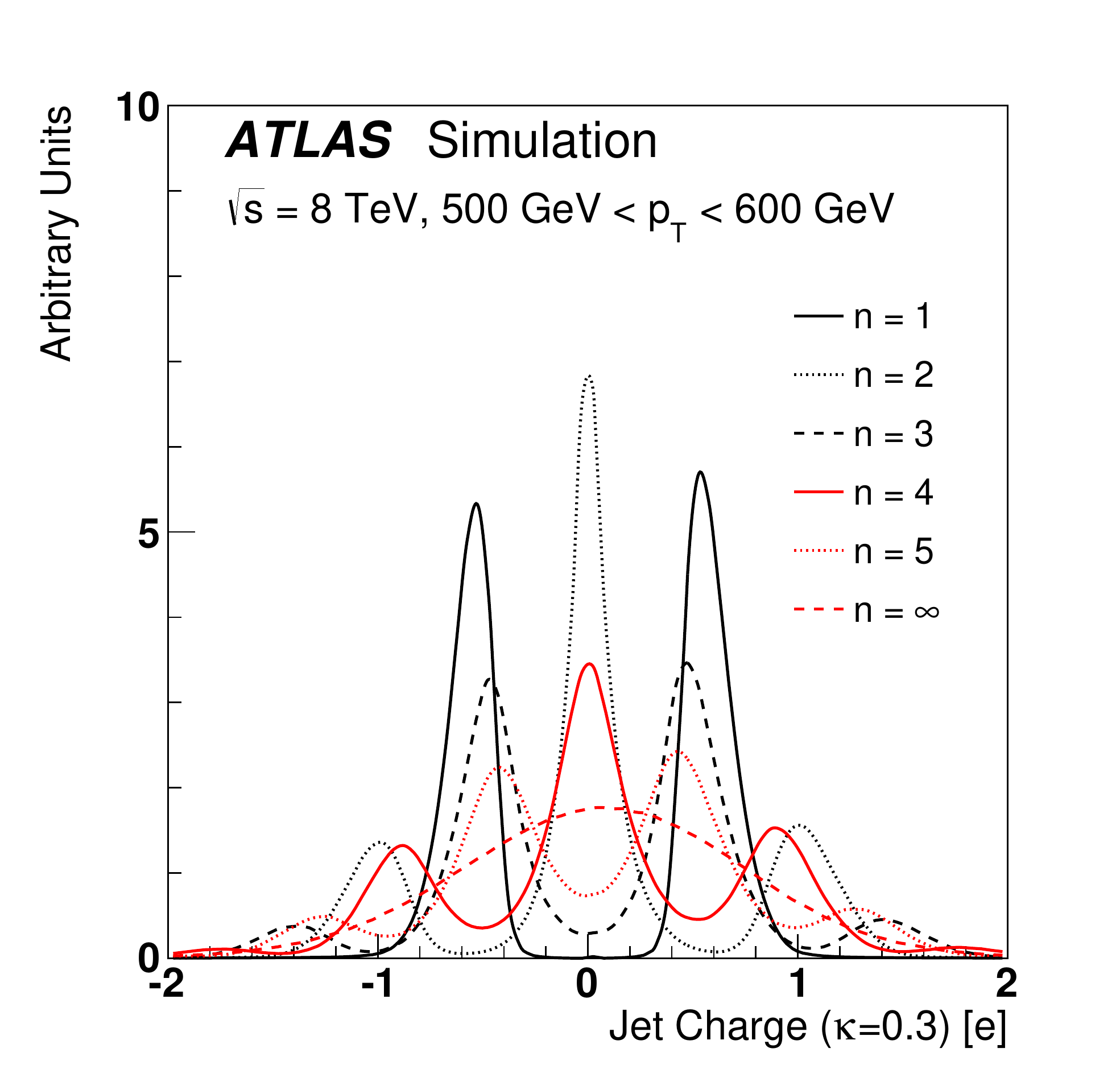}}\subfloat[]{\includegraphics[width=0.45\textwidth]{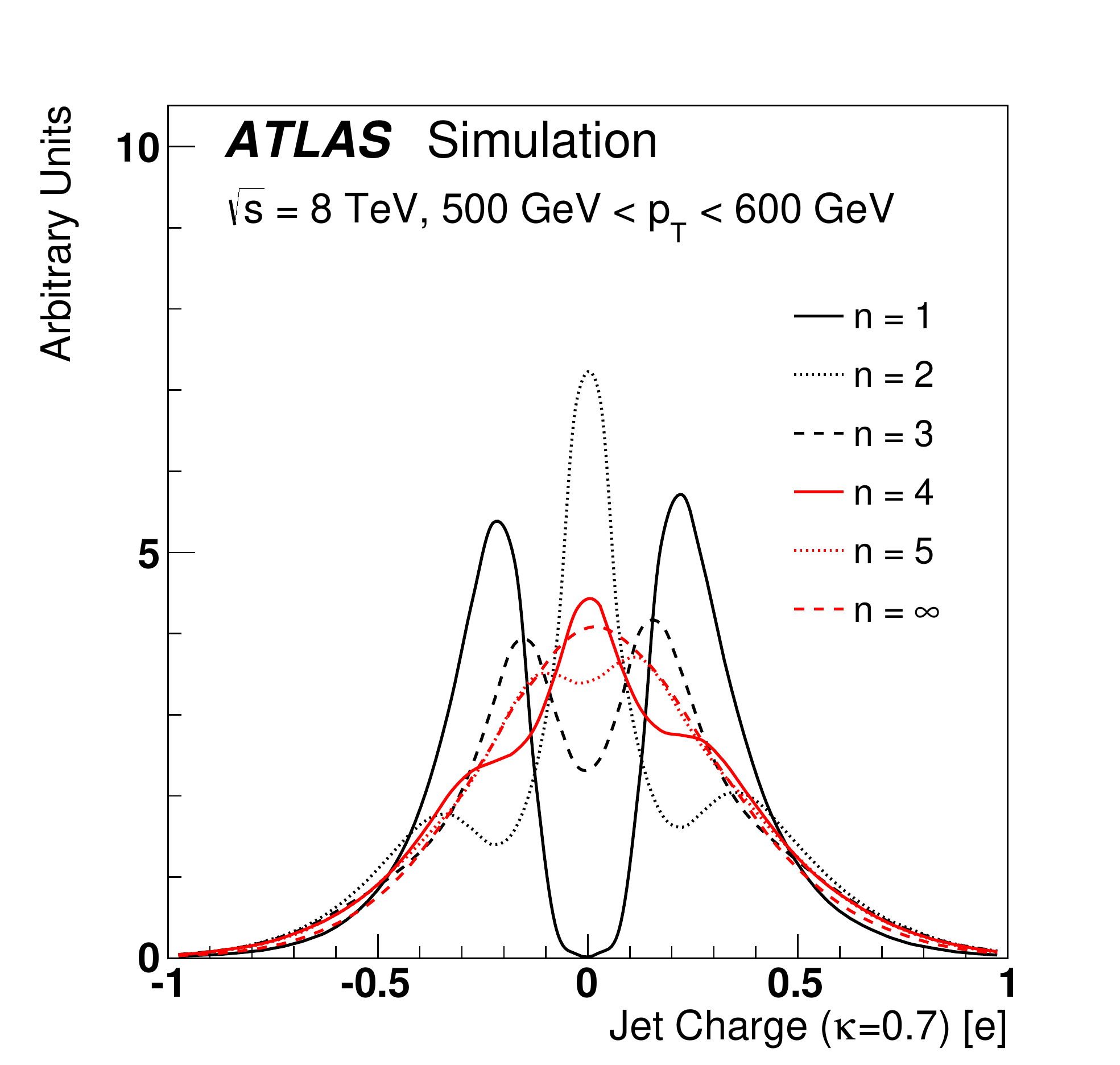}}
\end{center}	
\caption{The distribution of the jet charge built from the leading $n$ tracks ($Q_{J,n}$) for (a) $\kappa=0.3$ and (b) $\kappa=0.7$ in a sample of jets with 500 GeV $<p_\text{T}<$ 600 GeV. The horizontal axis ranges are not the same.  In this $p_\text{T}$ range, the median number of tracks is about 15.}
\label{fig:qn}
\end{figure}

The parameter $\kappa$ in Eq.~(\ref{chargedef}) regulates the sensitivity of the jet charge to soft radiation.  For $\kappa > 0$, the jet charge is infrared safe.\footnote{The jet charge is never collinear safe for $\kappa>0$ and not even the sign of the jet charge is Lorentz invariant, although it is clearly invariant under longitudinal boosts.}  Low values of $\kappa$ enhance the contribution to the jet charge from low-$p_\text{T}$ particles while in the $\kappa\rightarrow\infty$ limit, only the highest-$p_\text{T}$ track contributes to the sum in Eq.~(\ref{chargedef}).   The dependence on the highest-$p_\text{T}$ tracks is demonstrated with the plots in Fig.~\ref{fig:qn} with the variable $Q_{J,n}$, which is the jet charge in Eq.~(\ref{chargedef}), but built from the leading $n$ tracks.  The variable $Q_{J,1}$ is simply the weighted fragmentation function of the leading-track $p_\text{T}$ to the jet $p_\text{T}$ with weight $\kappa$.  The usual $Q_J$ is recovered in the limit $n\rightarrow\infty$.  Figure~\ref{fig:qn} shows the sequence $Q_{J,n}$ for $\kappa=0.3$ and $\kappa=0.7$.  For lower values of $\kappa$, many tracks are required for the sequence of distributions to converge to the full jet charge.  However, for $\kappa=0.7$, the distribution converges quickly, indicating that only the highest-$p_\text{T}$ tracks are contributing.  All reconstructed tracks are henceforth used when computing the jet charge, but the plots in Fig.~\ref{fig:qn} give an indication of the contribution of (relatively) high- and low-$p_\text{T}$ tracks.  Dedicated studies~\cite{ATLAS-CONF-2013-086} agree with theoretical predictions~\cite{Waalewijn2012} that suggest that $\kappa\sim0.5$ is the most sensitive to the charge of the parton initiating a jet.  Therefore, the measurement presented in this paper uses $\kappa=0.5$ in addition to $\kappa=0.3$ and $\kappa=0.7$ in order to maintain a broad sensitivity to both hard and soft radiation inside jets.

\begin{figure}
\begin{center}
\subfloat[]{\includegraphics[width=0.45\textwidth]{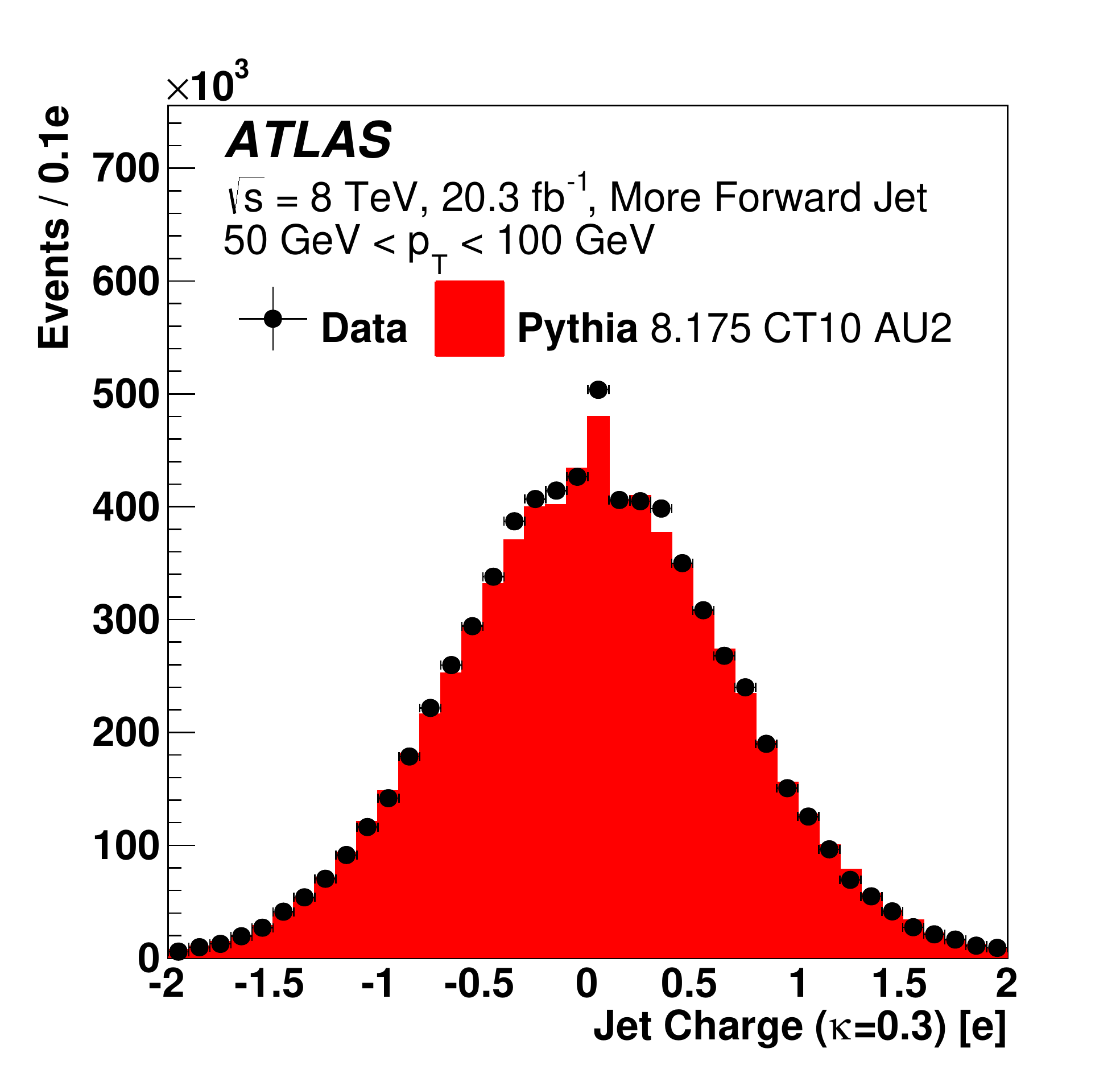}}\subfloat[]{\includegraphics[width=0.45\textwidth]{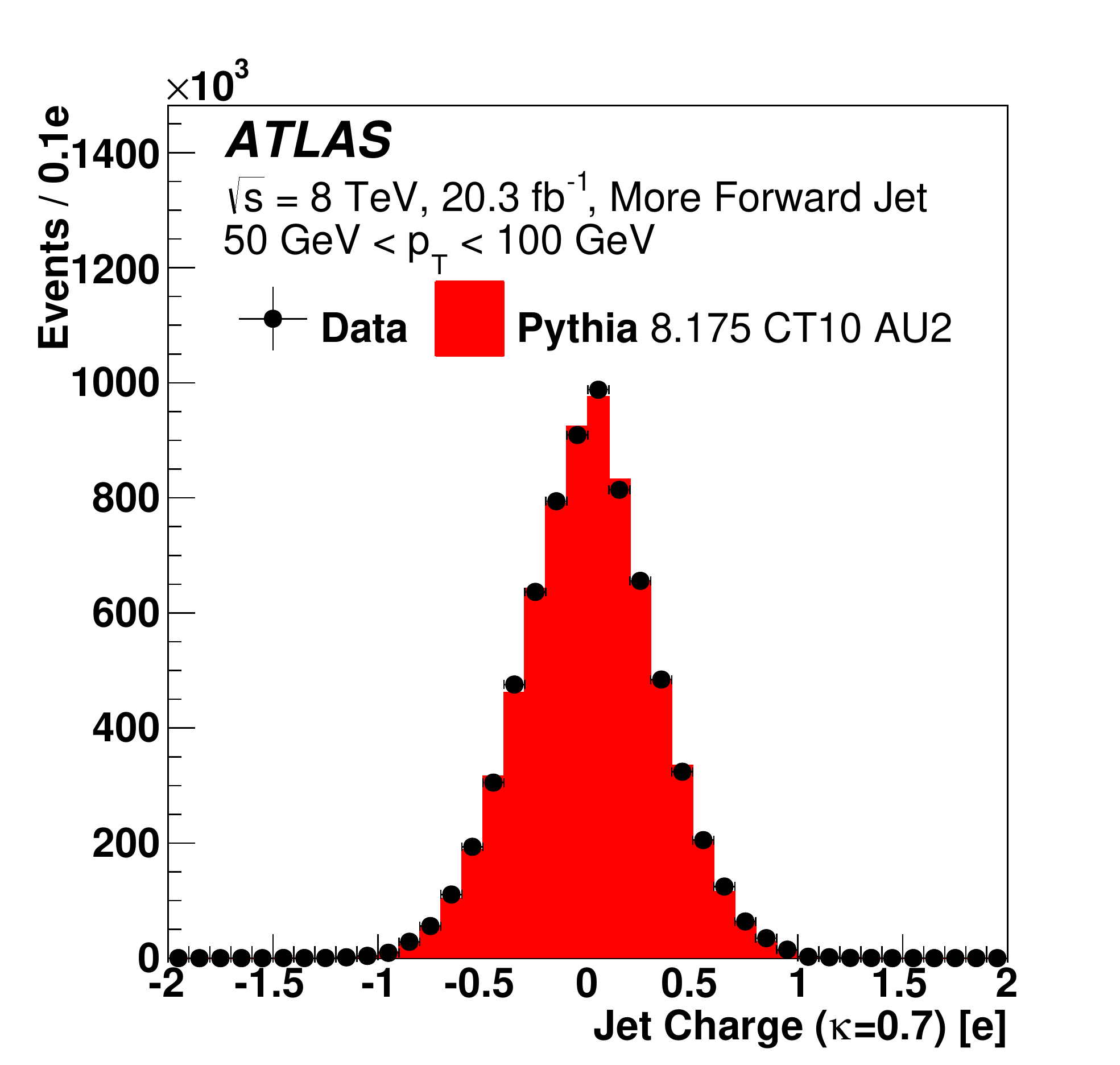}}\\
\subfloat[]{\includegraphics[width=0.45\textwidth]{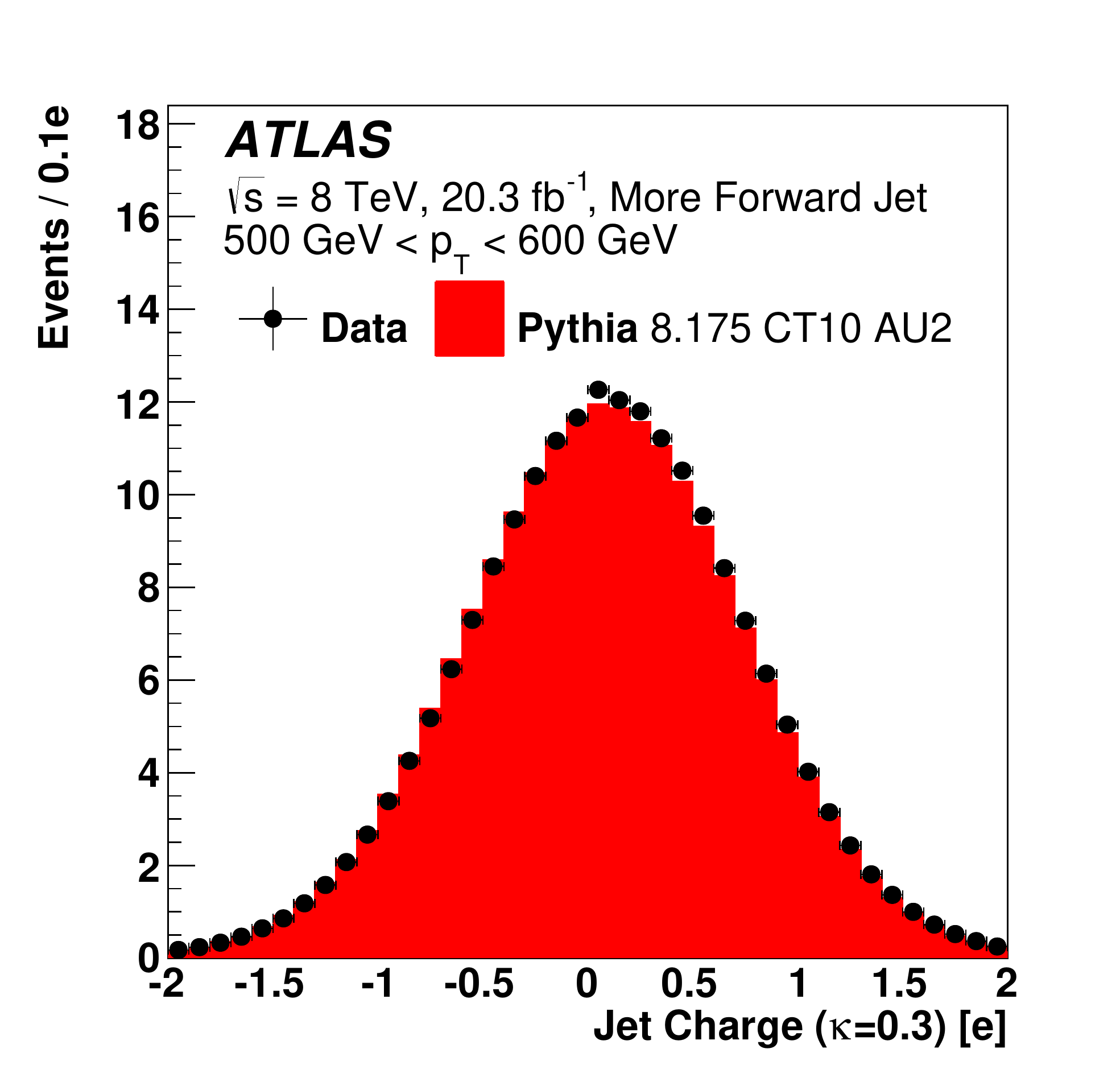}}\subfloat[]{\includegraphics[width=0.45\textwidth]{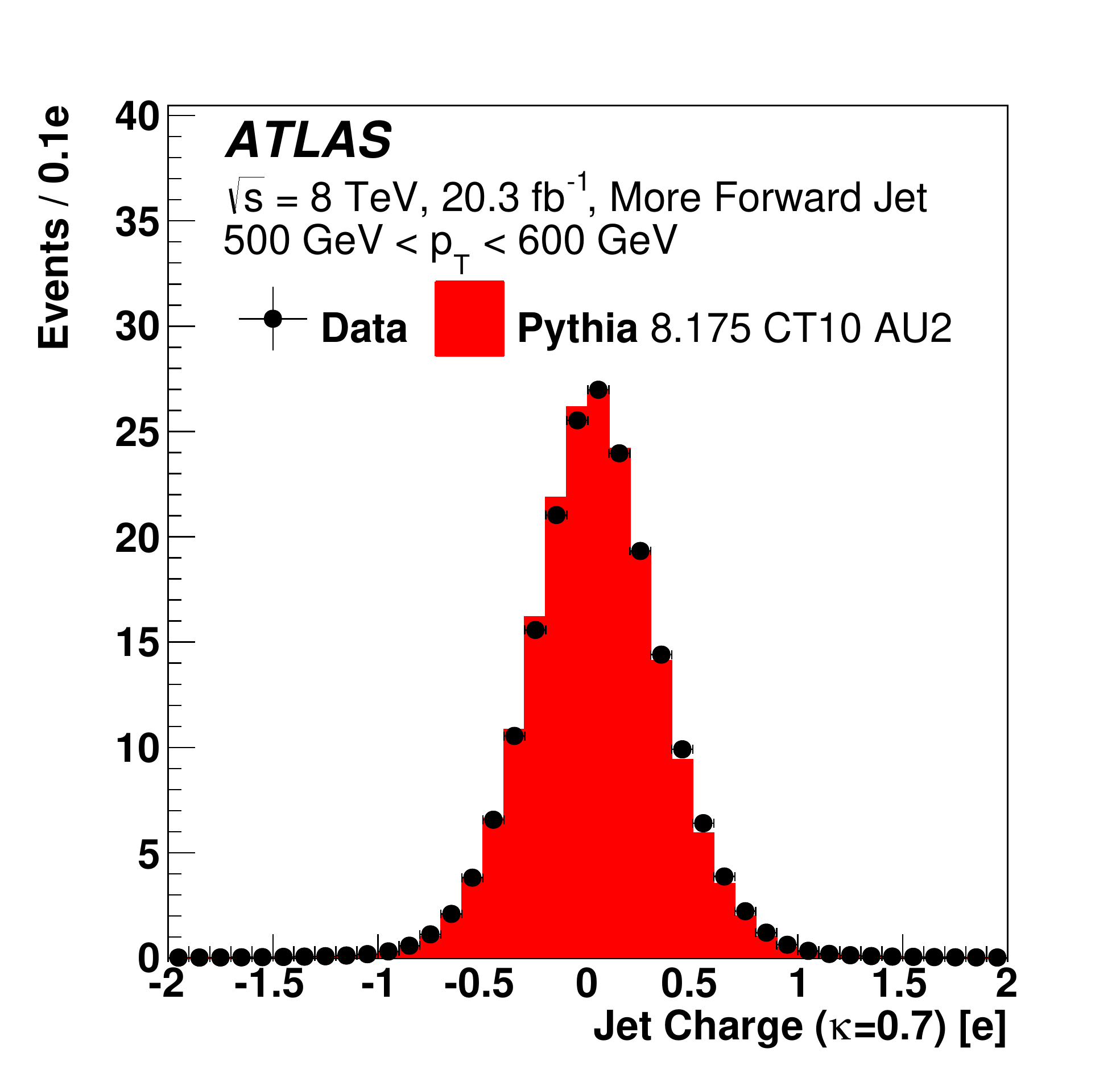}}
\end{center}	
\caption{The detector-level jet charge distributions for (a,b) 50 \GeV~$<p_\text{T}$ < 100 \GeV~and (c,d) 500 \GeV~$<p_\text{T}<600$ \GeV~for the more forward jet and (a,c) $\kappa=0.3$ and (b,d) $\kappa=0.7$.  The peak at zero in the top left plot is due to jets without any tracks.}
\label{fig:datamc}
\end{figure}

The reconstructed jet charge distributions for $\kappa=0.3$ and $\kappa=0.7$ are shown in Fig.~\ref{fig:datamc} for events passing the selection described in Sec.~\ref{sec:objects}.   Section~\ref{sec:unfolding} describes how the jet charge moments are corrected for detector resolution and acceptance effects through an unfolding procedure.  

\section{Unfolding}
\label{sec:unfolding}

\begin{figure}
\begin{center}
\subfloat[]{\includegraphics[width=0.5\textwidth]{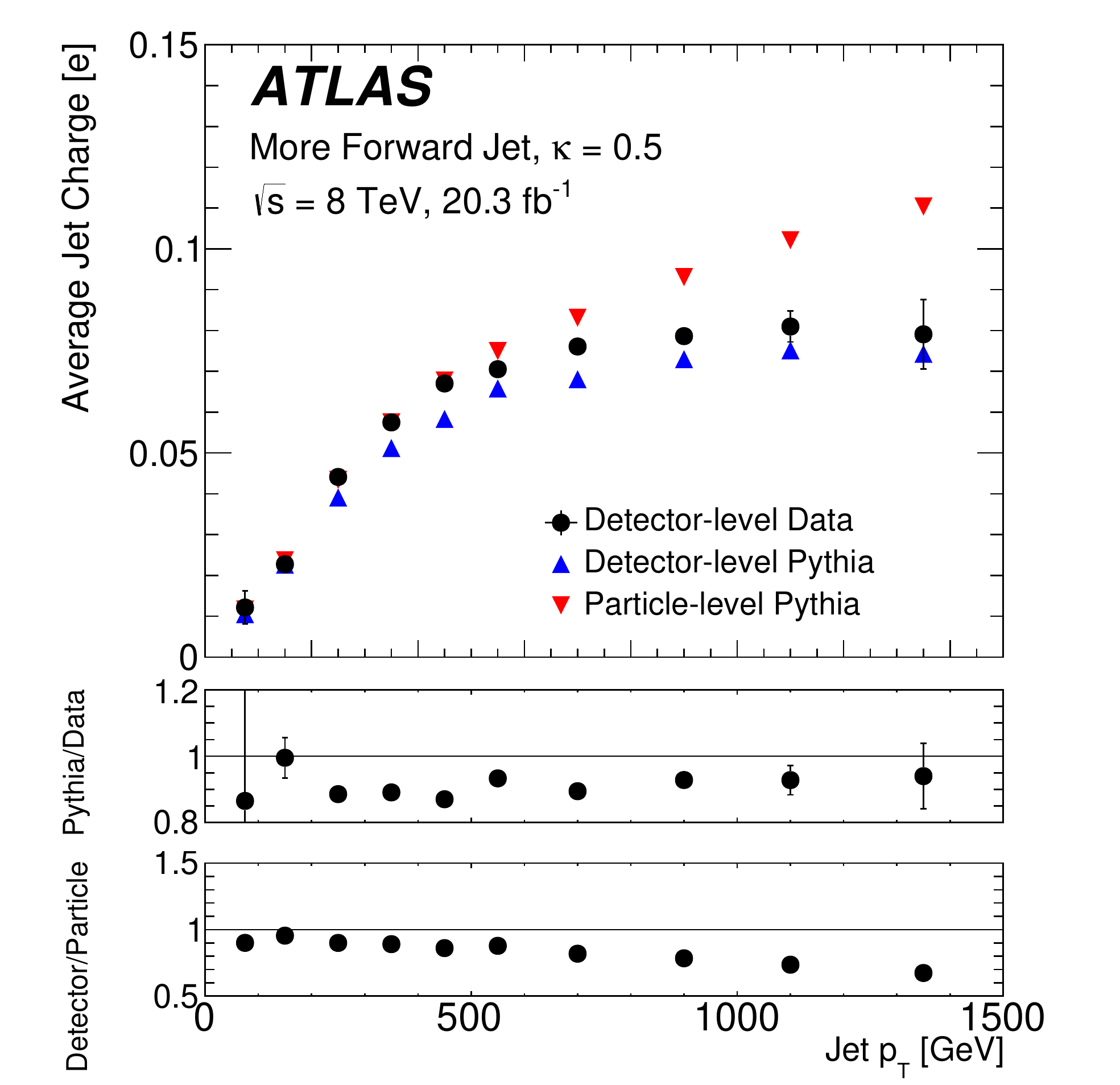}}\subfloat[]{\includegraphics[width=0.5\textwidth]{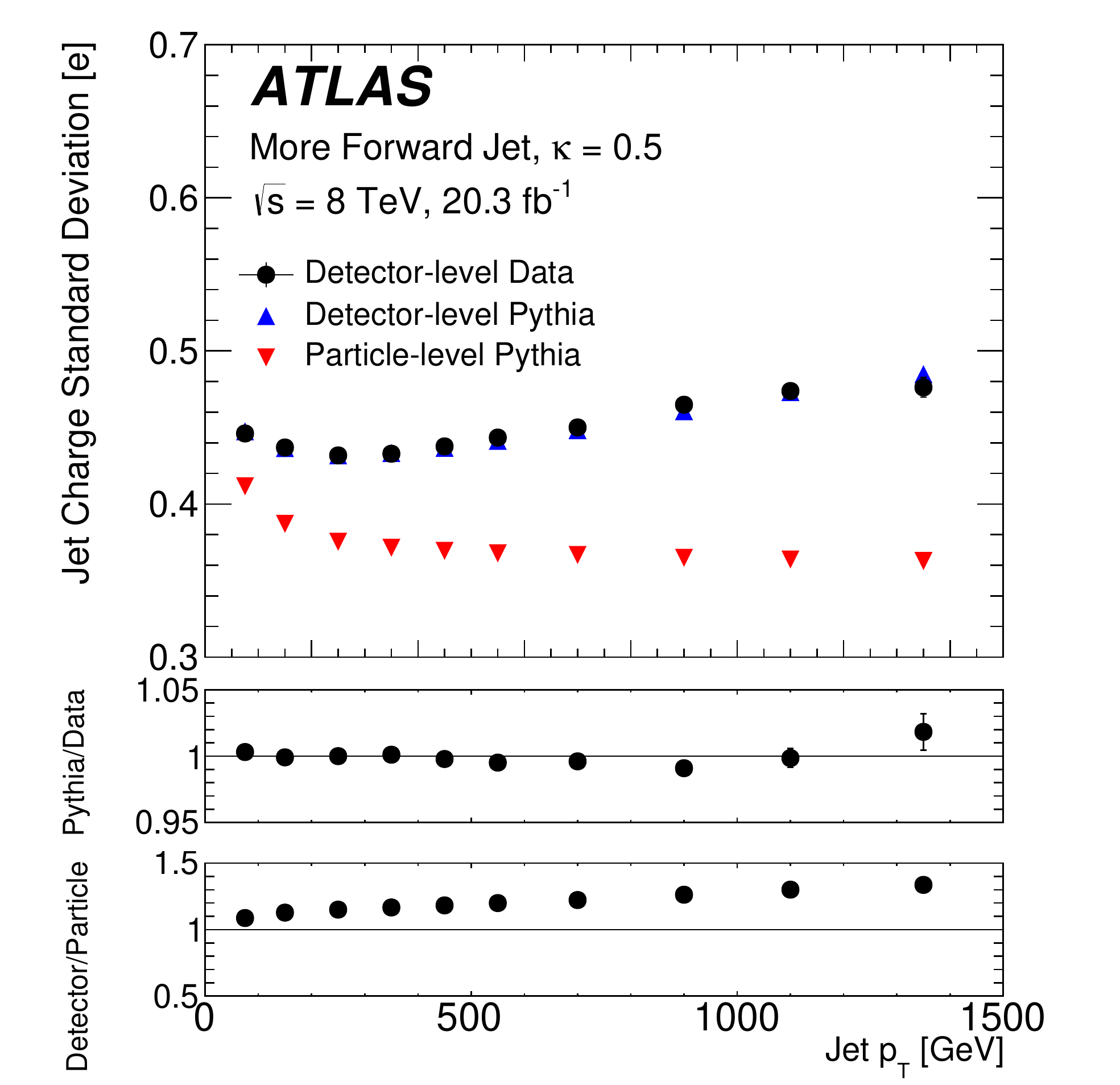}}
\end{center}	
\caption{The detector-level (data and simulation) and particle-level jet charge distribution's (a) average and (b) standard deviation as a function of the jet $p_\text{T}$ for the more forward jet.  The ratios in the bottom panel are constructed from the simulation, and show the prediction of detector-level {\sc Pythia} over the data (top ratio), and detector-level {\sc Pythia} over particle-level {\sc Pythia} (bottom ratio).   Bars on the data markers represent only the statistical uncertainties. For both (a) and (b), $\kappa=0.5$.}
\label{fig:raw}
\end{figure}

The particle-level jet charge distribution's mean and standard deviation are measured as a function of jet $p_\text{T}$.  This is accomplished by unfolding a discretized two-dimensional distribution of the jet charge and the jet $p_\text{T}$ and then computing the first two moments of the jet charge distribution in each bin of $p_\text{T}$.  The jet charge distribution is discretized into 15 bins in each of 10 bins of the jet $p_\text{T}$.  The jet charge mean is robust against the bias introduced from the discretization procedure, where fewer than five charge bins per $p_\text{T}$ bin is required to ensure negligible bias after recovering the mean from the discretized distribution.  However, the standard deviation of the jet charge distribution is sensitive to the discretization and requires about 15 charge bins\footnote{Fifteen is the number used in the unfolding and differs from the number shown for illustration in e.g. Fig.~\ref{fig:datamc}.} in order for the inherent bias due to discretization to be negligible. The jet charge spans the range $|Q_J|<1.8$ for $\kappa=0.3$, $|Q_J|<1.2$ for $\kappa=0.5$ and $|Q_J|<0.9$ for $\kappa=0.7$.  Events in the overflow of the jet charge distribution are placed in the first or last bins.  The $p_\text{T}$ binning is given by: [50,100), [100, 200), [200, 300), [300, 400), [400, 500), [500, 600), [600, 800), [800, 1000), [1000, 1200), and [1200, 1500] \GeV.  Figure~\ref{fig:raw} displays the $p_\text{T}$-dependence of the jet charge distribution's mean and standard deviation for detector-level data and simulation and for particle-level simulation.  The differences between the simulated detector- and particle-level distributions give a indication of the corrections required to account for detector acceptance and resolution effects in the unfolding procedure.  The growing difference between the particle- and detector-level average jet charge is due to the loss of charged-particle momentum inside jets as a result of track merging.  At particle level, the standard deviation of the jet charge distribution decreases with increasing $p_\text{T}$, but at detector level it increases with $p_\text{T}$ due to resolution effects.   

There is no unique way to extract the particle-level distribution of the jet charge and jet $p_\text{T}$ from the reconstructed distributions.  An iterative Bayesian (IB) technique~\cite{D'Agostini:1994zf}, implemented in the \texttt{RooUnfold} framework~\cite{Adye:2011gm}, is used to unfold the two-dimensional jet charge and jet $p_\text{T}$ distribution.   In the IB unfolding technique, the number of iterations and the prior distribution are the input parameters.  In the first step, the raw data are corrected using the simulation to account for events that pass the fiducial selection at detector level, but not the corresponding selection at particle level; this correction is the {\it fake factor}.  Then, the IB method iteratively applies Bayes' theorem using the \textit{response matrix} to connect the prior to posterior at each step, with the nominal {\sc Pythia} sample used for the initializing prior. The response matrix describes the bin migrations between the particle-level and detector-level two-dimensional jet charge and jet $p_\text{T}$ distributions.  While the response matrix is nearly diagonal, the resolution degrades at high $p_\text{T}$ where more bin-to-bin migrations from particle level to detector level occur.  The number of iterations in the IB method trades off unfolding bias with statistical fluctuations. An optimal value of four iterations is obtained by minimizing the bias when unfolding pseudo-data {\sc Herwig++} with {\sc Pythia} as a test of the methodology.  The last step of the unfolding applies another correction from simulation to the unfolded data to account for the differential rate of events passing the particle-level selection but not the detector-level selection; this correction is the {\it inefficiency factor}.  

\section{Systematic uncertainties}
\label{sec:uncerts}

All stages of the jet charge measurement are sensitive to sources of potential bias.  The three stages of the measurement are listed below, with an overview of the systematic uncertainties that impact the results at each stage:

\begin{description}
\item[Correction Factors:] Fake and inefficiency factors are derived from simulation to account for the fraction of events that pass either the detector-level or particle-level fiducial selection, but not both.  These factors are generally between $0.9$ and $1.0$ except in the first $p_\text{T}$ bin, where threshold effects introduce corrections that can be as large as 20\%.  Experimental uncertainties correlated with the detector-level selection acceptance, such as the jet energy scale uncertainty, result in uncertainties in these correction factors.  An additional source of uncertainty on the correction factors is due to the explicit dependence on the particle-level jet charge and jet $p_\text{T}$ spectra.  A comparison of particle-level models ({\sc Pythia} and {\sc Herwig++}) is used to estimate the impact on the correction factors.
\item[Response Matrix:] For events that pass both the detector-level and particle-level fiducial selections, the response matrix describes migrations between bins when moving between the detector level and the particle level.  The response matrix is taken from simulation and various experimental uncertainties on the jet charge and jet $p_\text{T}$ spectra result in uncertainties in the matrix.  Uncertainties can be divided into two classes: those impacting the calorimeter jet $p_\text{T}$ and those impacting track reconstruction inside jets.
\item[Unfolding Procedure:] A data-driven technique is used to estimate the potential bias from a given choice of prior and number of iterations in the IB method~\cite{Malaescu:2009dm}.  The particle-level spectrum is reweighted using the response matrix so that the simulated detector-level spectrum has significantly improved agreement with data.  The modified detector-level distribution is unfolded with the nominal response matrix and the difference between this and the reweighted particle-level spectrum is an indication of the bias due to the unfolding method (in particular, the choice of prior).
\end{description}

The following two subsections describe the impact of the detector-related sources of systematic uncertainty in more detail.  Uncertainties on the calorimeter jet $p_\text{T}$ are described in Sec.~\ref{sec:calojet} and the uncertainties related to tracking are described in Sec.~\ref{sec:tracking}.   Summaries of the systematic uncertainties for the more forward jet and $\kappa=0.5$ are found in Table~\ref{tab:systs_Mean_all_Forward_5aaa} and Table~\ref{tab:systs_RMS_all_Forward_5aaa} for the average jet charge and the jet charge distribution's standard deviation, respectively.\footnote{The uncertainties on the first $p_\text{T}$ bin of the average jet charge are much larger than on the other bins because the mean is small compared to the resolution.}  The uncertainties for the more central jet are similar.

{\renewcommand{\arraystretch}{1.3}
\begin{table}[h]
\begin{tabular}{cclclllllll}
 {\bf Average Jet Charge} & \multicolumn{10}{c}{Jet $p_\text{T}$ Range {[}100 GeV{]}} \\
\begin{tabular}[c]{@{}c@{}}Systematic\\ Uncertainty [\%]\end{tabular} & {[}0.5,1{]}          & {[}1,2{]}             & {[}2,3{]}            & {[}3,4{]}             & {[}4,5{]}             & {[}5,6{]}             & {[}6,8{]}             & {[}8,10{]}            & {[}10,12{]}           & {[}12,15{]}           \\ \hline
Total Jet Energy Scale & \multicolumn{1}{c}{${}^{+  8.4}_{- 13.6}$}
& \multicolumn{1}{c}{${}^{+  3.8}_{-  3.5}$}
& \multicolumn{1}{c}{${}^{+  0.9}_{-  5.0}$}
& \multicolumn{1}{c}{${}^{+  0.8}_{-  0.3}$}
& \multicolumn{1}{c}{${}^{+  1.1}_{-  1.6}$}
& \multicolumn{1}{c}{${}^{+  1.1}_{-  1.1}$}
& \multicolumn{1}{c}{${}^{+  0.7}_{-  1.0}$}
& \multicolumn{1}{c}{${}^{+  0.7}_{-  0.9}$}
& \multicolumn{1}{c}{${}^{+  0.4}_{-  0.7}$}
& \multicolumn{1}{c}{${}^{+  0.9}_{-  0.3}$}
\\
Jet Energy Resolution & \multicolumn{1}{c}{${}^{+  6.8}_{-  6.8}$}
& \multicolumn{1}{c}{${}^{+  2.3}_{-  2.3}$}
& \multicolumn{1}{c}{${}^{+  0.7}_{-  0.7}$}
& \multicolumn{1}{c}{${}^{+  0.7}_{-  0.7}$}
& \multicolumn{1}{c}{${}^{+  0.3}_{-  0.3}$}
& \multicolumn{1}{c}{${}^{+  0.3}_{-  0.3}$}
& \multicolumn{1}{c}{${}^{+  0.1}_{-  0.1}$}
& \multicolumn{1}{c}{${}^{+  0.1}_{-  0.1}$}
& \multicolumn{1}{c}{${}^{+  0.1}_{-  0.1}$}
& \multicolumn{1}{c}{${}^{+  0.3}_{-  0.3}$}
\\
Charged Energy Loss & \multicolumn{1}{c}{${}^{+  0.0}_{-  0.0}$}
& \multicolumn{1}{c}{${}^{+  0.0}_{-  0.0}$}
& \multicolumn{1}{c}{${}^{+  0.0}_{-  0.0}$}
& \multicolumn{1}{c}{${}^{+  0.0}_{-  0.0}$}
& \multicolumn{1}{c}{${}^{+  1.7}_{-  0.0}$}
& \multicolumn{1}{c}{${}^{+  1.5}_{-  0.0}$}
& \multicolumn{1}{c}{${}^{+  1.5}_{-  0.0}$}
& \multicolumn{1}{c}{${}^{+  1.5}_{-  0.0}$}
& \multicolumn{1}{c}{${}^{+  1.6}_{-  0.0}$}
& \multicolumn{1}{c}{${}^{+  3.6}_{-  0.0}$}
\\
Other Tracking & \multicolumn{1}{c}{${}^{+  3.3}_{-  1.6}$}
& \multicolumn{1}{c}{${}^{+  0.0}_{-  0.4}$}
& \multicolumn{1}{c}{${}^{+  0.9}_{-  0.2}$}
& \multicolumn{1}{c}{${}^{+  0.7}_{-  0.1}$}
& \multicolumn{1}{c}{${}^{+  0.5}_{-  0.4}$}
& \multicolumn{1}{c}{${}^{+  1.4}_{-  0.6}$}
& \multicolumn{1}{c}{${}^{+  0.7}_{-  0.9}$}
& \multicolumn{1}{c}{${}^{+  1.2}_{-  1.2}$}
& \multicolumn{1}{c}{${}^{+  1.1}_{-  1.3}$}
& \multicolumn{1}{c}{${}^{+  0.9}_{-  1.7}$}
\\
Track Multiplicity & \multicolumn{1}{c}{${}^{+  0.0}_{-  1.5}$}
& \multicolumn{1}{c}{${}^{+  0.1}_{-  0.0}$}
& \multicolumn{1}{c}{${}^{+  0.0}_{-  0.6}$}
& \multicolumn{1}{c}{${}^{+  0.0}_{-  1.1}$}
& \multicolumn{1}{c}{${}^{+  0.0}_{-  0.8}$}
& \multicolumn{1}{c}{${}^{+  0.0}_{-  0.6}$}
& \multicolumn{1}{c}{${}^{+  0.0}_{-  1.2}$}
& \multicolumn{1}{c}{${}^{+  0.0}_{-  1.4}$}
& \multicolumn{1}{c}{${}^{+  0.0}_{-  2.1}$}
& \multicolumn{1}{c}{${}^{+  0.0}_{-  2.9}$}
\\
Correction Factors & \multicolumn{1}{c}{${}^{+   23}_{-   23}$}
& \multicolumn{1}{c}{${}^{+  0.9}_{-  0.9}$}
& \multicolumn{1}{c}{${}^{+  0.8}_{-  0.8}$}
& \multicolumn{1}{c}{${}^{+  1.0}_{-  1.0}$}
& \multicolumn{1}{c}{${}^{+  0.3}_{-  0.3}$}
& \multicolumn{1}{c}{${}^{+  0.6}_{-  0.6}$}
& \multicolumn{1}{c}{${}^{+  0.1}_{-  0.1}$}
& \multicolumn{1}{c}{${}^{+  0.3}_{-  0.3}$}
& \multicolumn{1}{c}{${}^{+  0.2}_{-  0.2}$}
& \multicolumn{1}{c}{${}^{+  0.1}_{-  0.1}$}
\\
Unfolding Procedure& \multicolumn{1}{c}{${}^{+   28}_{-   28}$}
& \multicolumn{1}{c}{${}^{+  2.4}_{-  2.4}$}
& \multicolumn{1}{c}{${}^{+  0.3}_{-  0.3}$}
& \multicolumn{1}{c}{${}^{+  0.2}_{-  0.2}$}
& \multicolumn{1}{c}{${}^{+  0.2}_{-  0.2}$}
& \multicolumn{1}{c}{${}^{+  0.3}_{-  0.3}$}
& \multicolumn{1}{c}{${}^{+  1.1}_{-  1.1}$}
& \multicolumn{1}{c}{${}^{+  1.0}_{-  1.0}$}
& \multicolumn{1}{c}{${}^{+  1.6}_{-  1.6}$}
& \multicolumn{1}{c}{${}^{+  0.6}_{-  0.6}$}
\\
\hline
Total Systematic & \multicolumn{1}{c}{${}^{+   38}_{-   39}$}
& \multicolumn{1}{c}{${}^{+  5.1}_{-  4.9}$}
& \multicolumn{1}{c}{${}^{+  1.7}_{-  5.2}$}
& \multicolumn{1}{c}{${}^{+  1.6}_{-  1.7}$}
& \multicolumn{1}{c}{${}^{+  2.1}_{-  1.9}$}
& \multicolumn{1}{c}{${}^{+  2.4}_{-  1.6}$}
& \multicolumn{1}{c}{${}^{+  2.1}_{-  2.1}$}
& \multicolumn{1}{c}{${}^{+  2.3}_{-  2.3}$}
& \multicolumn{1}{c}{${}^{+  2.6}_{-  3.0}$}
& \multicolumn{1}{c}{${}^{+  3.8}_{-  3.4}$}
\\
Data Statistics & \multicolumn{1}{c}{   28}
& \multicolumn{1}{c}{  7.4}
& \multicolumn{1}{c}{  1.4}
& \multicolumn{1}{c}{  0.7}
& \multicolumn{1}{c}{  0.3}
& \multicolumn{1}{c}{  0.6}
& \multicolumn{1}{c}{  0.9}
& \multicolumn{1}{c}{  2.0}
& \multicolumn{1}{c}{  4.2}
& \multicolumn{1}{c}{  7.0}
\\
Total Uncertainty & \multicolumn{1}{c}{${}^{+   47}_{-   48}$}
& \multicolumn{1}{c}{${}^{+  9.0}_{-  8.9}$}
& \multicolumn{1}{c}{${}^{+  2.2}_{-  5.4}$}
& \multicolumn{1}{c}{${}^{+  1.8}_{-  1.9}$}
& \multicolumn{1}{c}{${}^{+  2.1}_{-  1.9}$}
& \multicolumn{1}{c}{${}^{+  2.5}_{-  1.7}$}
& \multicolumn{1}{c}{${}^{+  2.3}_{-  2.3}$}
& \multicolumn{1}{c}{${}^{+  3.0}_{-  3.0}$}
& \multicolumn{1}{c}{${}^{+  5.0}_{-  5.2}$}
& \multicolumn{1}{c}{${}^{+  8.0}_{-  7.8}$}
\\
\hline\hline
Measured Value [e] & \multicolumn{1}{c}{0.014}
& \multicolumn{1}{c}{0.024}
& \multicolumn{1}{c}{0.049}
& \multicolumn{1}{c}{0.065}
& \multicolumn{1}{c}{0.076}
& \multicolumn{1}{c}{0.082}
& \multicolumn{1}{c}{0.092}
& \multicolumn{1}{c}{0.100}
& \multicolumn{1}{c}{0.108}
& \multicolumn{1}{c}{0.115}
\\
\end{tabular}
\caption{A summary of all the systematic uncertainties and their impact on the mean jet charge for $\kappa=0.5$ and the more forward jet.   The correction factors are the fake and inefficiency corrections applied before/after the response matrix.  The Other Tracking category includes uncertainty on the track reconstruction efficiency, track momentum resolution, charge misidentification, and fake track rate.  All numbers are given in percent.}
\label{tab:systs_Mean_all_Forward_5aaa}
\end{table} 

{\renewcommand{\arraystretch}{1.3}
\begin{table}[h]
\begin{tabular}{cclclllllll}
 {\bf Standard Deviation} & \multicolumn{10}{c}{Jet $p_\text{T}$ Range {[}100 GeV{]}} \\
\begin{tabular}[c]{@{}c@{}}Systematic\\ Uncertainty [\%]\end{tabular} & {[}0.5,1{]}          & {[}1,2{]}             & {[}2,3{]}            & {[}3,4{]}             & {[}4,5{]}             & {[}5,6{]}             & {[}6,8{]}             & {[}8,10{]}            & {[}10,12{]}           & {[}12,15{]}           \\ \hline
Total Jet Energy Scale & \multicolumn{1}{c}{${}^{+  1.9}_{-  1.7}$}
& \multicolumn{1}{c}{${}^{+  1.5}_{-  1.3}$}
& \multicolumn{1}{c}{${}^{+  1.1}_{-  1.1}$}
& \multicolumn{1}{c}{${}^{+  1.1}_{-  1.0}$}
& \multicolumn{1}{c}{${}^{+  0.9}_{-  0.8}$}
& \multicolumn{1}{c}{${}^{+  1.0}_{-  0.7}$}
& \multicolumn{1}{c}{${}^{+  0.8}_{-  0.8}$}
& \multicolumn{1}{c}{${}^{+  0.7}_{-  0.8}$}
& \multicolumn{1}{c}{${}^{+  0.5}_{-  0.5}$}
& \multicolumn{1}{c}{${}^{+  0.5}_{-  0.5}$}
\\
Jet Energy Resolution & \multicolumn{1}{c}{${}^{+  1.3}_{-  1.3}$}
& \multicolumn{1}{c}{${}^{+  0.3}_{-  0.3}$}
& \multicolumn{1}{c}{${}^{+  0.1}_{-  0.1}$}
& \multicolumn{1}{c}{${}^{+  0.2}_{-  0.2}$}
& \multicolumn{1}{c}{${}^{+  0.3}_{-  0.3}$}
& \multicolumn{1}{c}{${}^{+  0.4}_{-  0.4}$}
& \multicolumn{1}{c}{${}^{+  0.2}_{-  0.2}$}
& \multicolumn{1}{c}{${}^{+  0.2}_{-  0.2}$}
& \multicolumn{1}{c}{${}^{+  0.2}_{-  0.2}$}
& \multicolumn{1}{c}{${}^{+  0.2}_{-  0.2}$}
\\
Charged Energy Loss & \multicolumn{1}{c}{${}^{+  0.0}_{-  0.0}$}
& \multicolumn{1}{c}{${}^{+  0.0}_{-  0.0}$}
& \multicolumn{1}{c}{${}^{+  0.0}_{-  0.0}$}
& \multicolumn{1}{c}{${}^{+  0.0}_{-  0.0}$}
& \multicolumn{1}{c}{${}^{+  0.2}_{-  0.0}$}
& \multicolumn{1}{c}{${}^{+  0.3}_{-  0.0}$}
& \multicolumn{1}{c}{${}^{+  0.3}_{-  0.0}$}
& \multicolumn{1}{c}{${}^{+  0.3}_{-  0.0}$}
& \multicolumn{1}{c}{${}^{+  0.4}_{-  0.0}$}
& \multicolumn{1}{c}{${}^{+  1.1}_{-  0.0}$}
\\
Other Tracking & \multicolumn{1}{c}{${}^{+  0.0}_{-  0.3}$}
& \multicolumn{1}{c}{${}^{+  0.1}_{-  0.3}$}
& \multicolumn{1}{c}{${}^{+  0.2}_{-  0.4}$}
& \multicolumn{1}{c}{${}^{+  0.3}_{-  0.4}$}
& \multicolumn{1}{c}{${}^{+  0.4}_{-  0.5}$}
& \multicolumn{1}{c}{${}^{+  0.5}_{-  0.4}$}
& \multicolumn{1}{c}{${}^{+  0.5}_{-  0.5}$}
& \multicolumn{1}{c}{${}^{+  0.5}_{-  0.5}$}
& \multicolumn{1}{c}{${}^{+  0.5}_{-  0.4}$}
& \multicolumn{1}{c}{${}^{+  0.4}_{-  0.4}$}
\\
Track Multiplicity & \multicolumn{1}{c}{${}^{+  0.0}_{-  0.2}$}
& \multicolumn{1}{c}{${}^{+  0.0}_{-  0.3}$}
& \multicolumn{1}{c}{${}^{+  0.0}_{-  0.2}$}
& \multicolumn{1}{c}{${}^{+  0.0}_{-  0.1}$}
& \multicolumn{1}{c}{${}^{+  0.0}_{-  0.0}$}
& \multicolumn{1}{c}{${}^{+  0.1}_{-  0.0}$}
& \multicolumn{1}{c}{${}^{+  0.2}_{-  0.0}$}
& \multicolumn{1}{c}{${}^{+  0.2}_{-  0.0}$}
& \multicolumn{1}{c}{${}^{+  0.3}_{-  0.0}$}
& \multicolumn{1}{c}{${}^{+  0.2}_{-  0.0}$}
\\
Correction Factors & \multicolumn{1}{c}{${}^{+  0.9}_{-  0.9}$}
& \multicolumn{1}{c}{${}^{+  0.1}_{-  0.1}$}
& \multicolumn{1}{c}{${}^{+  0.0}_{-  0.0}$}
& \multicolumn{1}{c}{${}^{+  0.1}_{-  0.1}$}
& \multicolumn{1}{c}{${}^{+  0.0}_{-  0.0}$}
& \multicolumn{1}{c}{${}^{+  0.1}_{-  0.1}$}
& \multicolumn{1}{c}{${}^{+  0.0}_{-  0.0}$}
& \multicolumn{1}{c}{${}^{+  0.0}_{-  0.0}$}
& \multicolumn{1}{c}{${}^{+  0.0}_{-  0.0}$}
& \multicolumn{1}{c}{${}^{+  0.0}_{-  0.0}$}
\\
Unfolding Procedure& \multicolumn{1}{c}{${}^{+  1.9}_{-  1.9}$}
& \multicolumn{1}{c}{${}^{+  0.4}_{-  0.4}$}
& \multicolumn{1}{c}{${}^{+  0.0}_{-  0.0}$}
& \multicolumn{1}{c}{${}^{+  0.1}_{-  0.1}$}
& \multicolumn{1}{c}{${}^{+  0.2}_{-  0.2}$}
& \multicolumn{1}{c}{${}^{+  0.0}_{-  0.0}$}
& \multicolumn{1}{c}{${}^{+  0.1}_{-  0.1}$}
& \multicolumn{1}{c}{${}^{+  0.3}_{-  0.3}$}
& \multicolumn{1}{c}{${}^{+  0.4}_{-  0.4}$}
& \multicolumn{1}{c}{${}^{+  1.7}_{-  1.7}$}
\\
\hline
Total Systematic & \multicolumn{1}{c}{${}^{+  3.1}_{-  3.0}$}
& \multicolumn{1}{c}{${}^{+  1.6}_{-  1.5}$}
& \multicolumn{1}{c}{${}^{+  1.1}_{-  1.2}$}
& \multicolumn{1}{c}{${}^{+  1.2}_{-  1.1}$}
& \multicolumn{1}{c}{${}^{+  1.1}_{-  1.0}$}
& \multicolumn{1}{c}{${}^{+  1.2}_{-  0.9}$}
& \multicolumn{1}{c}{${}^{+  1.0}_{-  0.9}$}
& \multicolumn{1}{c}{${}^{+  1.0}_{-  1.0}$}
& \multicolumn{1}{c}{${}^{+  1.0}_{-  0.8}$}
& \multicolumn{1}{c}{${}^{+  2.1}_{-  1.8}$}
\\
Data Statistics & \multicolumn{1}{c}{  0.9}
& \multicolumn{1}{c}{  0.3}
& \multicolumn{1}{c}{  0.1}
& \multicolumn{1}{c}{  0.1}
& \multicolumn{1}{c}{  0.0}
& \multicolumn{1}{c}{  0.1}
& \multicolumn{1}{c}{  0.1}
& \multicolumn{1}{c}{  0.3}
& \multicolumn{1}{c}{  0.6}
& \multicolumn{1}{c}{  1.0}
\\
Total Uncertainty & \multicolumn{1}{c}{${}^{+  3.2}_{-  3.1}$}
& \multicolumn{1}{c}{${}^{+  1.6}_{-  1.5}$}
& \multicolumn{1}{c}{${}^{+  1.1}_{-  1.2}$}
& \multicolumn{1}{c}{${}^{+  1.2}_{-  1.1}$}
& \multicolumn{1}{c}{${}^{+  1.1}_{-  1.0}$}
& \multicolumn{1}{c}{${}^{+  1.2}_{-  0.9}$}
& \multicolumn{1}{c}{${}^{+  1.0}_{-  1.0}$}
& \multicolumn{1}{c}{${}^{+  1.1}_{-  1.0}$}
& \multicolumn{1}{c}{${}^{+  1.2}_{-  1.0}$}
& \multicolumn{1}{c}{${}^{+  2.4}_{-  2.1}$}
\\
\hline\hline
Measured Value [e] & \multicolumn{1}{c}{0.410}
& \multicolumn{1}{c}{0.387}
& \multicolumn{1}{c}{0.375}
& \multicolumn{1}{c}{0.372}
& \multicolumn{1}{c}{0.370}
& \multicolumn{1}{c}{0.369}
& \multicolumn{1}{c}{0.368}
& \multicolumn{1}{c}{0.367}
& \multicolumn{1}{c}{0.362}
& \multicolumn{1}{c}{0.355}
\\
\end{tabular}
\caption{A summary of all the systematic uncertainties and their impact on the jet charge distribution's standard deviation for $\kappa=0.5$ and the more forward jet.   The correction factors are the fake and inefficiency corrections applied before/after the response matrix.  The Other Tracking category includes uncertainty on the track reconstruction efficiency, track momentum resolution, charge misidentification, and fake track rate.  All numbers are given in percent.}
\label{tab:systs_RMS_all_Forward_5aaa}
\end{table} 

\subsection{Calorimeter jet uncertainties}
\label{sec:calojet}

Jets are calibrated so that the detector-level $p_\text{T}$ is an unbiased measurement of the particle-level jet $p_\text{T}$ and various data-driven techniques are used to derive {\it in situ} estimates of the difference in this calibration between the data and the simulation.  Uncertainties in the energy scale and resolution of calibrated jets impact the jet charge in the normalization of Eq.~(\ref{chargedef}) (but preserve the jet charge sign) as well as the binning for the 2D distribution.  Complete details of this source of uncertainty can be found in Ref.~\cite{Aad:2014bia}.  There are many components of the jet energy scale uncertainty.  The {\it in situ} correction is derived from data using the momentum balance in events with $Z$ bosons (low $p_\text{T}$) or photons (moderate $p_\text{T}$) produced in association with jets as well as the balance of multijet (high $p_\text{T}$) and dijet (high $|\eta|$) systems.  Uncertainties on this method stem from the modeling of these processes in simulation.  There is also a contribution from the response to single hadrons~\cite{Aad:2012vm}.  Additional sources of uncertainty are due to the modeling of the in-time and out-of-time pileup corrections to the jet energy scale as well as differences in the response due to the flavor of the jet.  To assess the impact of each component of the jet energy scale uncertainty, the jet energies in simulation are shifted according to the $p_\text{T}$- and $\eta$-dependent $\pm 1\sigma$ variations.  For a fixed variation, the response matrix, and fake and inefficiency factors are recomputed and the unfolding procedure is repeated.  The resulting uncertainty on the jet charge distribution's mean and standard deviation is about 1\% or less for jet $p_\text{T}$ above $200$ \GeV.  The jet energy resolution uncertainty is derived using data-driven techniques in dijet events~\cite{Aad:2012ag}.  To assess the impact of a slightly larger jet energy resolution, jet energies are smeared according to $p_\text{T}$- and $\eta$-dependent factors and propagated through the entire unfolding procedure, as for the jet energy scale uncertainty.  The jet energy resolution uncertainty is subdominant to the jet energy scale uncertainty.

\subsection{Tracking uncertainties}
\label{sec:tracking}

Uncertainties on tracking are broken down into contributions related to (i) the efficiency of reconstructing tracks and (ii) measurements of those tracks that are successfully reconstructed.  The uncertainty on the inclusive track reconstruction efficiency is dominated by the uncertainty in the material in the ID.  The amount of material is known to within $\sim 5\%$~\cite{Aad:2011cxa}.  Simulated detector geometries with various levels of material in the ID within the measured uncertainties are used to estimate the track reconstruction efficiency uncertainty.  These uncertainties are $\eta$- and $p_\text{T}$-dependent, ranging from $\lesssim 1\%$ for $|\eta|<2.1$ to $\lesssim 4\%$ for $2.1\leq|\eta|<2.3$ and $\lesssim 7\%$ for $2.3\leq|\eta|<2.5$.  The impact of the uncertainty is estimated by randomly removing tracks within the $p_\text{T}$- and $\eta$-dependent probabilities, leading to a $\lesssim 0.5\%$ uncertainty on the jet charge distribution's mean and standard deviation.  An additional uncertainty accounts for the difference in efficiency between data and simulation due to the modeling of the track $\chi^2$ per number of degrees of freedom (NDF) requirement.  A requirement of $\chi^2/\text{NDF}<3$ is more than 99\% efficient across jet and track $p_\text{T}$, but the efficiency is generally higher in simulation than in data.  The difference in the efficiency between data and simulation is $\lesssim 10\%$ of the inefficiency.  The impact of this mismodeling is evaluated by independently removing tracks with a probability that is 10\% of the $\chi^2/\text{NDF}<3$ requirement inefficiency.  As a result of this procedure, the jet charge distribution's mean and standard deviation change by $\lesssim 0.1\%$ in most $p_\text{T}$ bins.

In addition to the loss of tracks due to the material in the ID, tracks can be lost due to the busy environment inside the cores of jets.  This loss can be studied in simulation by comparing the reconstructed charged particle momentum with the charged particle momentum inside the corresponding particle-level jet.  In order to remove the impact of the tracking resolution and the contribution from fake tracks already accounted for separately, reconstructed tracks in the simulation are matched with charged particles.  The matching is performed by considering the energy deposited in the various layers of the ID by charged particles due to material interactions modeled with GEANT4.  Weights are assigned to charged particles based on the energy deposited in detector elements that were used to reconstruct a given track.  A match is declared if the weight for one charged particle is sufficiently high.  Figure~\ref{fig:tide}(a) shows the ratio in simulation of the sum of the $p_\text{T}$ of charged particles that were matched to reconstructed tracks to the sum of the $p_\text{T}$ of all the charged particles as a function of the jet $p_\text{T}$.  At low $p_\text{T}$, the ratio increases with $p_\text{T}$ due to losses as a result of hadronic interactions with the material in the ID.   Beyond about 200 \GeV, the fraction monotonically decreases due to the loss of tracks in the core of the jet.  A related quantity is $\langle \Sigma \text{ track $p_\text{T}$ }/\text{ Jet $p_\text{T}$}\rangle$, where the denominator is the reconstructed calorimeter jet $p_\text{T}$ and the numerator is a sum over tracks associated with the jet.  Since the particle-level charged-to-neutral fraction of the energy is independent of $p_\text{T}$, a degradation in this ratio can provide information about the loss of tracks inside the core of a jet in data.  Figure~\ref{fig:tide}(b) shows the distribution of $\langle \Sigma \text{ track $p_\text{T}$ }/\text{ Jet $p_\text{T}$}\rangle$ as a function of jet $p_\text{T}$.  It exhibits trends very similar to those in Fig.~\ref{fig:tide}(a), and in fact the relative loss (fraction with respect to the peak) is similar.  The MC underestimates the loss by $\lesssim 1\%$.  The impact of the charged-particle momentum loss inside the cores of jets is estimated by randomly removing tracks with a $p_\text{T}$-dependent probability such that the relative loss in the simulation matches that in the data.  This uncertainty is negligible for jets with $p_\text{T}<400$ \GeV, but is non-negligible for higher-$p_\text{T}$ jets, resulting in a $\lesssim 4\%$ uncertainty on the average jet charge in the highest $p_\text{T}$ bin.

\begin{figure}
\begin{center}
\subfloat[]{\includegraphics[width=0.5\textwidth]{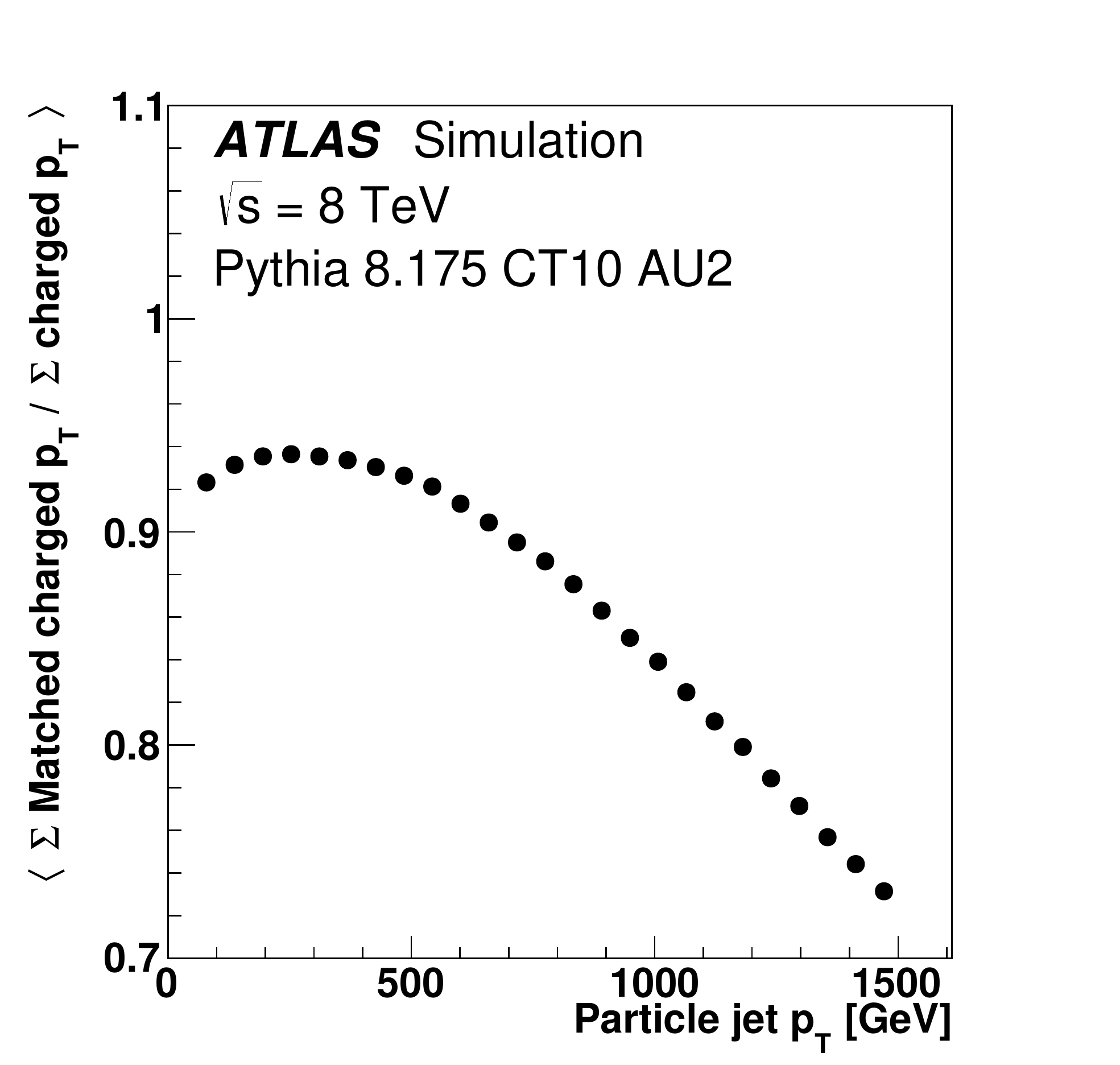}}\subfloat[]{\includegraphics[width=0.5\textwidth]{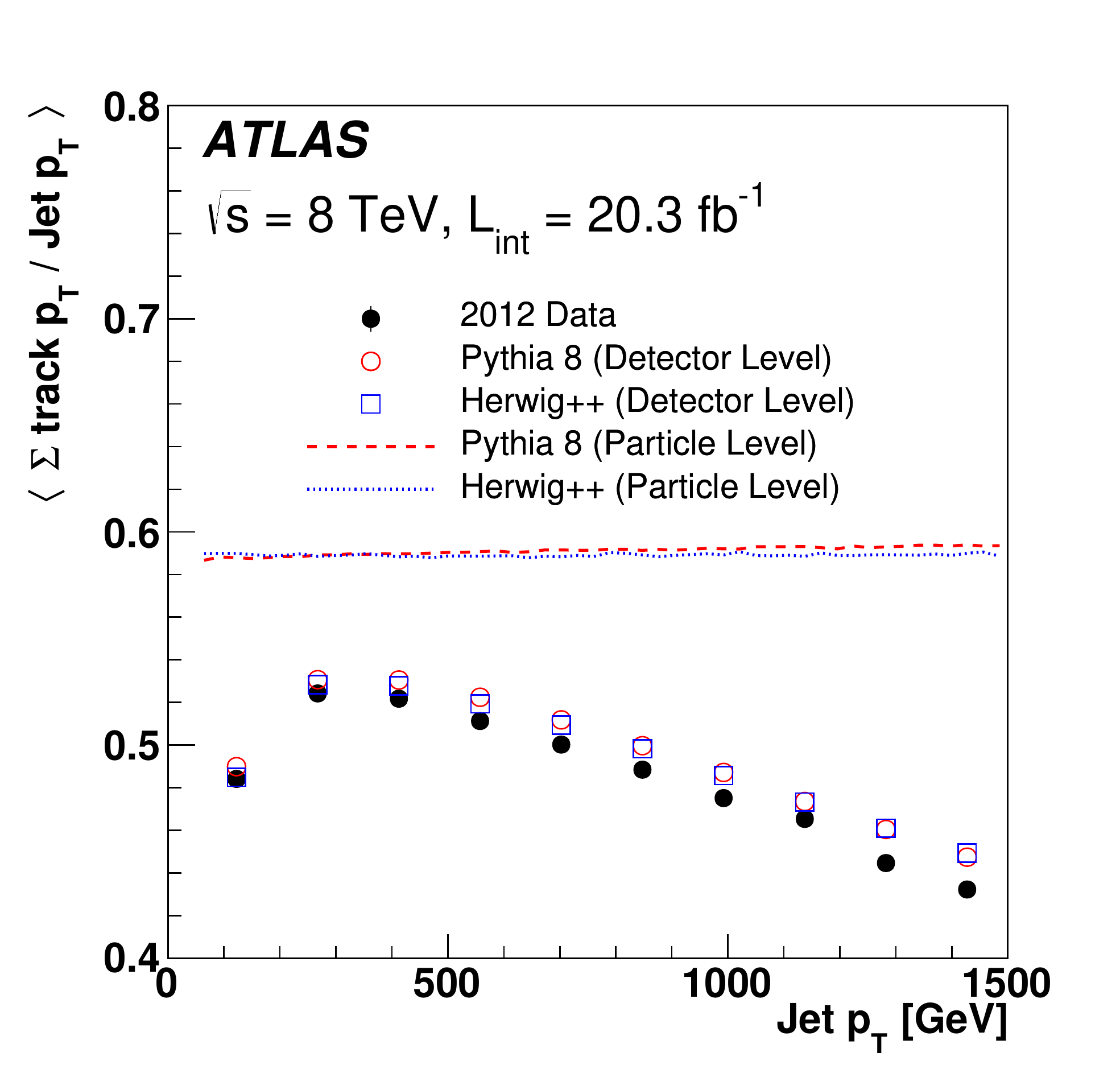}}
\end{center}	
\caption{(a): The average of the $p_\text{T}$-weighted ratio of charged particles that were matched to reconstructed tracks ($\Sigma\text{ Matched charged $p_\text{T}$}$) to all the charged particles ($\Sigma\text{ charged $p_\text{T}$}$) as a function of the particle-level jet $p_\text{T}$.  (b): The average of the ratio of the sum of $p_\text{T}$ from tracks to the calorimeter jet $p_\text{T}$ ($\Sigma \text{ track $p_\text{T}$ }/\text{ Jet $p_\text{T}$}$) as a function of jet $p_\text{T}$ in both data and simulation.  The momentum ratio of charged particles to all particles is nearly $2/3$ due to the number of pion species (as indicated by the straight lines for {\sc Herwig++} and {\sc Pythia} predictions at particle level), but is not exactly 2/3 due to the presence of photons and kaons in the jet.}
\label{fig:tide}
\end{figure}

The momentum resolution of isolated tracks has been well measured in $J/\psi\rightarrow\mu\mu$ and $Z\rightarrow\mu\mu$ events~\cite{Aad:2014rra}.  The scale and resolution of reconstructed muon candidates are shifted and smeared in the MC simulation to account for differences between the data and the simulation for $m_{\mu\mu}$.  Generic tracks are not corrected in the same way as muon candidates reconstructed from the ID (and ignoring the muon spectrometer).  Thus, the correction factors are taken here as the systematic uncertainty on the momentum resolution.  The momentum resolution is parameterized as a sum in quadrature of a $p_\text{T}^{-1}$-dependent term, a constant term, and a term linear in the track $p_\text{T}$, with coefficients $r_0$, $r_1$, and $r_2$, respectively.  The first term accounts for fluctuations in the energy loss in the detector material, the second term captures effects due to multiple scattering, and the third term accounts for the intrinsic resolution caused by misalignment and the finite spatial resolution of ID hits.  Unlike muon spectrometer tracks, ID tracks do not traverse a significant amount of material and so the energy-loss coefficient $r_0$ and its uncertainty are neglected.  The uncertainties on $r_1$, $r_2$ and the momentum scale $s$ are estimated by randomly smearing the $p_\text{T}$ of every track with $|\eta|$-dependent factors that are $\lesssim 2\%$ for track $p_\text{T}<100$ \GeV~and increase to 10--20\% at 1 \TeV~depending on $|\eta|$.  Propagating these variations through the unfolding procedure results in uncertainties that are subdominant to other uncertainties, but non-negligible ($\sim 2\%$) in the highest $p_\text{T}$ bins for the average jet charge.

Aside from the track $p_\text{T}$, the other track parameter that is relevant for the jet charge is the track charge.  Especially at high $p_\text{T}$ when the tracks are very straight, the probability for misidentifying the track charge can become non-negligible.  Simulated particles matched to generator-level information, as described above, are used to study the charge flipping probability (charge misidentification rate) for nonfake tracks originating from charged particles inside jets (mostly pions).  The rate predicted from the simulation is $<0.1\%$ for track $p_\text{T}<100$ \GeV, $0.5\%$ for $100$ \GeV~$\leq p_\text{T}<200$ \GeV,  $1\%$ for $200$ \GeV~$\leq p_\text{T}<300$ \GeV, $2\%$ for $300$ \GeV~$\leq p_\text{T}<400$ \GeV~and $4\%$ for $p_\text{T}\geq 400$ \GeV.  Dedicated studies of charge flipping in searches for same-sign leptons~\cite{ATLAS:2014kca} suggest that the mismodeling of the charge-flipping rate is (much) less than 50\%. Therefore, the impact of charge flipping on the jet charge measurement is conservatively estimated by randomly flipping the charge of tracks at 50\% of the charge misidentification rate.  The impact on the measured jet charge mean and standard deviation is negligible.

Random combinations of hits in the detector can be combined to form a reconstructed track.  Tracks resulting in particular from multiparticle trajectories that have kinks can result in very large reconstructed track $p_\text{T}$.   The quality criteria are effective at mitigating the presence of fake tracks, which constitute less than $\lesssim 0.1\%$ of all reconstructed tracks.  To determine the impact of fake tracks on the jet charge, fake tracks are randomly removed with a probability that is $50\%$ of the rate in simulation.   This results in a negligibly small uncertainty on the mean jet charge and a $\lesssim 0.5\%$ uncertainty on the standard deviation of the jet charge distribution.

The tracking uncertainties described so far take into account the resolution and efficiency of reconstruction of charged-particle momenta.  One last source of systematic uncertainty is the number of charged particles.  The unfolding procedure uncertainty takes into account the uncertainty on the prior due to the charged-particle multiplicity, but the jet charge resolution also changes with the charged-particle multiplicity.  To assess the impact on the response matrix of the mismodeled charged-particle multiplicity, the distribution of $n_\text{track}$ is reweighted in the simulation per $p_\text{T}$ bin and the relative difference when unfolding the nominal {\sc Pythia} distribution with the reweighted {\sc Pythia} distribution is taken as a systematic uncertainty.\footnote{Since the prior is also changed, this uncertainty at least partially includes the unfolding procedure uncertainty.}  This uncertainty is subdominant for the standard deviation across $p_\text{T}$ and for the mean at low to moderate jet $p_\text{T}$.  For the mean jet charge, the largest uncertainty is with the smallest $\kappa$ and for large $p_\text{T}$, where it is 3--4\% percent in the highest $p_\text{T}$ bin for $\kappa=0.3$ and $\kappa=0.5$.

\section{Results}
\label{sec:results}

The data satisfying the event selection criteria described in Sec.~\ref{sec:objects} are unfolded according to the procedure in Sec.~\ref{sec:unfolding} and the average and standard deviation of the jet charge distribution are computed as a function of the jet $p_\text{T}$.  These results, along with the systematic uncertainties detailed in Sec.~\ref{sec:uncerts}, are discussed in Sec.~\ref{sec:rawunfolded}.  The PDF uncertainty and jet formation uncertainties in the theory predictions are compared to the unfolded data in Secs.~\ref{sec:PDFsensitivity} and~\ref{sec:NPmodeling}, respectively.  Using PDF information as input, the average charge per jet flavor is extracted in Sec.~\ref{sec:updownextract} and its $p_\text{T}$-dependence is studied in Sec.~\ref{sec:scaleviolation}.

\subsection{Unfolded Jet Charge Spectrum}
\label{sec:rawunfolded}

The unfolded jet charge mean is shown as a function of the jet $p_\text{T}$ in the top plots of Fig.~\ref{fig:mean} for $\kappa=0.3$, $0.5$ and $0.7$.  The average charge increases with jet $p_\text{T}$ due to the increase in up-flavor jets from PDF effects.  The average charge increases from $0.01e$ at $p_\text{T}\sim 100$ \GeV~to $0.15e$ at $p_\text{T}\sim 1.5$ \TeV.  Systematic uncertainties are generally a few percent, except at low jet $p_\text{T}$ where the fractional uncertainty is large because the average jet charge in the denominator is small, and at high $p_\text{T}$ where the tracking uncertainties are not negligible.  The statistical uncertainty is estimated by repeating the measurement on an ensemble of bootstrapped datasets: each event is used in each pseudodataset $n$ times, where $n$ is a random number distributed according to a Poission distribution with mean one.  The first bin suffers from large statistical uncertainties (up to 170\%), but for the higher $p_\text{T}$ bins the systematic uncertainty is dominant, except at the highest $p_\text{T}$ bin where statistical and systematic uncertainty are of similar size (about 7\%).  The jet charge distributions of the more forward and more central jet differ in shape, in particular at low $p_\text{T}$, due to the different shape of the up/down flavor fractions in those bins as shown in Fig.~\ref{fig:flavorfrac}(b).

Analogous results for the standard deviation of the jet charge distribution are shown in the bottom plots of Fig.~\ref{fig:mean}.   Even though the standard deviation of the reconstructed jet charge distribution increases with jet $p_\text{T}$ (Fig.~\ref{fig:raw}), the particle-level value decreases and approaches an asymptote for $p_\text{T}\gtrsim 300$ \GeV.

\begin{figure}
\begin{center}
\includegraphics[width=0.45\textwidth]{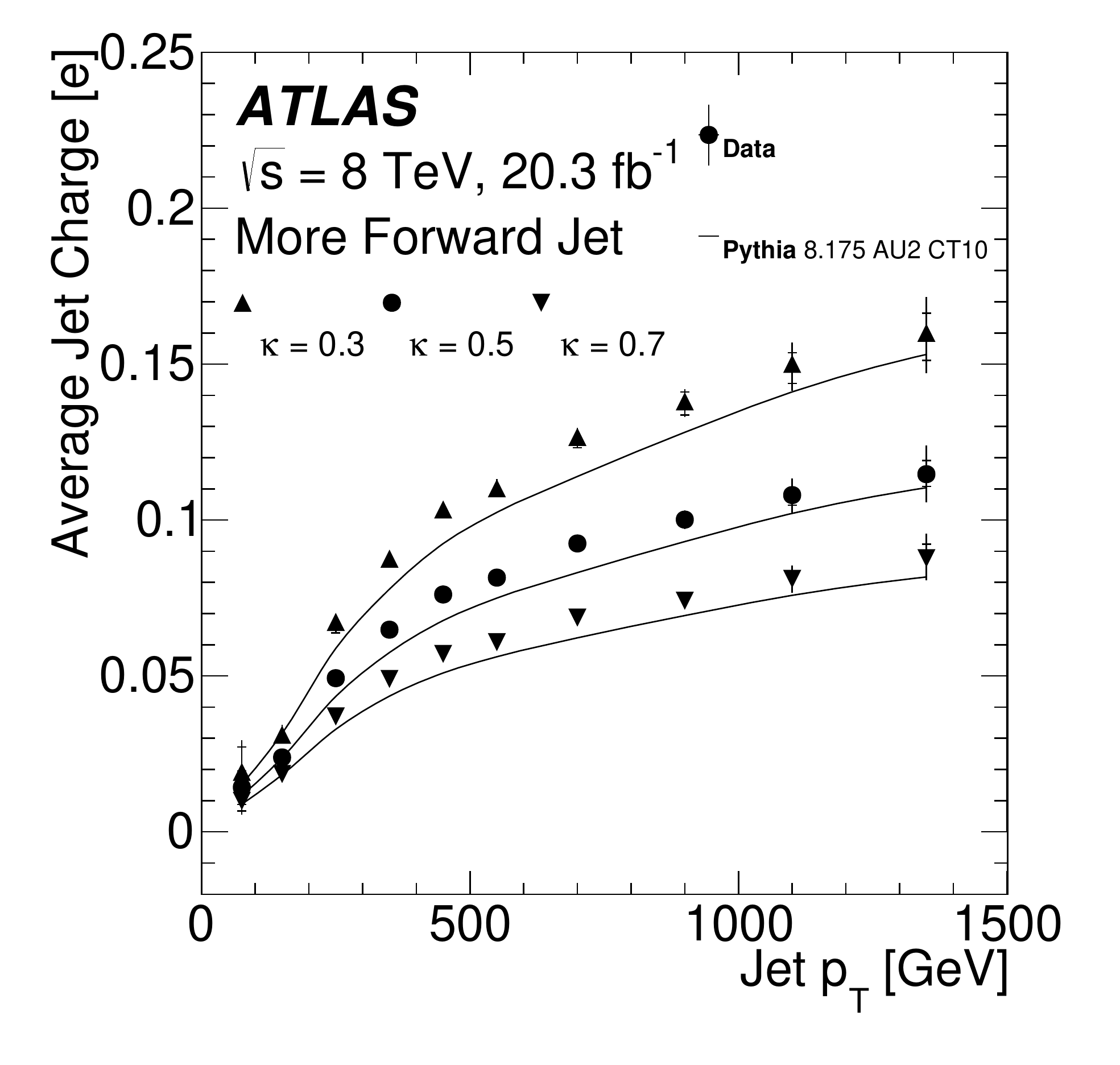}
\includegraphics[width=0.45\textwidth]{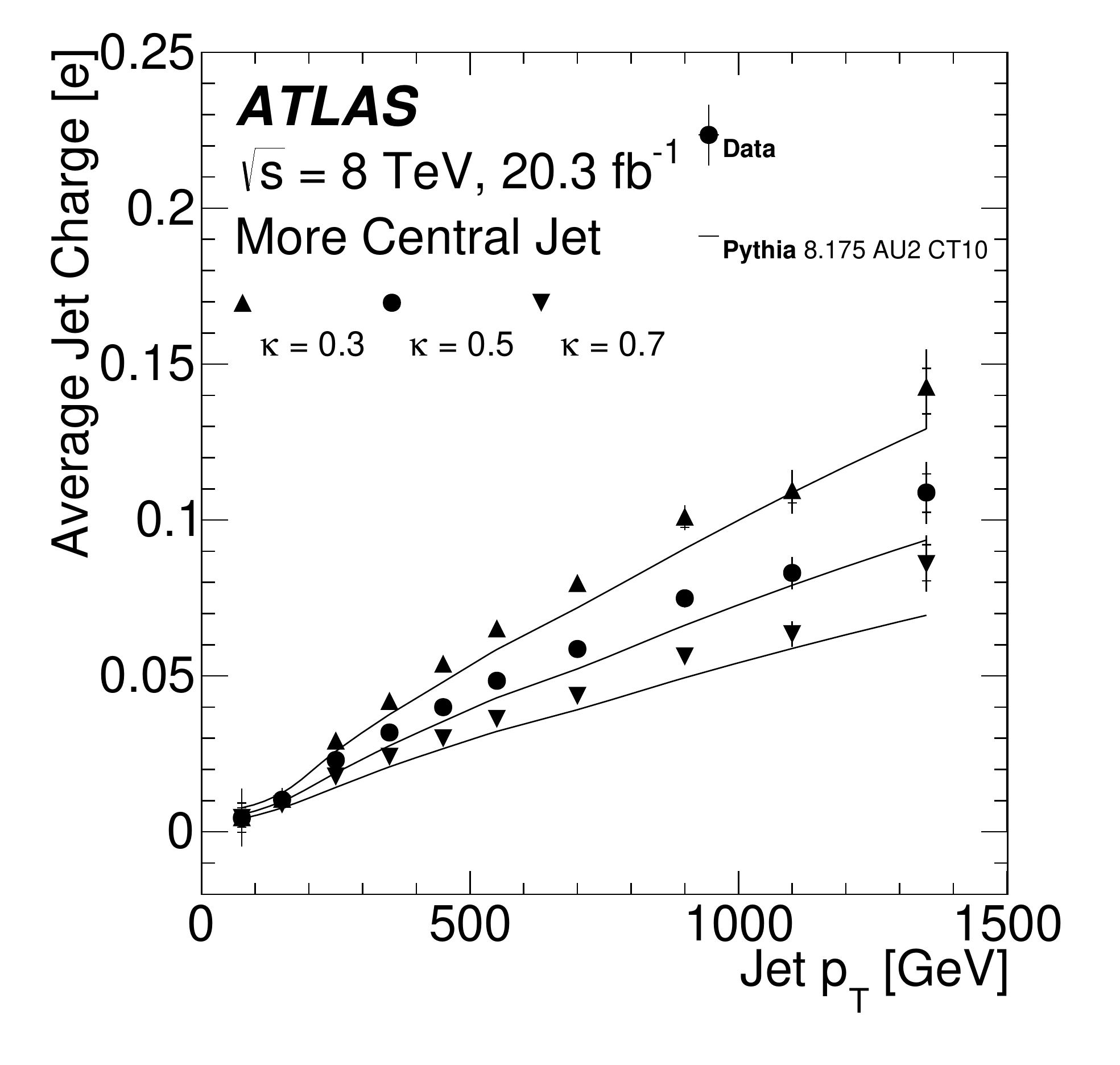}\\
\includegraphics[width=0.45\textwidth]{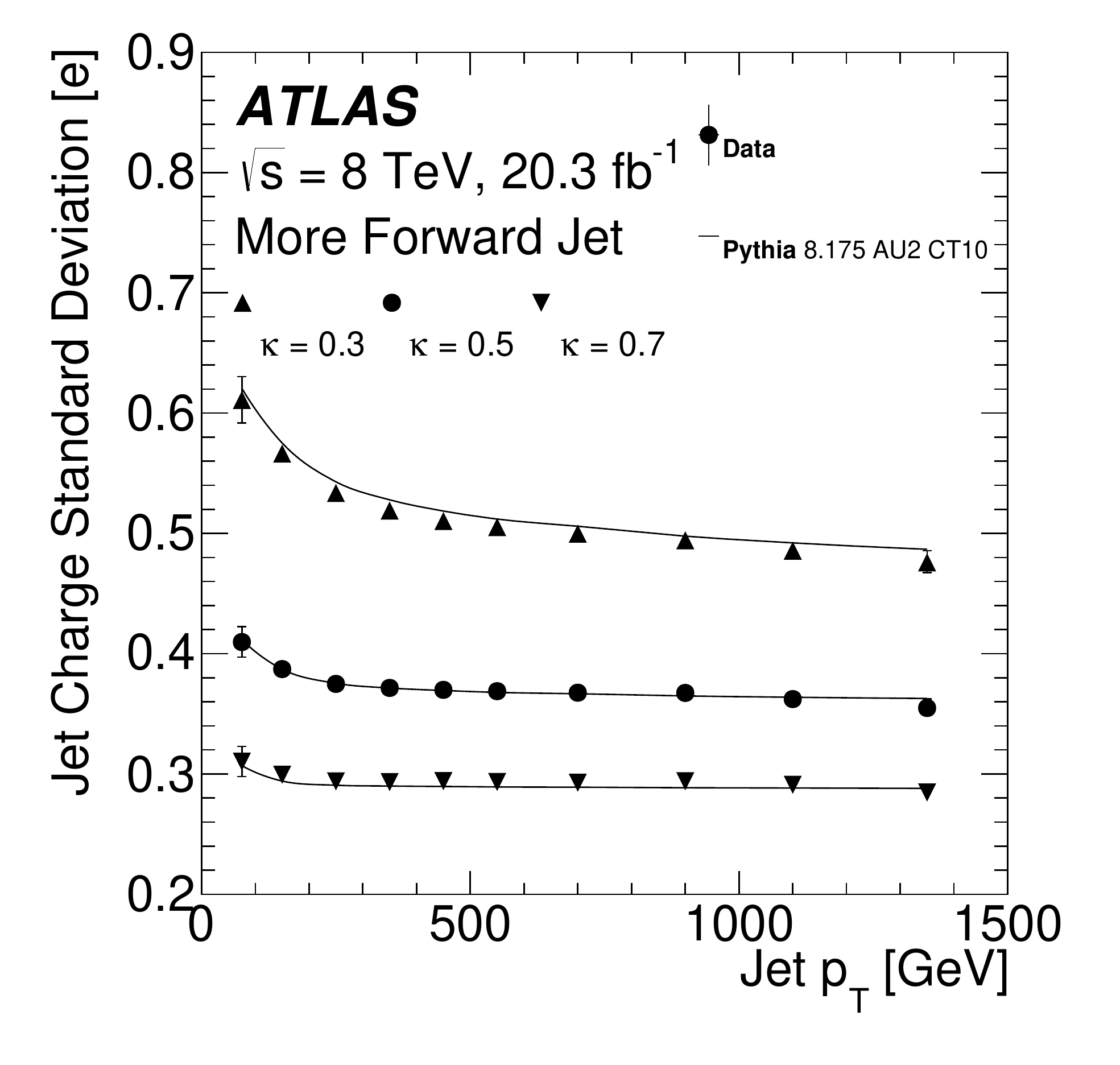}
\includegraphics[width=0.45\textwidth]{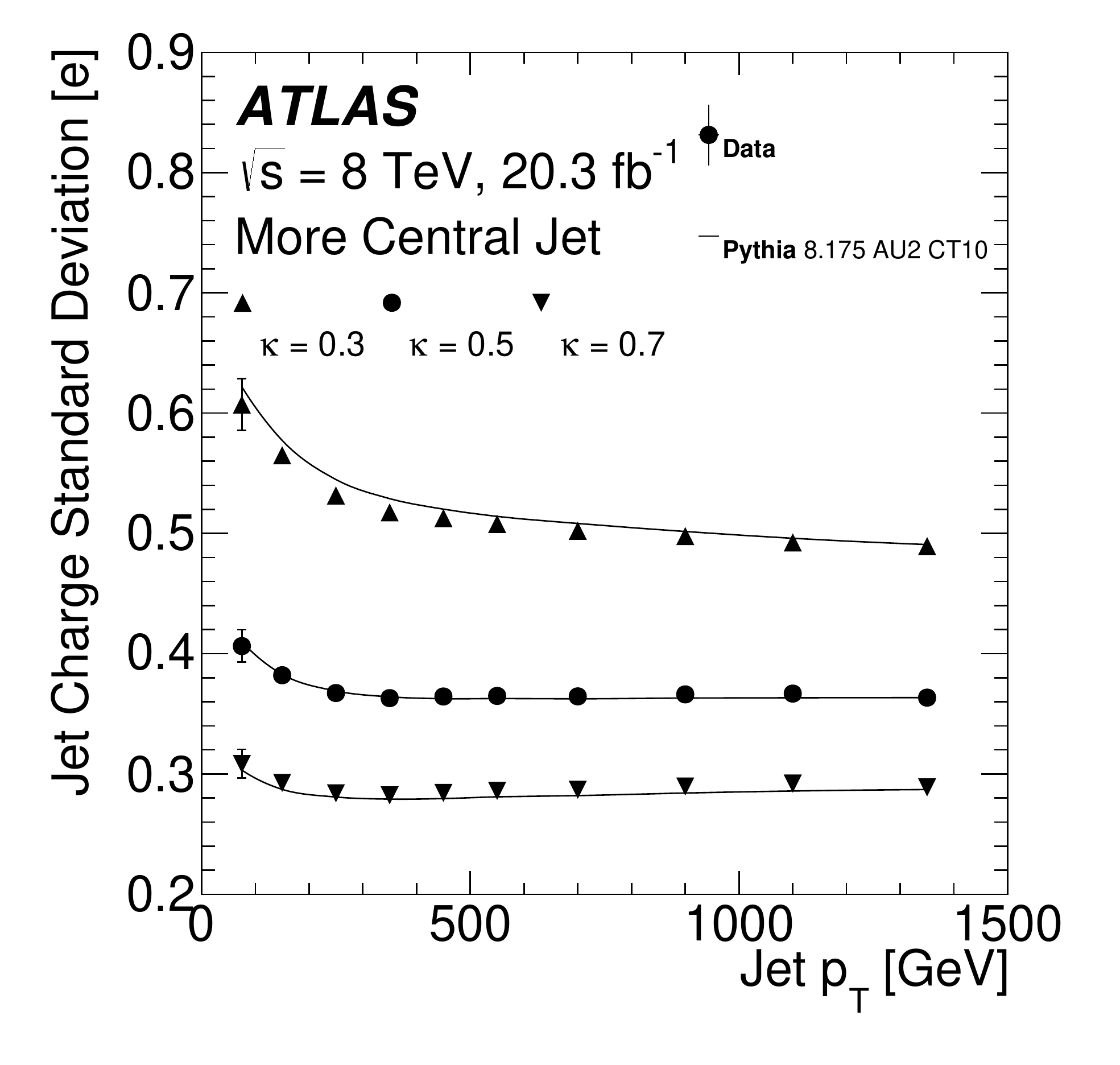}
\end{center}	
\caption{The measured average (standard deviation) of the jet charge distribution on top (bottom) in units of the positron charge as a function of the jet $p_\text{T}$ for $\kappa=0.3, 0.5,$ and $0.7$ for the more forward jet (left) and the more central jet (right).   The crossed lines in the bars on the data indicate the systematic uncertainty and the full extent of the bars is the sum in quadrature of the statistical and systematic uncertainties.}
\label{fig:mean}
\end{figure}

\subsection{Sensitivity of PDF Modeling}
\label{sec:PDFsensitivity}

Variations in the PDF set impact the relative flavor fractions and thus in turn change the jet charge distribution.  Such changes do not vary much with $\kappa$, since the PDF impacts the jet charge distribution mostly through the flavor fractions.  Figures~\ref{fig:PDFnom} and~\ref{fig:PDFnomrms} compare the unfolded distributions of the jet charge distribution's average and standard deviation with several PDF sets, with tuned predictions for {\sc Pythia} for each PDF, and with the same AU2 family of tunes.   The sampling of PDF sets results in a significant spread for the average jet charge, but has almost no effect on the standard deviation. CTEQ6L1 describes the data best, although the data/MC ratio has a stronger $p_\text{T}$ dependence. In particular, the data/MC differences with CTEQ6L1 are up to 10\% (15\%) at moderate $p_\text{T}$ for the more forward (central) jet.  For high $p_\text{T}$, differences between data and simulation are less significant.  NLO PDFs such as CT10 are consistently below the data by about 10\%-15\%.

\begin{figure}
\begin{center}
\subfloat[]{\includegraphics[width=0.5\textwidth]{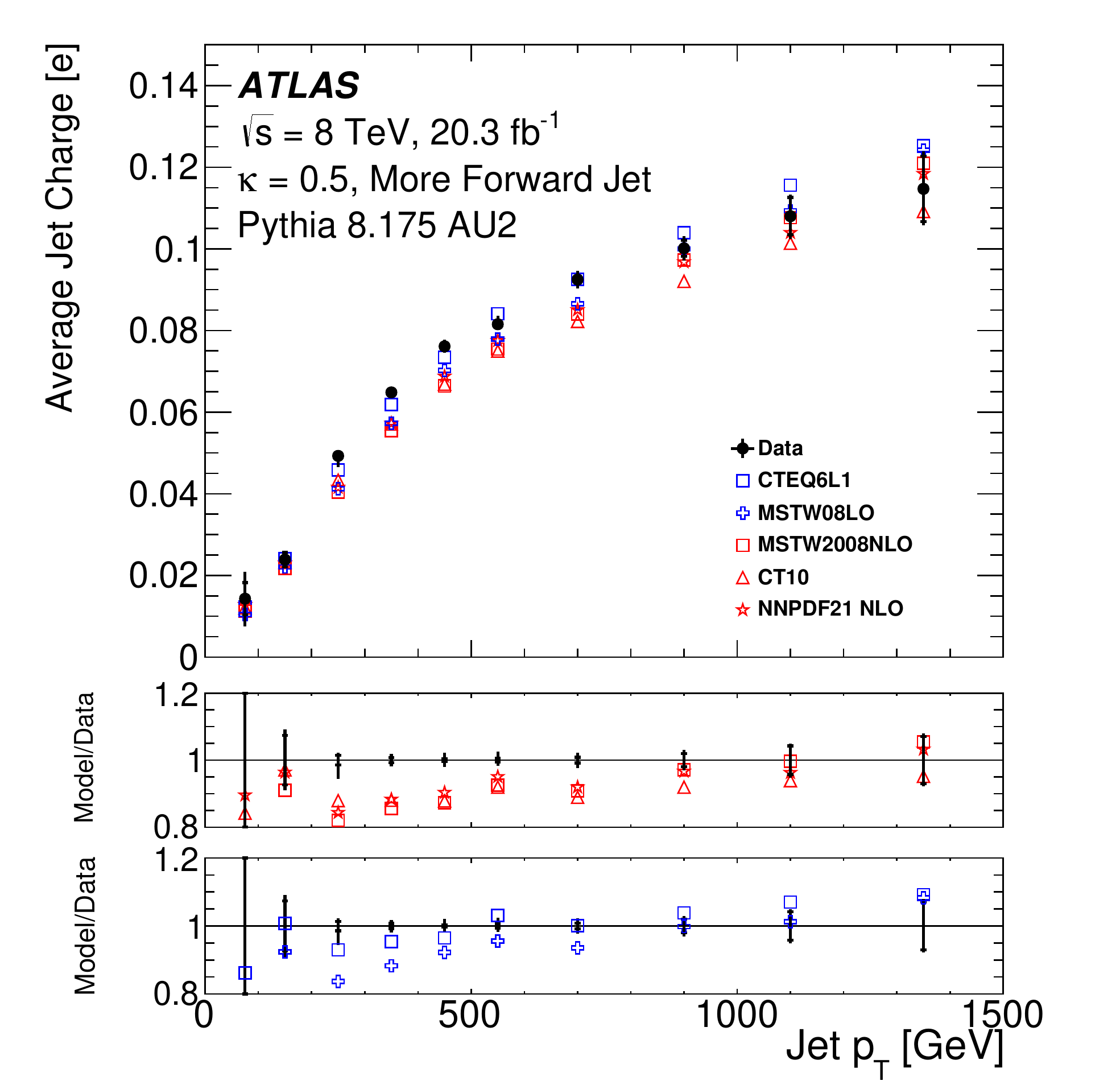}}\subfloat[]{\includegraphics[width=0.5\textwidth]{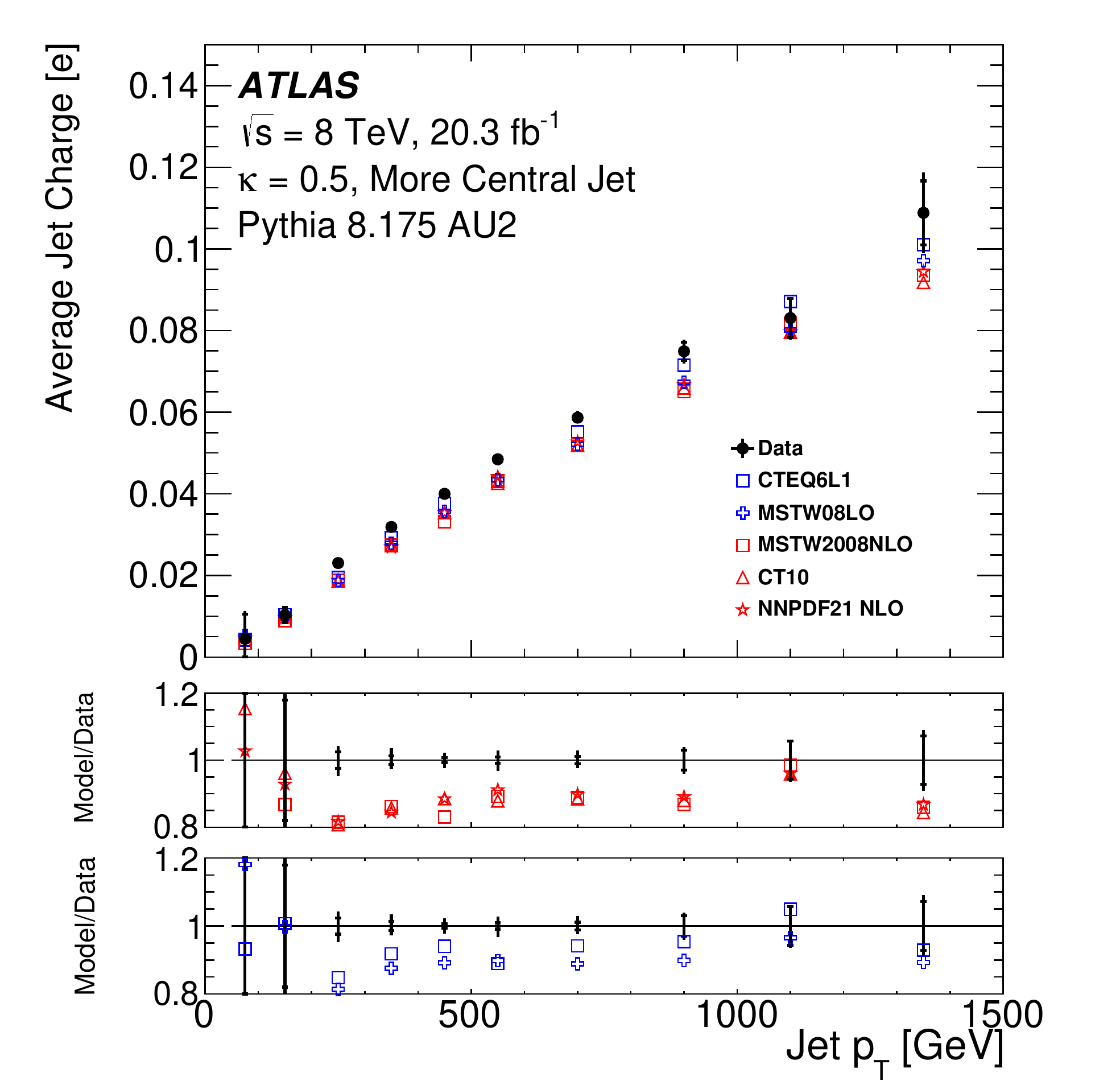}}
\end{center}	
\caption{The average jet charge ($\kappa=0.5$) in units of the positron charge for (a) the more forward jet and (b) the more central jet compared with theory predictions due to various PDF sets. The crossed lines in the bars on the data indicate the statistical uncertainty and the full extent of the bars is the sum in quadrature of the statistical and systematic uncertainties.}
\label{fig:PDFnom}
\end{figure}

\begin{figure}
\begin{center}
\subfloat[]{\includegraphics[width=0.5\textwidth]{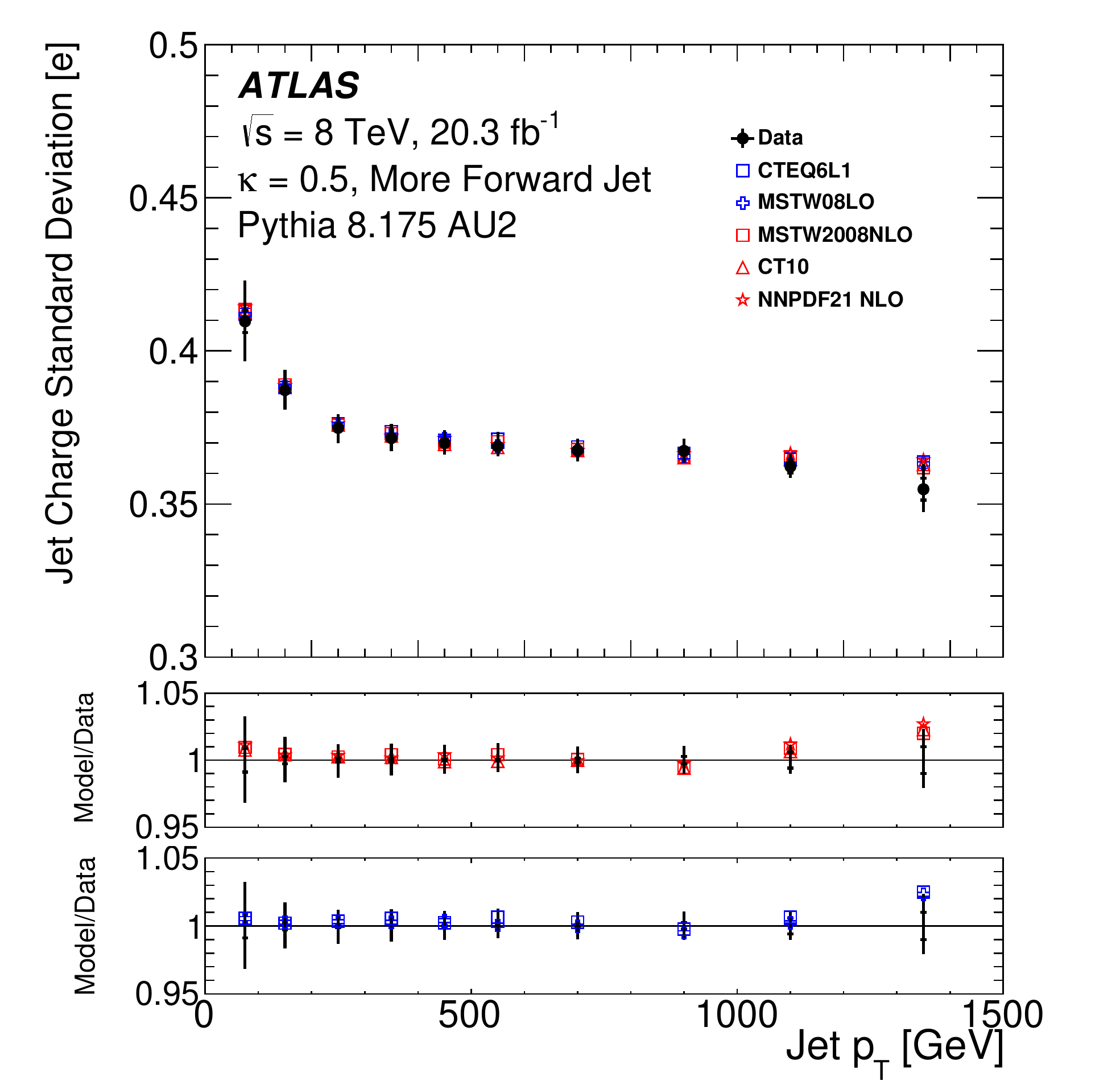}}\subfloat[]{\includegraphics[width=0.5\textwidth]{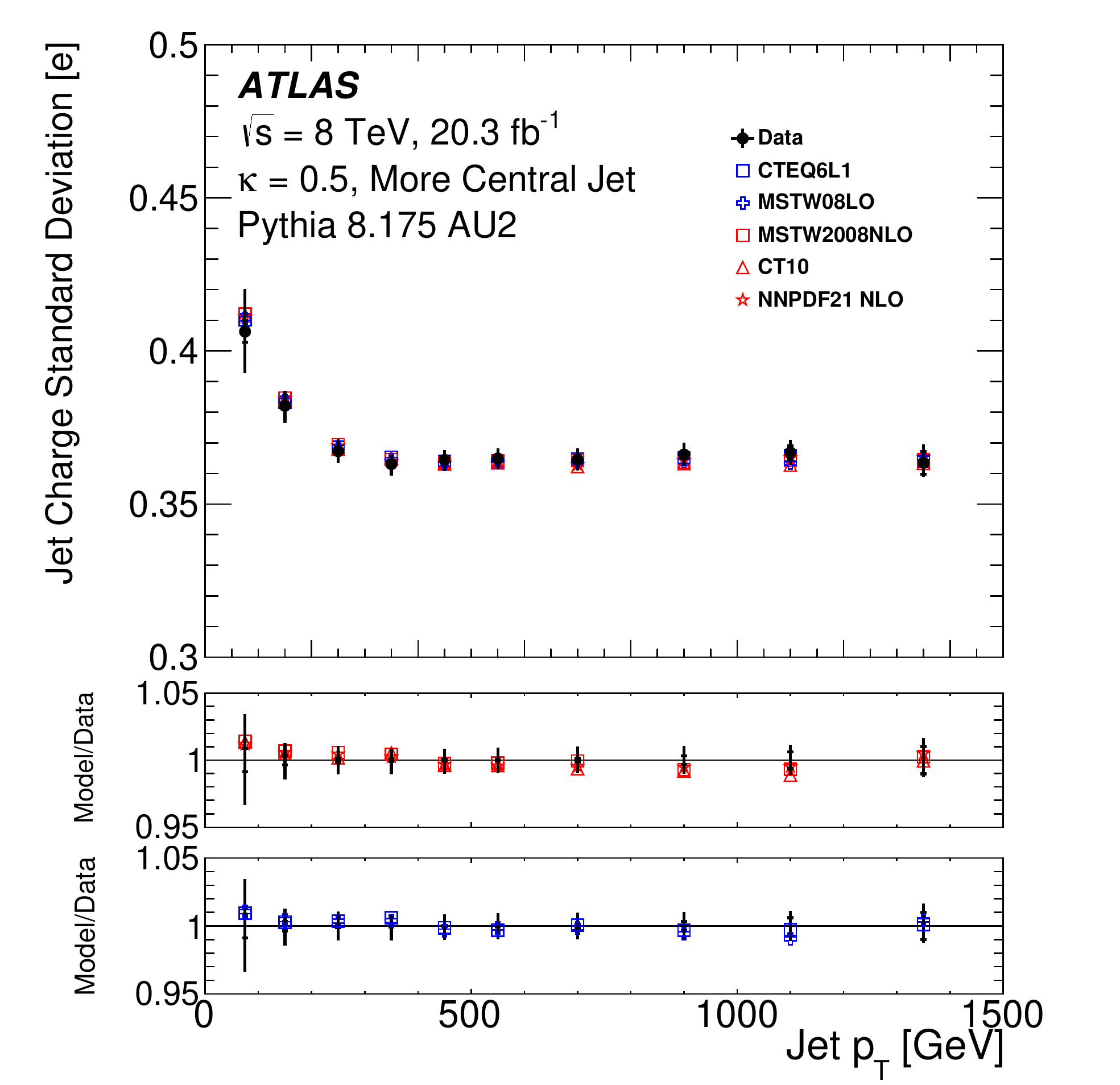}}
\end{center}	
\caption{The standard deviation of the jet charge ($\kappa=0.5$) distribution in units of the positron charge for (a) the more forward jet and (b) the more central jet compared with theory predictions due to various PDF sets. The crossed lines in the bars on the data indicate the statistical uncertainty and the full extent of the bars is the sum in quadrature of the statistical and systematic uncertainties. }
\label{fig:PDFnomrms}
\end{figure}

\subsection{Sensitivity of QCD Models and Tunes}
\label{sec:NPmodeling}

The measurements presented in Sec.~\ref{sec:rawunfolded} show that there are qualitative differences between the data and the MC simulations, and comparisons in Sec.~\ref{sec:PDFsensitivity} suggest that variations in the PDF set cannot fully explain the differences. Differences in Sec.~\ref{sec:rawunfolded} between {\sc Pythia} and {\sc Herwig++} suggest that some aspect of the modeling of fragmentation could lead to the observed differences between the simulation and the data.  One possible source is the hadronization modeling, which differs between {\sc Pythia} (Lund-string fragmentation) and {\sc Herwig++} (cluster fragmentation).  The modeling of final-state radiation (FSR) is expected to have an impact on the jet charge distribution because variations in the radiation lead to different energy flow around the initial parton and hence different fragmentation of the jet.  The plots in Fig.~\ref{fig:nprms} and Fig.~\ref{fig:nprm2} show the measured average jet charge and the jet charge distribution's standard deviation, respectively, for $\kappa=0.3,$ 0.5, and 0.7, compared to various models for a fixed PDF set (CTEQ6L1).   In addition to {\sc Pythia} 8 and {\sc Herwig++} model predictions, Figs.~\ref{fig:nprms} and~\ref{fig:nprm2} contain the predictions from {\sc Pythia} 6 using the Perugia 2012 tune~\cite{Skands:2010ak} and the radHi and radLo Perugia 2012 tune variations.  These Perugia tune variations test the sensitivity to higher/lower amounts of initial- and final-state radiation (via the scaling of $\alpha_\text{s}$), although only variations of the FSR are important for the jet charge distribution. For the mean jet charge, {\sc Pythia} 6 with the P2012 radLo tune is very similar to {\sc Pythia} 8 with the AU2 tune.  The spread in the average jet charge due to the difference between the radHi and radLo tunes increases with $\kappa$, since suppression of soft radiation makes the jet charge distribution more sensitive to the modeling of the energy fraction of the leading emissions.  For the jet charge distribution's standard deviation, the sensitivity to the $\alpha_\text{s}$ scaling is large at both high and low $\kappa$.  However, the sensitivity is inverted: radHi gives a larger standard deviation for $\kappa=0.3$, but a lower standard deviation for $\kappa=0.7$.  Other Perugia 2012 tunes have been studied, testing the sensitivity to color-reconnection and multiple parton interactions, but the differences in the jet charge distribution's mean and standard deviation are small.  The Perugia 2012 tunes may not fully capture the spread in nonperturbative effects, which is also suggested by the increasing difference between {\sc Pythia} 8 and {\sc Herwig++} for decreasing $\kappa$.

\begin{figure}
\begin{center}
\subfloat[]{\includegraphics[width=0.5\textwidth]{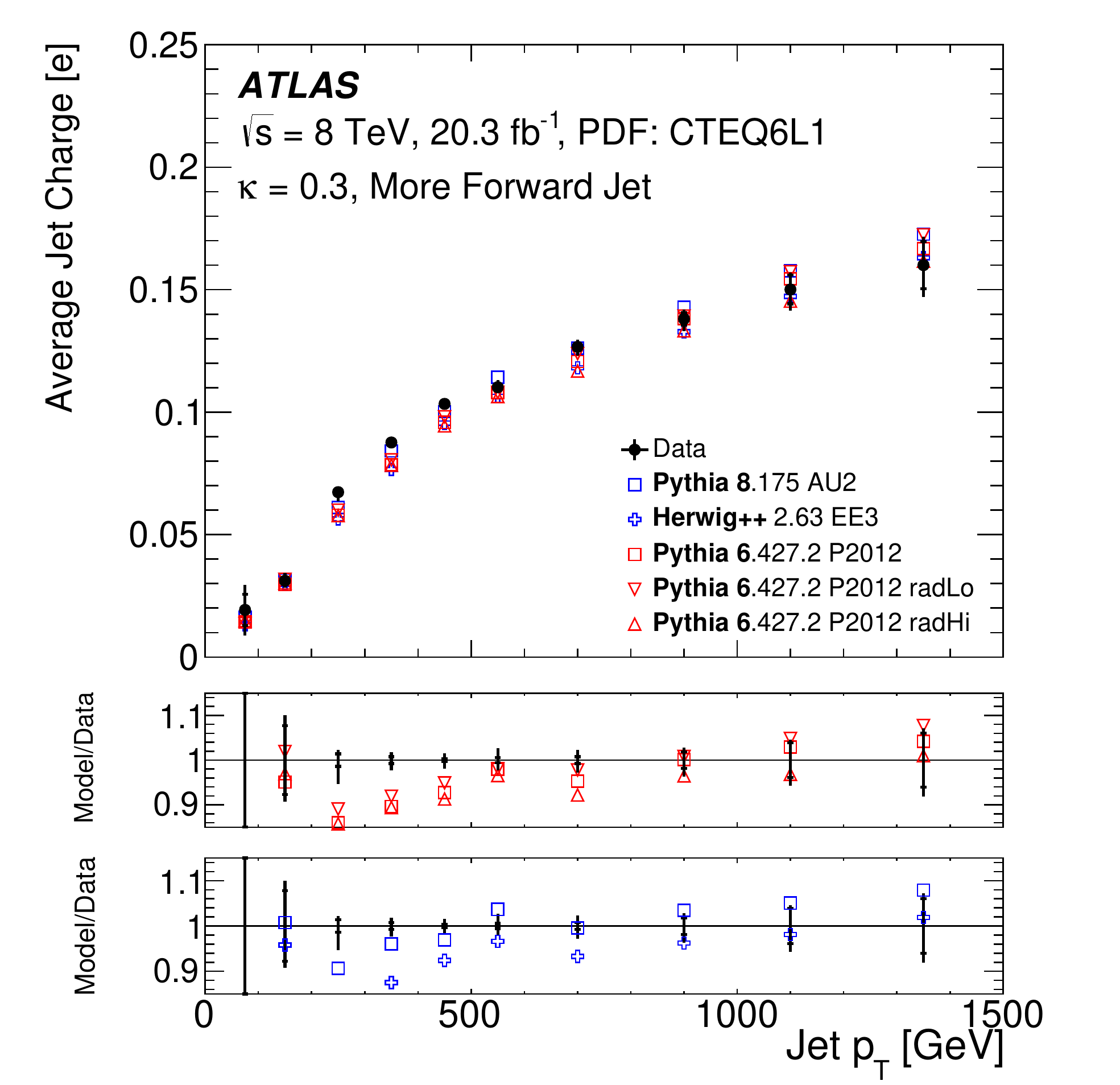}}\subfloat[]{\includegraphics[width=0.5\textwidth]{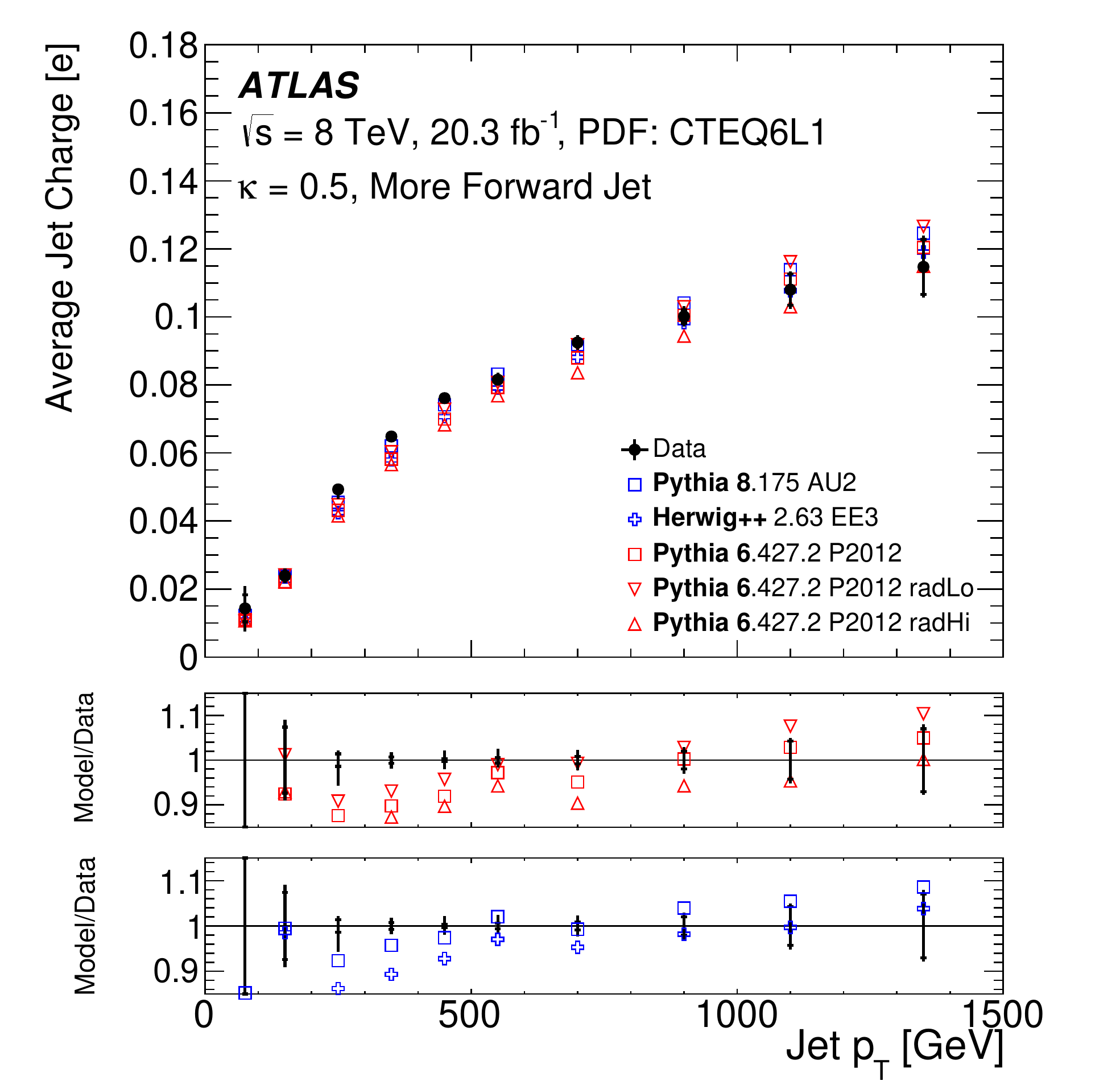}}\\
\subfloat[]{\includegraphics[width=0.5\textwidth]{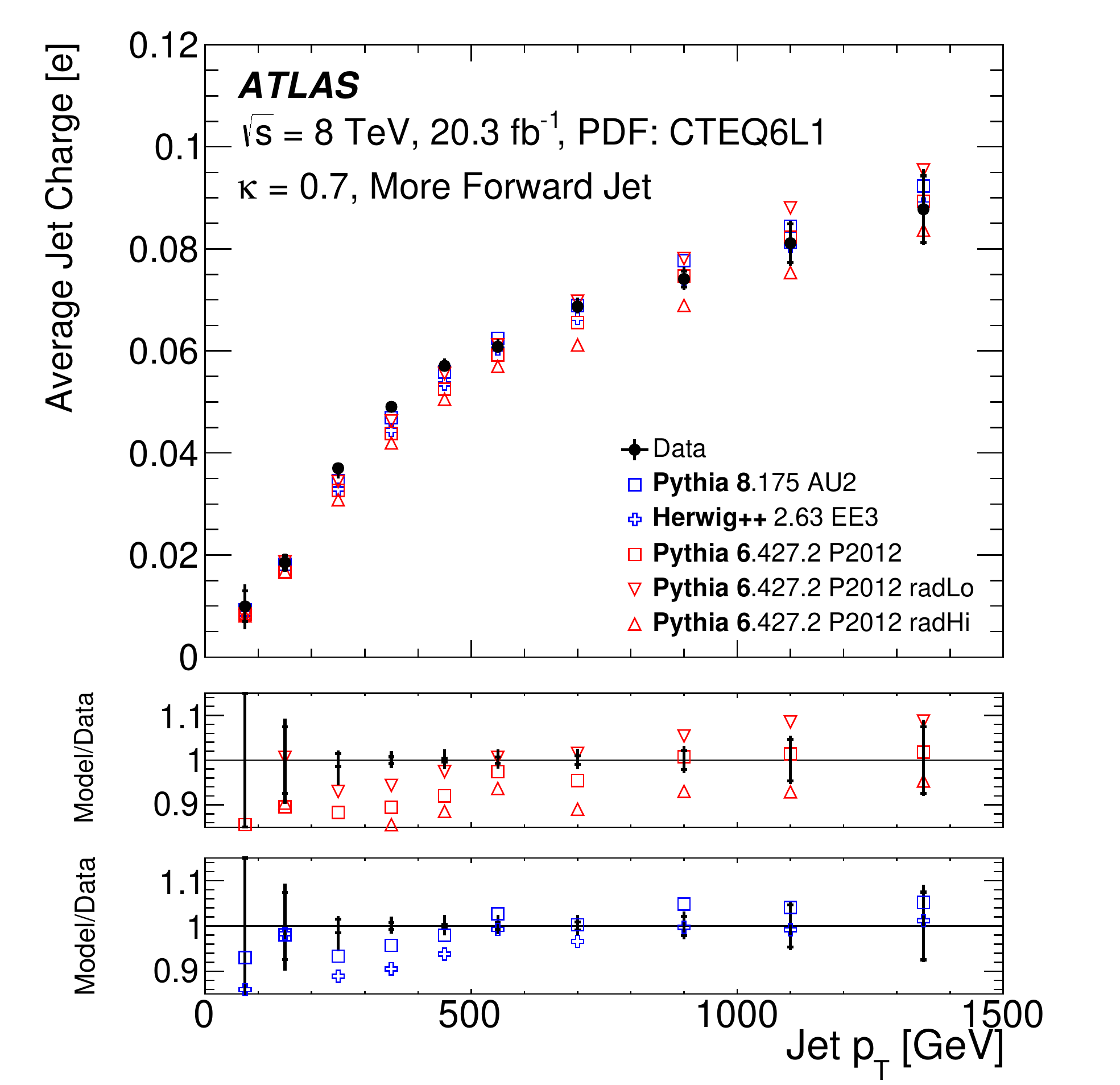}}
\end{center}	
\caption{The average of the jet charge distribution in units of the positron charge for (a) $\kappa=0.3$, (b) 0.5, and (c) 0.7 comparing various QCD MC models and tunes for the more forward jet.  The crossed lines in the bars on the data indicate the statistical uncertainty and the full extent of the bars is the sum in quadrature of the statistical and systematic uncertainties.}
\label{fig:nprms}
\end{figure}

\begin{figure}
\begin{center}
\subfloat[]{\includegraphics[width=0.5\textwidth]{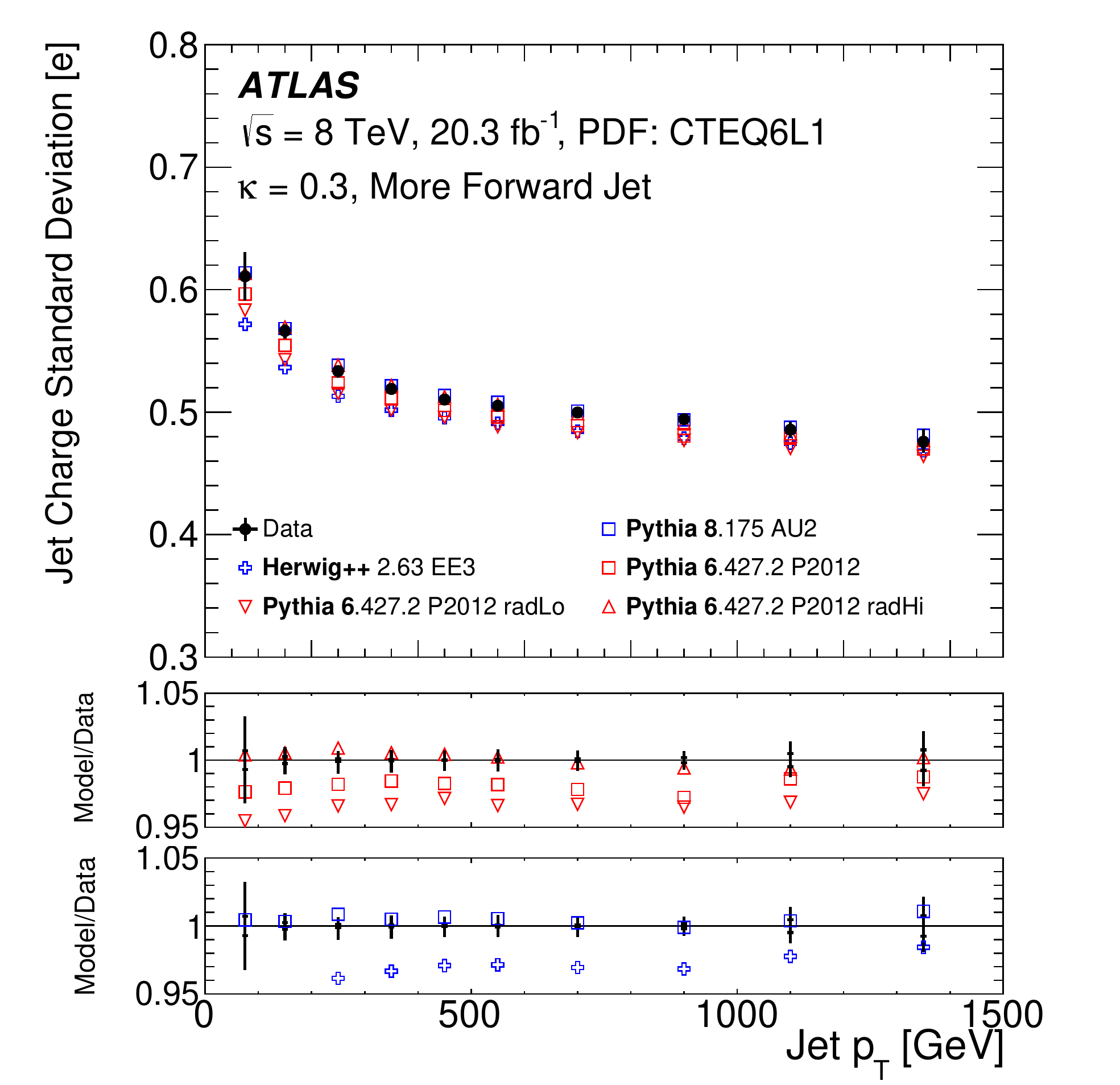}}\subfloat[]{\includegraphics[width=0.5\textwidth]{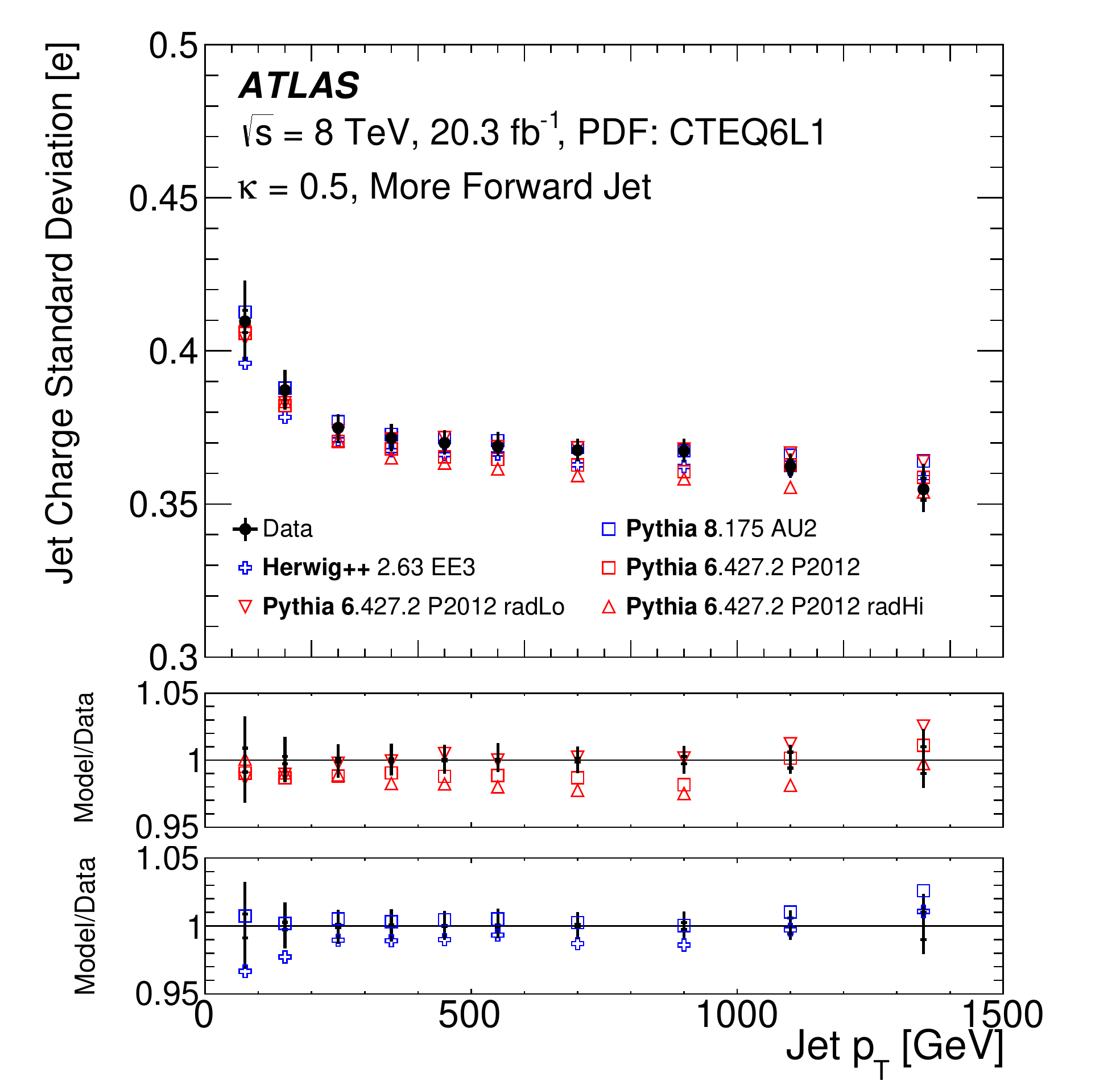}}\\
\subfloat[]{\includegraphics[width=0.5\textwidth]{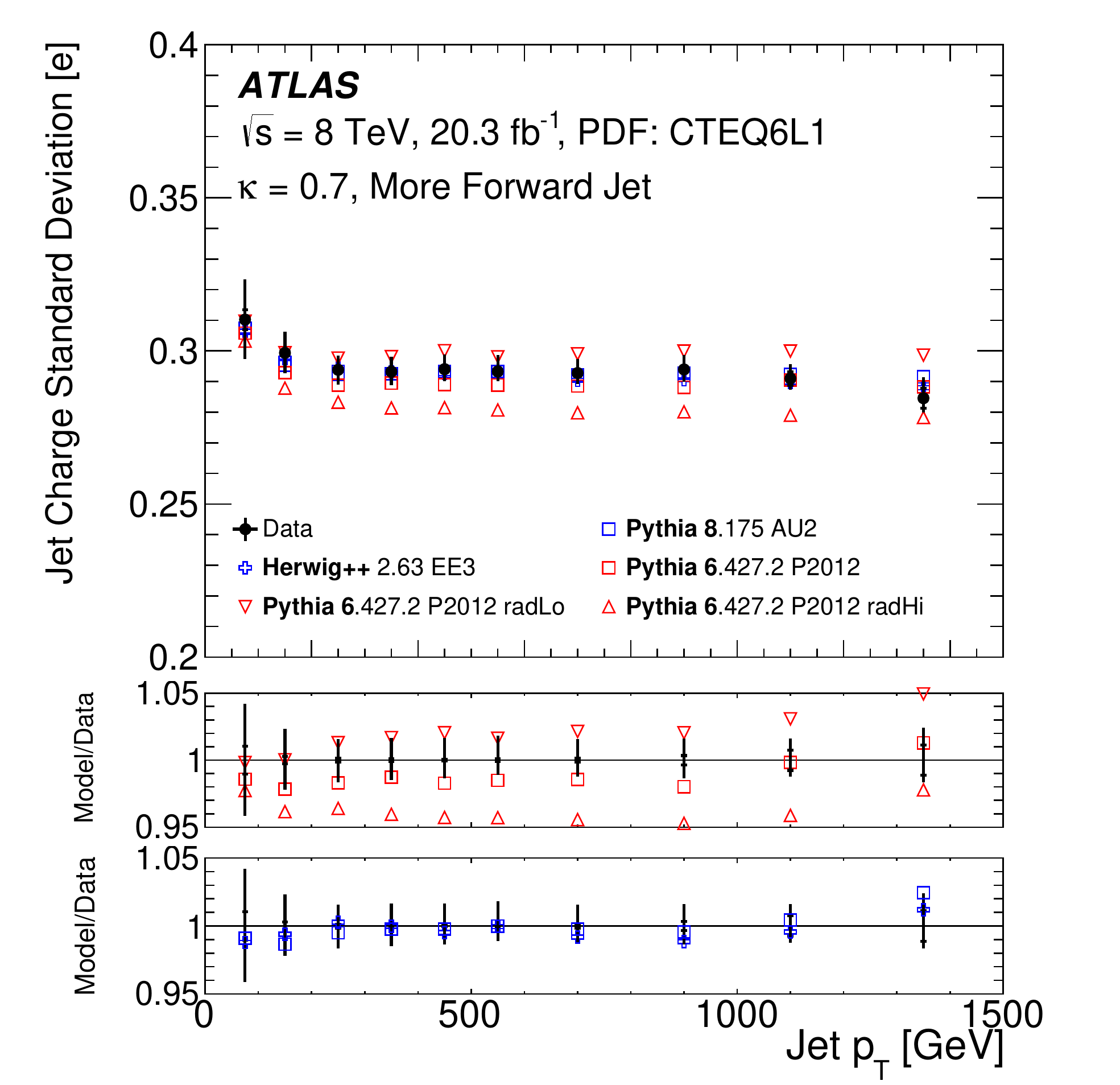}}
\end{center}	
\caption{The standard deviation of the jet charge distribution in units of the positron charge for (a) $\kappa=0.3$, (b) 0.5, and (c) 0.7 comparing various QCD MC models and tunes for the more forward jet.  The crossed lines in the bars on the data indicate the statistical uncertainty and the full extent of the bars is the sum in quadrature of the statistical and systematic uncertainties.}
\label{fig:nprm2}
\end{figure}

\clearpage
\newpage

\subsection{Model comparison overview}

Figures~\ref{fig:meana} and~\ref{fig:mean2} show comparisons of the unfolded jet charge distribution's mean and standard deviation for different QCD simulations using LO and NLO PDF sets. The predictions using the CT10 NLO PDF set as shown in Fig.~\ref{fig:meana} are generally about 10\% below the data.  Consistent with the expectation that the PDF and (nearly collinear) fragmentation are responsible for the jet charge distribution's mean and standard deviation, there does not seem to be an effect from the {\sc Powheg} NLO matrix element.  For the jet charge distribution's standard deviation and $\kappa=0.3$, the data falls between {\sc PYTHIA} (larger standard deviation) and {\sc Herwig++} (smaller standard deviation), but this trend is less evident for larger $\kappa$ values, suggesting a difference due to soft tracks.  As seen in Sec.~\ref{sec:PDFsensitivity}, comparisons with CTEQ6L1 show it be to a better model for the $p_\text{T}$-dependence of the mean jet charge than CT10.  The analogous plots to Fig.~\ref{fig:mean} but using CTEQ6L1 instead of CT10 are shown in Fig.~\ref{fig:mean2}.  Generally, there is agreement between the simulation and the data with only a $\lesssim 5\%$ difference in the lower $p_\text{T}$ bins.

\begin{figure}
\begin{center}
\includegraphics[width=0.48\textwidth]{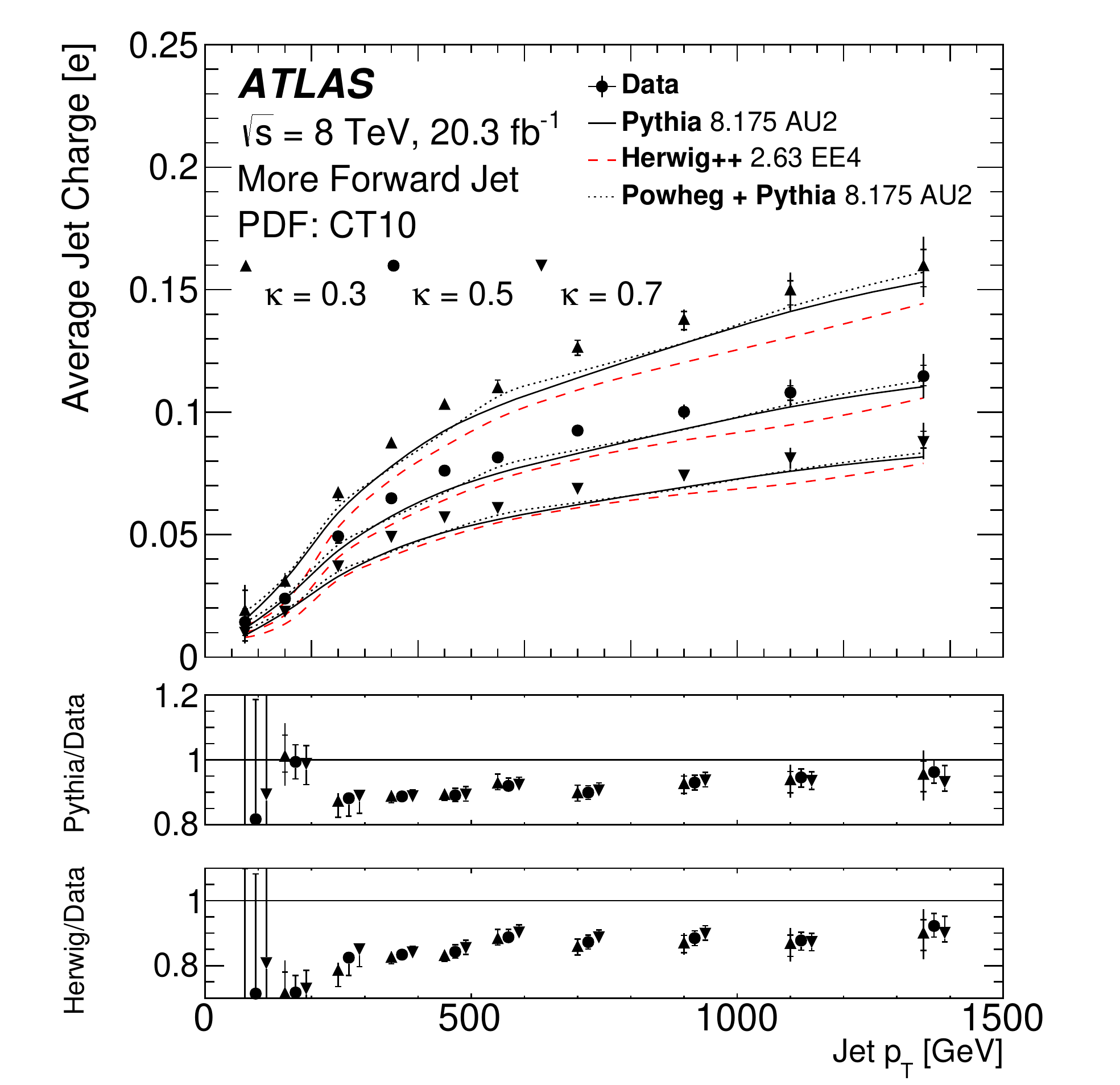}
\includegraphics[width=0.48\textwidth]{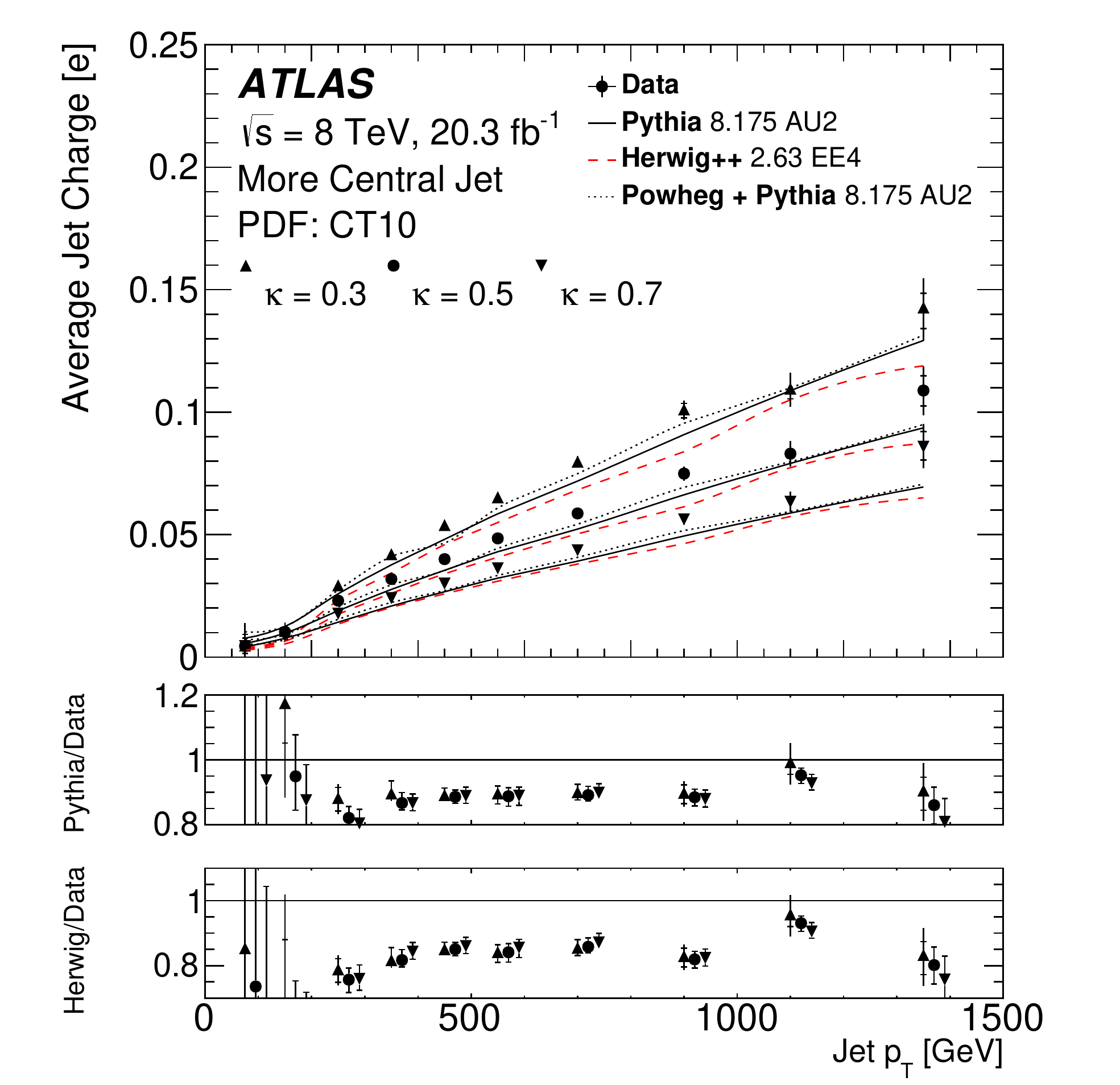}\\
\includegraphics[width=0.48\textwidth]{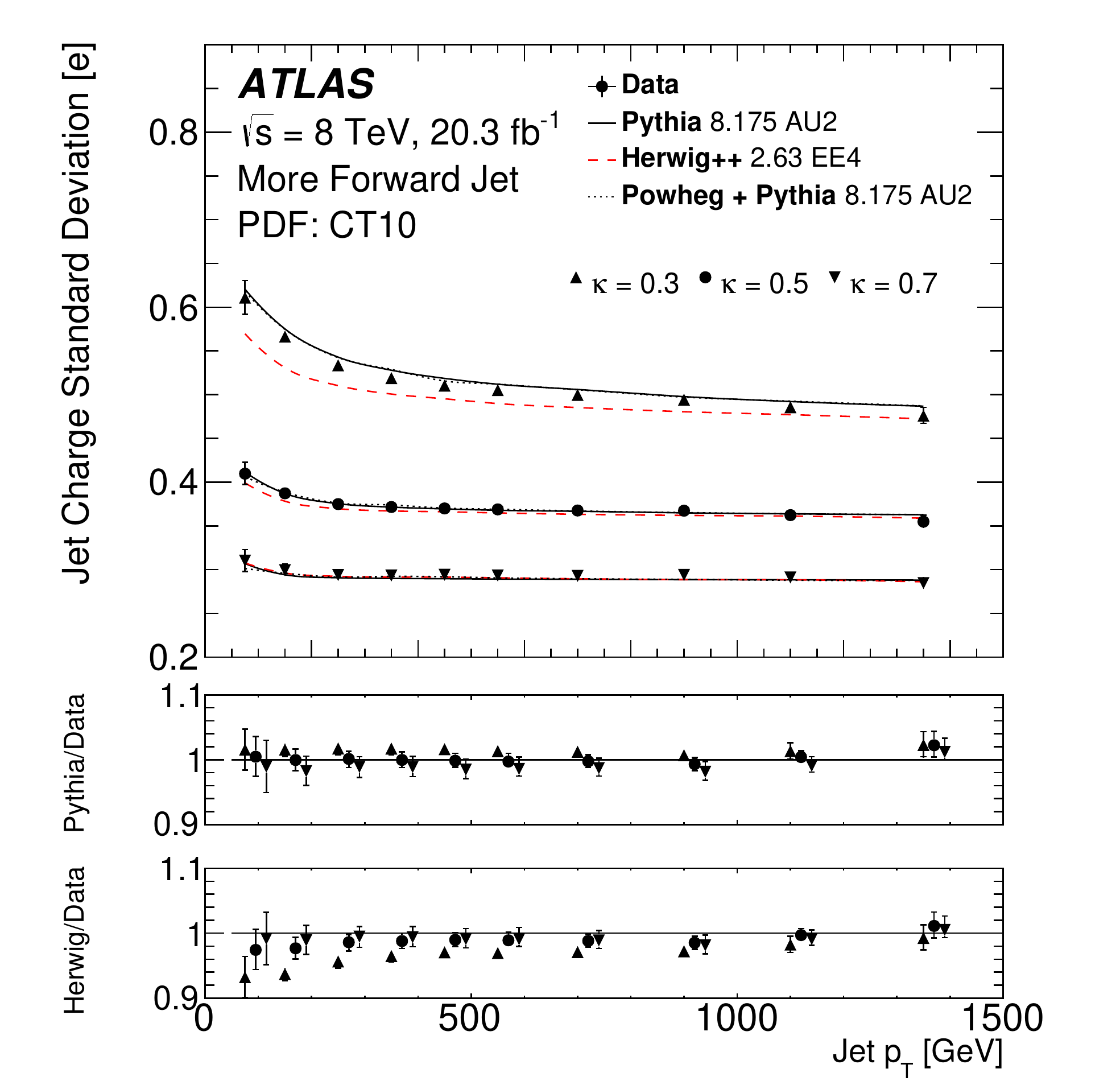}
\includegraphics[width=0.48\textwidth]{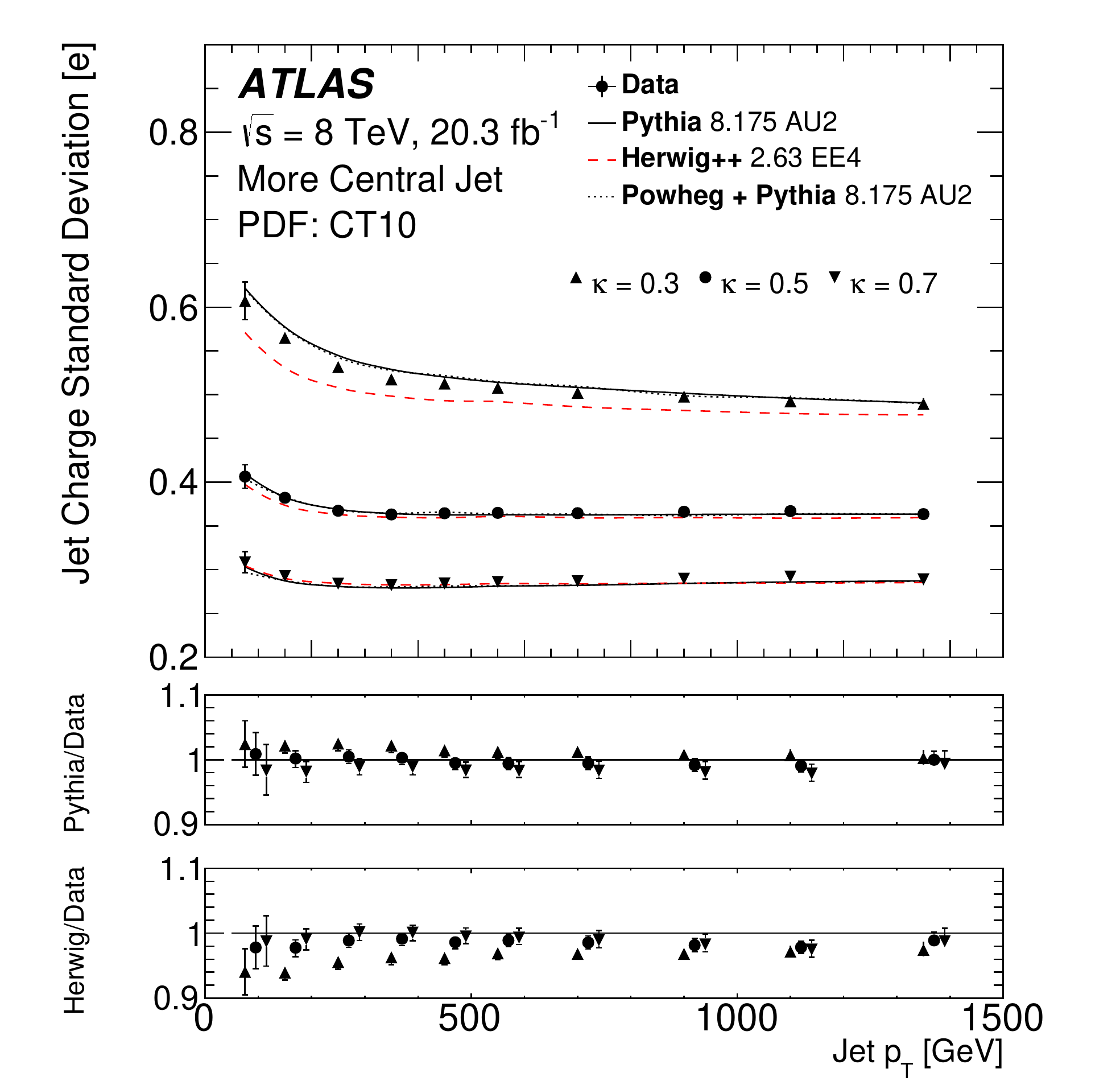}
\end{center}	
\caption{The measured average of the jet charge distribution (top), and the standard deviation (bottom), in units of the positron charge as a function of the jet $p_\text{T}$ for $\kappa=0.3, 0.5,$ and $0.7$ for the more forward jet (left) and the more central jet (right) using CT10 as the PDF set.  The markers in the lower panel are artificially displaced horizontally to make distinguishing the three $\kappa$ values easier.  The {\sc Powheg}+{\sc Pythia} curves are nearly on top of the {\sc Pythia} curves.  The crossed lines in the bars on the data indicate the systematic uncertainty and the full extent of the bars is the sum in quadrature of the statistical and systematic uncertainties.}
\label{fig:meana}
\end{figure}

\begin{figure}
\begin{center}
\includegraphics[width=0.48\textwidth]{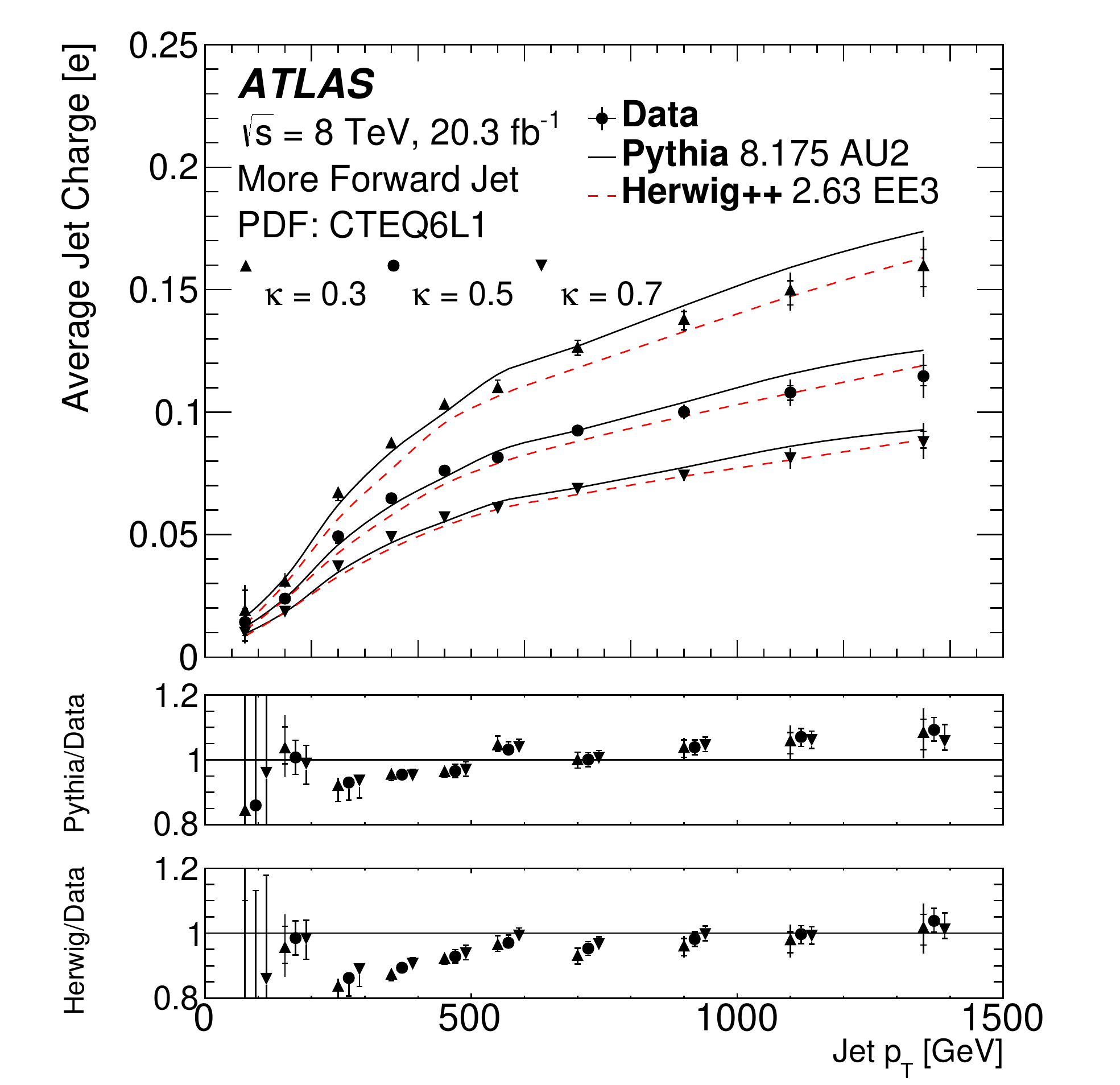}
\includegraphics[width=0.48\textwidth]{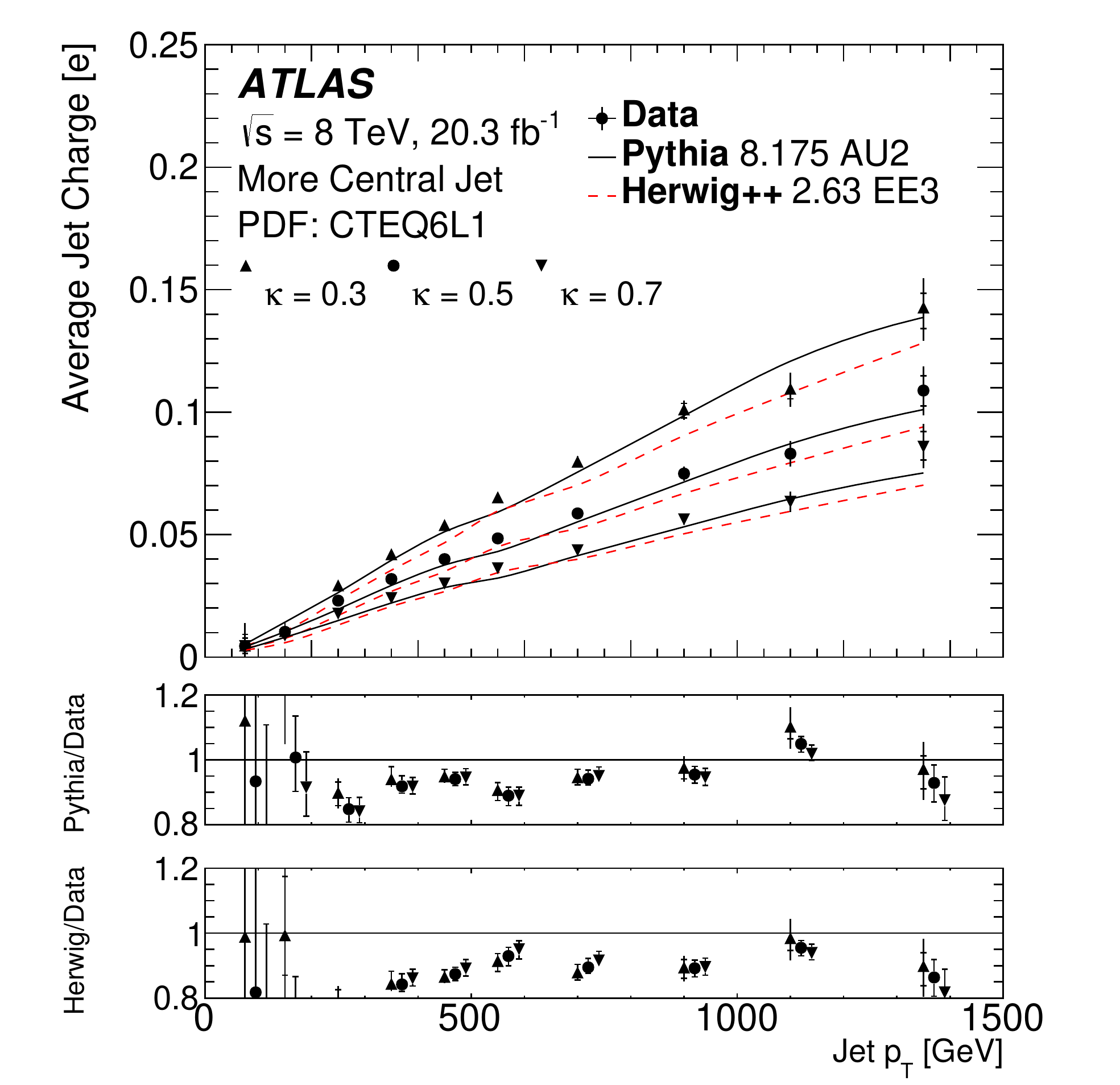}\\
\includegraphics[width=0.48\textwidth]{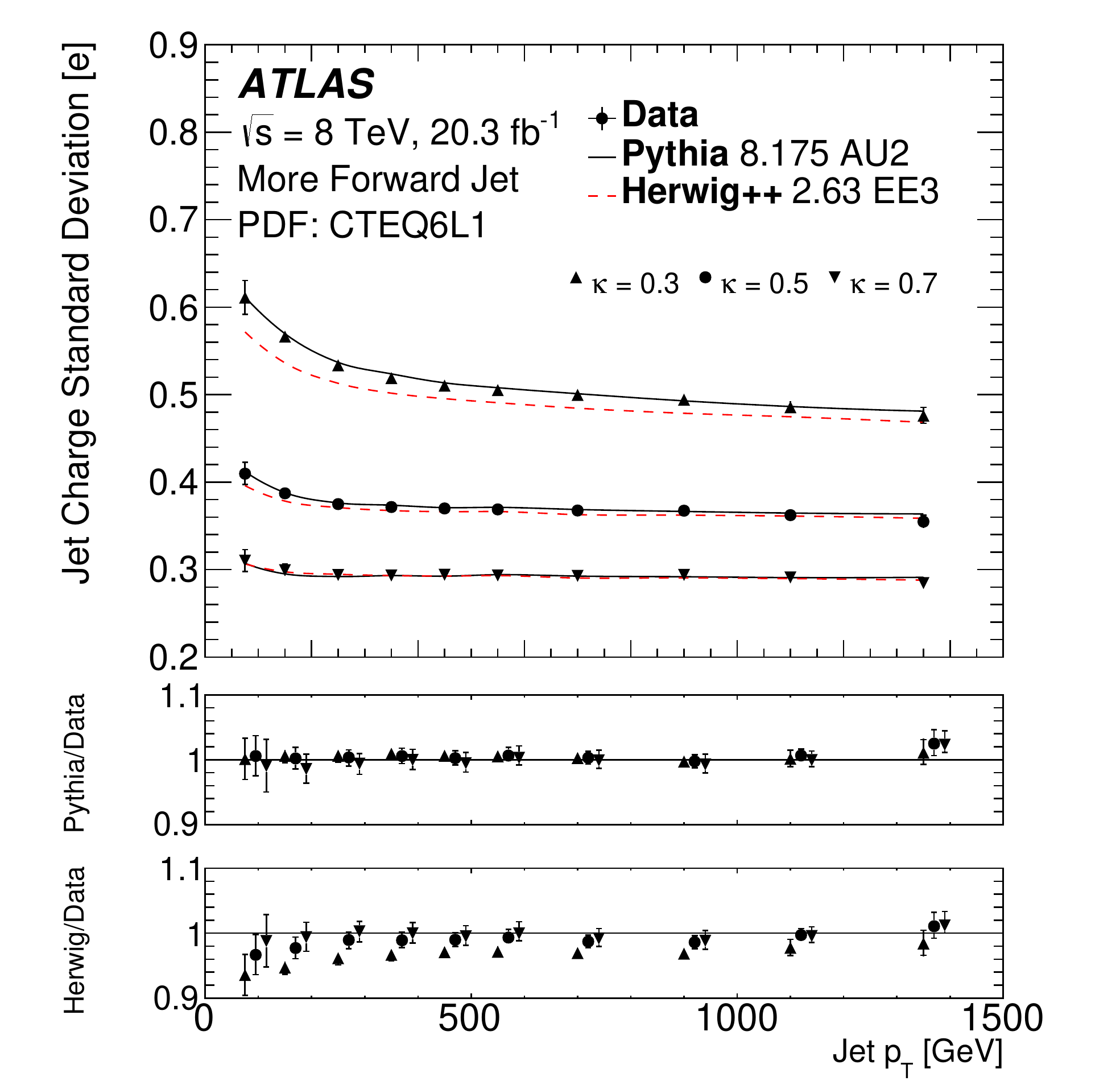}
\includegraphics[width=0.48\textwidth]{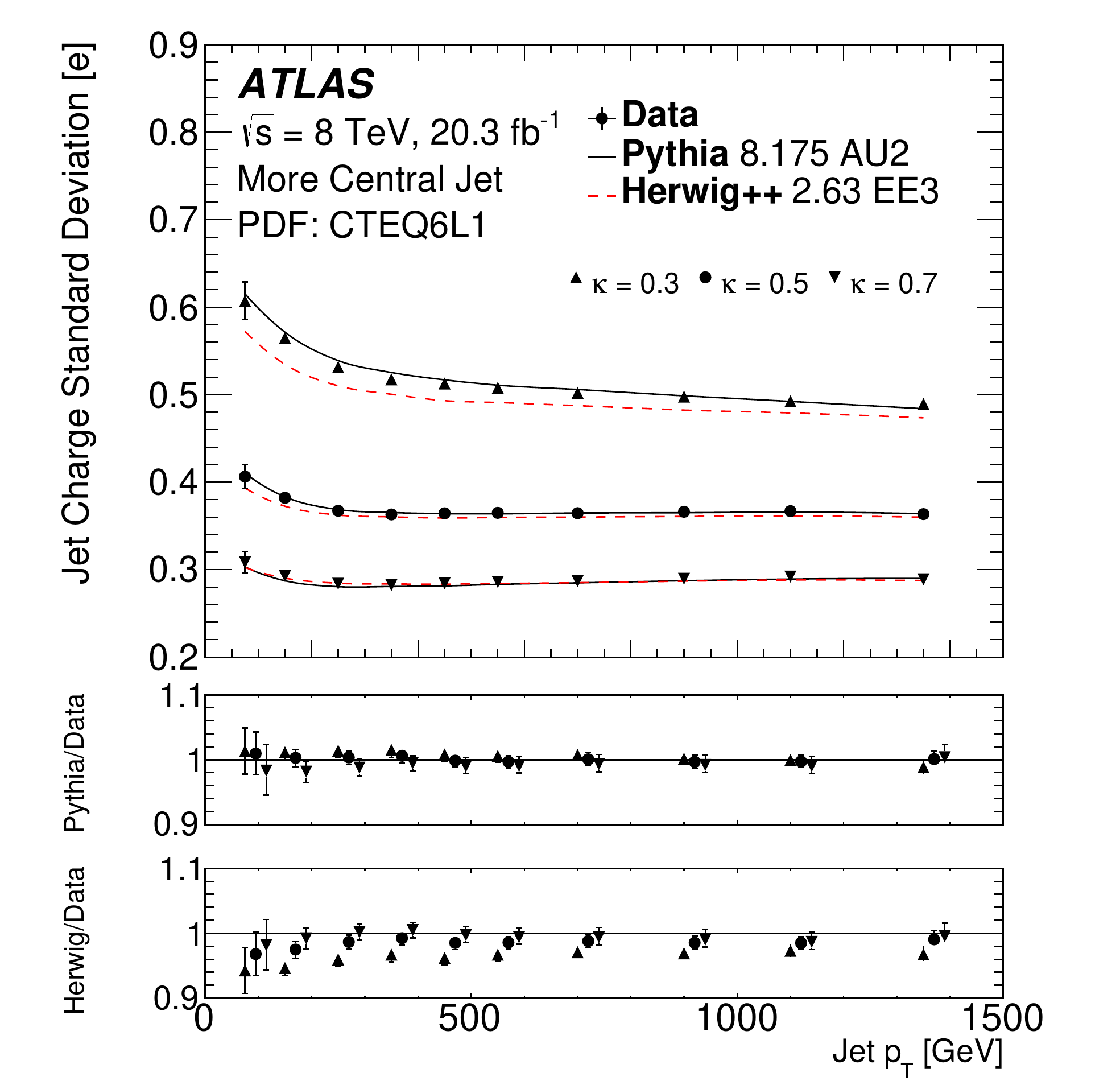}
\end{center}	
\caption{The measured average of the jet charge distribution (top), and the standard deviation (bottom), in units of the positron charge as a function of the jet $p_\text{T}$ for $\kappa=0.3, 0.5,$ and $0.7$ for the more forward jet (left) and the more central jet (right) using CTEQ6L1 as the PDF set.  The markers in the lower panel are artificially displaced horizontally to make distinguishing the three $\kappa$ values easier. The crossed lines in the bars on the data indicate the systematic uncertainty and the full extent of the bars is the sum in quadrature of the statistical and systematic uncertainties.}
\label{fig:mean2}
\end{figure}

\clearpage

\subsection{Extraction of the average up-quark and down-quark jet charges}
\label{sec:updownextract}

In addition to understanding the trends in the jet charge distribution from PDFs, one can use PDFs to extract information about jets of a particular flavor.  These {\it exclusive} interpretations rely on flavor-fraction information in PDFs and matrix element calculations to extract the jet charge distribution for particular jet (anti-)flavors in each $p_\text{T}$ bin.  The required nonperturbative information is summarized in Fig.~\ref{fig:flavorfrac}(a).  Jets with flavors other than up/down/anti-up/anti-down/gluon are not included in Fig.~\ref{fig:flavorfrac}(a) and give a negligible contribution ($\lesssim 2\%$) in the highest $p_\text{T}$ bins.

One way of extracting the up- and down-flavor average jet charges is to exploit the difference in flavor fractions shown in Fig.~\ref{fig:flavorfrac}(a) between the more forward and the more central jets.  Due to the $p_\text{T}$-balance requirement between the leading and subleading jet in the event selection, to a good approximation, the $p_\text{T}$ spectrum is the same for the more forward and the more central jet.  Assuming that the average jet charge of the sum of flavors that are not up/down/anti-up/anti-down is zero, in each bin $i$ of $p_\text{T}$:

\begin{align}
\label{eq:syst}
\langle Q_J^\text{forward}\rangle_i &= \left(f_\text{up,i}^\text{forward}-f_\text{anti-up,i}^\text{forward}\right)Q_i^\text{up}+(f_\text{down,i}^\text{forward}-f_\text{anti-down,i}^\text{forward})Q_i^\text{down}\\\nonumber
\langle Q_J^\text{central}\rangle_i &= \left(f_\text{up,i}^\text{central}-f_\text{anti-up,i}^\text{central}\right)Q_i^\text{up}+(f_\text{down,i}^\text{central}-f_\text{anti-down,i}^\text{central})Q_i^\text{down},
\end{align}

where $Q_J$ is the jet charge from Eq.~(\ref{chargedef}), $f_{y,i}^x$ is the fraction of flavor $y$ in $p_\text{T}$ bin $i$ for the jet $x\in\{\text{more forward, more central}\}$ and $Q_i^y$ is the average jet charge for such jets.  The values $f_{y,i}^x$ are taken from simulation ({\sc Pythia} with CT10 PDF and AU2 tune), which then allows an extraction of $Q_i^y$ by solving the system of equations in Eq.~(\ref{eq:syst}).  This extraction is performed separately in each $p_\text{T}$ bin.  Figure~\ref{fig:extracedupdown} shows the extracted up- and down-flavor jet charges in bins of jet $p_\text{T}$.  At very high jet $p_\text{T}$, the absolute quark flavor fractions are large (Fig.~\ref{fig:flavorfrac}), but the difference between the more forward and more central jets is small and the statistical uncertainty is large.  At low jet $p_\text{T}$, the difference between the more forward and more central jets is large (Fig.~\ref{fig:flavorfrac}), but the absolute quark flavor fraction is small and the statistical uncertainty is once again large because the mean jet charge is close to zero.  In the limit that the flavor fractions are identical for the more forward and more central jet, the equations become degenerate and it is not possible to simultaneously extract the average up- and down-flavor jet charges.  The uncertainties on the flavor fractions and on the measured average jet charges are propagated through the solutions of Eq.~(\ref{eq:syst}).  Generally, the uncertainty is larger for the down-flavor jets because the fraction of these jets is smaller than the fraction of up-flavor jets.

\begin{figure}
\begin{center}
\includegraphics[width=0.8\textwidth]{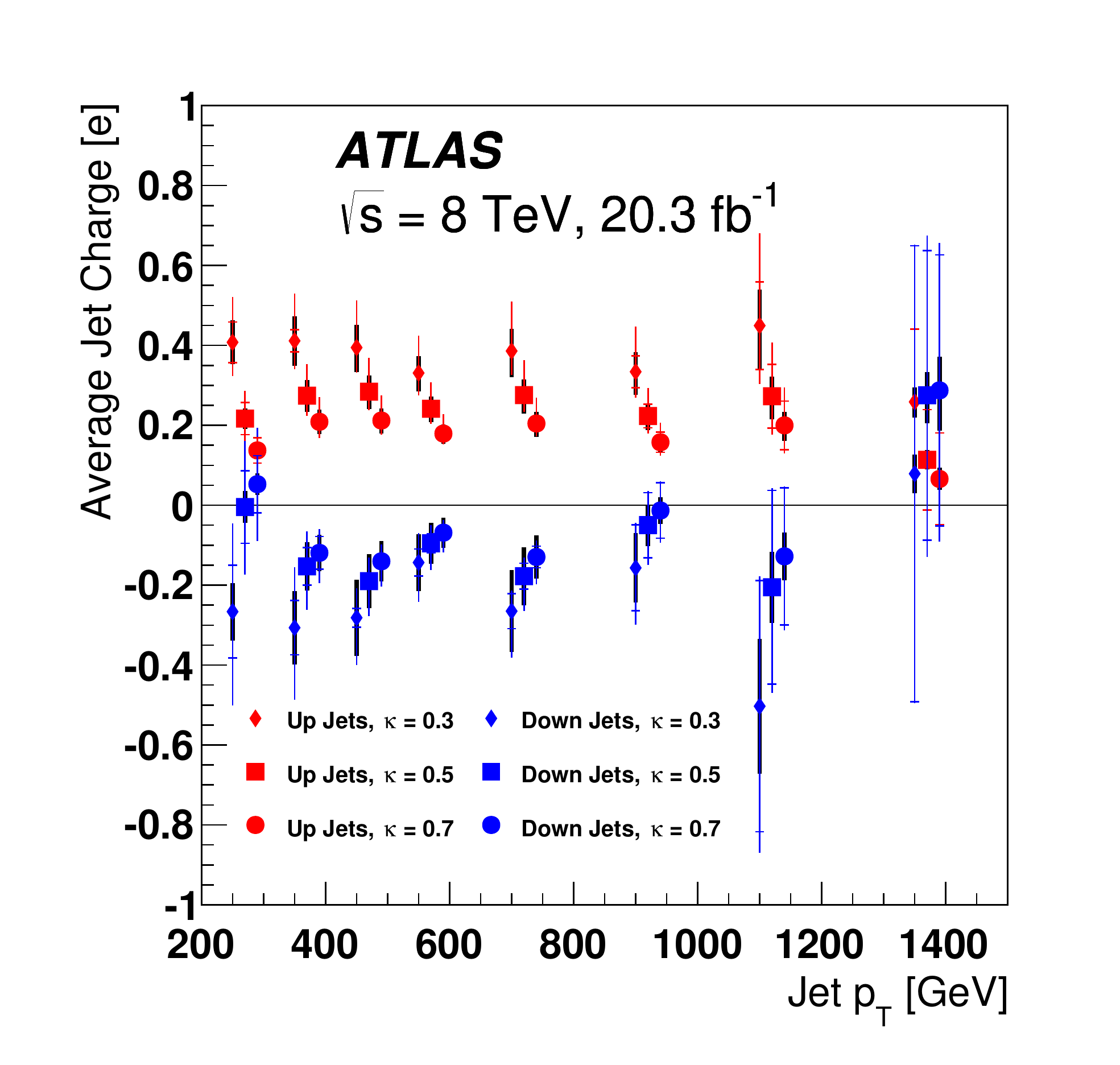}
\end{center}	
\caption{The extracted value of up- and down-quark jet charges in units of the positron charge in bins of jet $p_\text{T}$ for $\kappa=0.3, 0.5,$ and $0.7$.  The error bars include statistical, experimental systematic, and CT10 PDF uncertainties added in quadrature. The thick part of the error bar indicates the PDF contribution to the total uncertainty and the horizontal line on each error bar indicates the contribution from the statistical uncertainty.  The first two $p_\text{T}$ bins are excluded due to their very large uncertainties.}
\label{fig:extracedupdown}
\end{figure} 

\subsection{Dependence of the up-quark and down-quark jet charge on $p_\text{T}$}
\label{sec:scaleviolation}

Using the methods of Sec.~\ref{sec:updownextract}, one can examine the residual $p_\text{T}$-dependence of the average jet charge {\it after} accounting for PDF effects.  The inclusive jet charge has been shown to increase with $p_\text{T}$ due to a mixing of jet flavors and the following subsection investigates the $p_\text{T}$-dependence of a fixed jet flavor.   Section~\ref{sec:theory} describes the theory predictions and the extracted $p_\text{T}$-dependence from the data is discussed in Sec.~\ref{sec:scaleviolatedata}.

\subsubsection{Theory prediction}
\label{sec:theory}

Recent theoretical studies have shown that the energy dependence of jet charge moments is calculable perturbatively~\cite{Waalewijn2012,Krohn2012}.  At leading power (treating $m_\text{jet}/E_\text{jet}$ as the expansion parameter), the charge defined with jet and track energies is equivalent to the one given in Eq.~(\ref{chargedef}), since the jet opening angle is small.  For jets defined by a radius $R$ at an energy $E$, the average jet charge $ \langle Q_J \rangle $ is given by

\begin{align}
\label{eq:qj}
 \langle Q_J \rangle  = \left[1+ \mathcal{O}(\alpha_\text{s})\right] \sum_h Q_h \widetilde{D}_q^h(\kappa,E\times R),
\end{align}

where $Q_h$ is the charge of hadron $h$ and the functions $\widetilde{D}_q^h(\kappa,\mu)$ are the Mellin moments of the fragmentation functions

\begin{align}
 \widetilde{D}_q^h (\kappa,\mu)= \int_0^1 d x\, x^\kappa D_q^h(x,\mu).
\end{align}

The fragmentation functions $D_q^h(x,\mu)$ describe the probability for a hadron $h$ to carry a momentum fraction $x$ of a quark $q$ at the energy scale $\mu$~\cite{Collins:1981uw,Collins:1981uk}.  The $\mathcal{O}(\alpha_\text{s})$ correction is small and is dominated by the uncertainty on the fragmentation functions~\cite{Waalewijn2012}.  Ratios of $\langle Q_J\rangle$ at different energies result in a cancellation of the leading corrections and associated uncertainties.  Soft corrections to the average jet charge are small not only because of the additional
suppression at $\kappa>0$ compared to collinear radiation, but also because the leading soft emissions are made of gluons, which carry no electric charge. Although gluons can split into quark and anti-quark pairs, the same number of quarks and antiquarks go into the jet so that  on average the jet charge is unchanged.\footnote{This is not strictly true, as the soft radiation
pattern depends on the partonic colors, which are correlated with electric charge, but these correlations are negligible compared to the other uncertainties.}

The leading energy dependence of the average jet charge is due to the derivative of the fragmentation functions with respect to $E$, which is determined by the renormalization group equations for $\widetilde{D}_q^h(\kappa,\mu)$:

\begin{equation}
\frac{d}{d \ln \mu} \widetilde{D}_q^h(\kappa,\mu) = \frac{\alpha_\text{s}}{\pi}\widetilde{P}_{qq}(\kappa)\widetilde{D}_q^h(\kappa,\mu) + \mathcal{O}(\alpha_\text{s}^2),
\end{equation}

where $\widetilde{P}_{qq}(\kappa)$ is the moment of the leading-order splitting function:
\begin{align}
\widetilde{P}_{qq}(\kappa)&=  C_F\int_0^1 dz (z^\kappa-1)  \frac{1+z^2}{1-z}.
\end{align}

Recalling that $E/p_\text{T}$ is constant at leading power for all particles in the jet, the prediction for the {\it scale violation} of a quark jet is given by

\begin{equation}
\label{eq:finalformula}
\frac{p_\text{T}}{\langle \mathcal{Q}_\kappa \rangle} 
\frac{d}{d p_\text{T}} \langle \mathcal{Q}_\kappa \rangle  = \frac{\alpha_\text{s}}{\pi}  \widetilde{P}_{qq}(\kappa)\equiv c_\kappa \approx
 \begin{cases} -0.024 \pm 0.004 & \kappa = 0.3\\  -0.038 \pm 0.006 & \kappa = 0.5 \\  -0.049 \pm 0.008 & \kappa=0.7 \end{cases}
\end{equation}

where the last form consists of numerical approximations setting $\mu= E\times R$ equal to $50$ \GeV~and $500$ \GeV~with their average giving the central value, and using $\alpha_\text{s}(50 \text{ \GeV~}) = 0.130$ and $\alpha_\text{s}(500 \text{ \GeV~}) = 0.094$.  Experimentally, one measures combinations of quark and gluon jets with the fractions of different partonic flavors varying with energy. The next subsection discusses how the scaling violation parameter, defined in Eq.~(\ref{eq:finalformula}), can be extracted from the data.
 
\subsubsection{Extraction from the data} 
\label{sec:scaleviolatedata}
 
Since $c_\kappa\ll 1$ from Eq.~(\ref{eq:finalformula}), one can approximate a linear dependence on $c_\kappa$:

\begin{align}
\langle Q_J\rangle(p_\text{T})=\bar{Q} (1+c_\kappa\ln(p_\text{T}/\bar{p}_\text{T}))+\mathcal{O}(c_\kappa^2),
\end{align}

\noindent where $\bar{Q}=\langle Q_J\rangle(\bar{p}_\text{T})$ for some fixed (but arbitrary) transverse momentum, $\bar{p}_\text{T}$.  Therefore, for a fixed $p_\text{T}$ bin $i$, the measured charge is given as a superposition of the average jet charge for various jet flavors:

\begin{align}
\label{eq:scaleviolate}
\langle Q_i\rangle \approx \sum_f \beta_{f,i}\bar{Q}_f(1+c_\kappa\ln(p_{\text{T},i}/\bar{p}_\text{T})),
\end{align}

\noindent where $\beta_{f,i}$ is the fraction of flavor $f$ in bin $i$, $\bar{Q}_f$ is the average jet charge of flavor $f$ and $\bar{p}_\text{T}$ is a fixed transverse momentum.   Fitting the model in Eq.~(\ref{eq:scaleviolate}) directly to the data to extract $\bar{Q}_f$ is not practical because there are three parameters and only 10 $p_\text{T}$ bins, some of which have very little sensitivity due to low fractions $\beta$ or large uncertainties on $\langle Q_J\rangle$.  One way around this is to extract $\bar{Q}_f$ in one fixed bin of transverse momentum (denoted $\bar{p}_\text{T}$) as described in Sec.~\ref{sec:updownextract}.  Then Eq.~(\ref{eq:scaleviolate}) is highly constrained, with only one parameter for which each other bin of $p_\text{T}$ gives an estimate.  The systematic uncertainties are propagated through the fit treated as fully correlated between bins and the statistical uncertainty is treated coherently by bootstrapping\footnote{Pseudo-datasets are generated by adding each event in the nominal dataset $j$ times where $j$ is a Poisson random variable with mean $1$.  Since events are coherently added, this respects the correlations in the statistical uncertainty for the more forward and central jet charges.}.  A weighted average is performed across all $p_\text{T}$ bins and for both the more forward and the more central jet.  The procedure is summarized below:

\begin{enumerate}
\item In the bin 600~$\GeV<p_\text{T}<$ 800~$\GeV$,  extract the values $\bar{Q}_\text{up}$ and $\bar{Q}_\text{down}$.  These values can be seen in the fifth $p_\text{T}$ bin of Fig.~\ref{fig:extracedupdown}.
\item With $\bar{Q}_\text{up}$ and $\bar{Q}_\text{down}$ fixed, extract the scale violation parameter estimate $c_{\kappa,i}$ in each $p_\text{T}$ bin $i$ by solving
\begin{align}
\langle Q_i\rangle_\text{measured}=\sum_f \beta_{f,i}\bar{Q}_f(1+c_{\kappa,i}\ln(p_{\text{T},i}/\bar{p}_\text{T}))
\end{align}
\noindent where $\bar{p}_\text{T}=700$~\GeV~is the bin center from the previous step.  
\item Repeat the above procedure for all systematic variations and for all bootstrap pseudo-datasets to arrive at estimates of the uncertainty $\sigma(c_{\kappa,i})$ for each $p_\text{T}$ bin $i$.
\item The central value for the extracted scale violation parameter is $c_\kappa=(\sum_i c_{\kappa,i}/\sigma(c_{\kappa,i}))/\sum_i(1/\sigma(c_{\kappa,i}))$.
\item The uncertainty $\sigma(c_\kappa)$ is determined by repeating step (3) with the nominal values $c_{\kappa,i}$ replaced by their systematic varied versions or the bootstrap pseudo-data values for the statistical uncertainty estimate.
\end{enumerate}

The results are presented in Fig.~\ref{fig:money}.  The data support the prediction that $c_\kappa<0$ and $\partial c_\kappa/\partial\kappa < 0$.  Linear correlations between $\kappa$ values can be determined using the bootstrapped datasets: about $0.9$ between $c_{0.3}$ and $c_{0.5}$ as well as between $c_{0.5}$ and $c_{0.7}$, while the correlation is about 0.7 between $c_{0.3}$ and $c_{0.7}$.  Thus, the three points are quite correlated, but there is additional information from considering more than one $\kappa$ value.
 
\begin{figure}
\begin{center}
\includegraphics[width=0.8\textwidth]{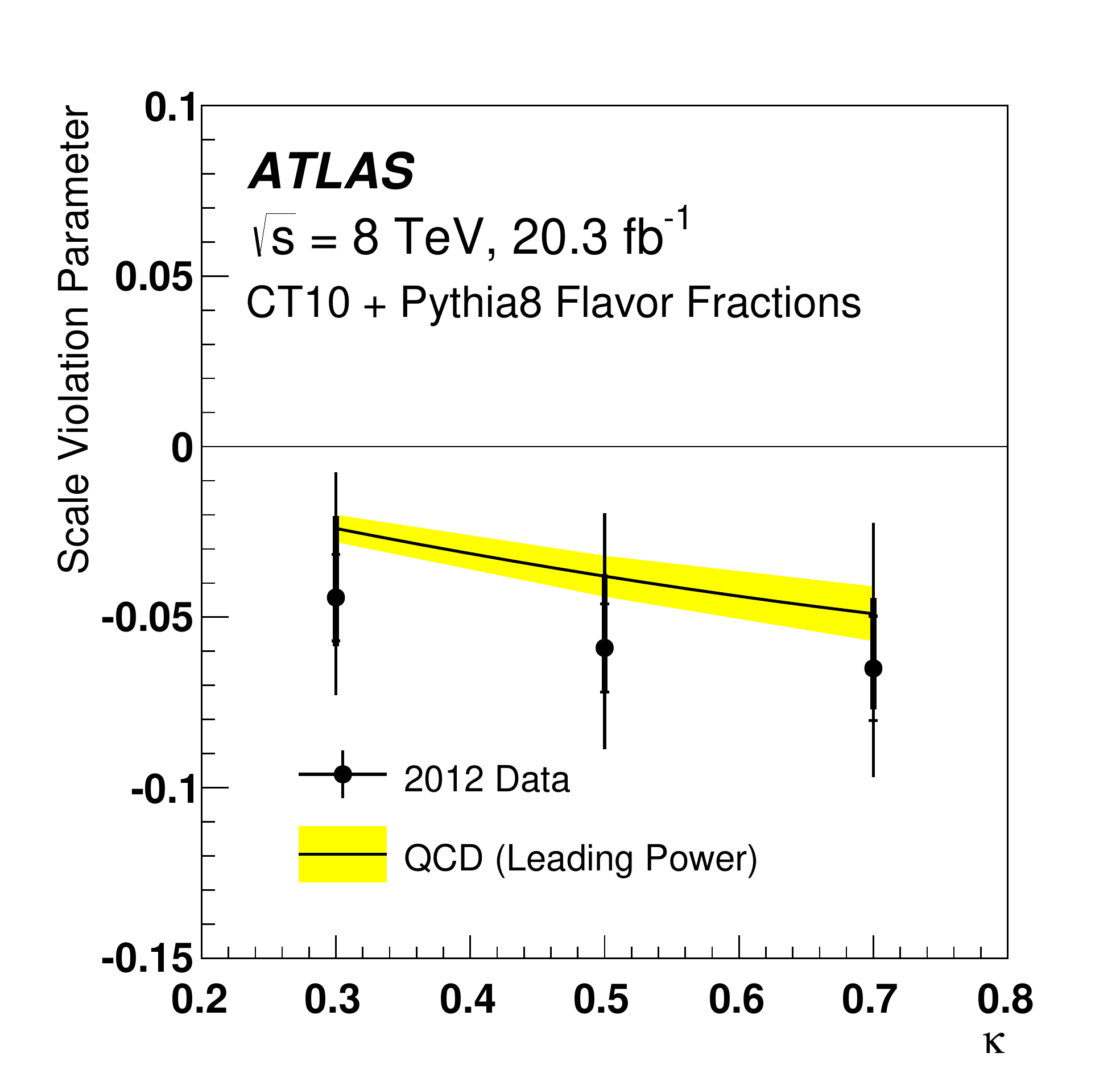}
\end{center}	
\caption{The extracted values of the scale violation parameter $c_\kappa$ from the data compared to theoretical calculations~\cite{Waalewijn2012,Krohn2012}.  The error bars include statistical, experimental systematic, and PDF uncertainties added in quadrature. The thick part of the error bar indicates the PDF contribution to the total uncertainty and the horizontal line on each error bar indicates the contribution from the statistical uncertainty (each shown without adding in quadrature any other source of uncertainty).}
\label{fig:money}
\end{figure} 
 
 \clearpage
 
\section{Summary}
\label{sec:summary}

This paper presents a measurement of the particle-level $p_\text{T}$-dependence of the jet charge distribution's mean and standard deviation in dijet events from 20.3 fb${}^{-1}$ of $\sqrt{s}=8$ \TeV~$pp$ collision data recorded by the ATLAS detector at the LHC.  The measured jet charge distribution is unfolded to correct for the detector acceptance and resolution for direct comparison to particle-level models.  Comparisons are made at particle level between the measured jet charge distribution and various PDF sets and models of jet formation.  Simulations with Pythia 8 using the CTEQ6L1 PDF set describe the average jet charge of the more forward jet within about 5\% and the more central jet within about 10\%. The jet charge distribution's standard deviation is described within 2\%.  {\sc Herwig++} shows a similarly good agreement for $p_\text{T}>500$~\GeV~and $\kappa=0.7$.  However, the {\sc Herwig++} predictions decrease systematically for both the average and the standard deviation for decreasing $\kappa$.  Predictions with the CT10 NLO PDF are systematically below the data across jet $p_\text{T}$ for the average jet charge and systematically above for the jet charge distribution's standard deviation.  Taking the PDFs as inputs, the average up- and down-flavor jet charges are extracted as a function of $p_\text{T}$ and are compared with predictions for scale violation.  The data show that the average up- and down-quark jet charges decrease slightly with $p_\text{T}$ and this decrease increases with $\kappa$, as predicted.  The particle-level spectra are publicly available~\cite{hepdata} for further interpretation and can serve as a benchmark for future measurements of the evolution of nonperturbative jet observables to validate QCD MC predictions and tune their free model parameters.

\section*{Acknowledgments}



We thank CERN for the very successful operation of the LHC, as well as the
support staff from our institutions without whom ATLAS could not be
operated efficiently.

We acknowledge the support of ANPCyT, Argentina; YerPhI, Armenia; ARC, Australia; BMWFW and FWF, Austria; ANAS, Azerbaijan; SSTC, Belarus; CNPq and FAPESP, Brazil; NSERC, NRC and CFI, Canada; CERN; CONICYT, Chile; CAS, MOST and NSFC, China; COLCIENCIAS, Colombia; MSMT CR, MPO CR and VSC CR, Czech Republic; DNRF, DNSRC and Lundbeck Foundation, Denmark; IN2P3-CNRS, CEA-DSM/IRFU, France; GNSF, Georgia; BMBF, HGF, and MPG, Germany; GSRT, Greece; RGC, Hong Kong SAR, China; ISF, I-CORE and Benoziyo Center, Israel; INFN, Italy; MEXT and JSPS, Japan; CNRST, Morocco; FOM and NWO, Netherlands; RCN, Norway; MNiSW and NCN, Poland; FCT, Portugal; MNE/IFA, Romania; MES of Russia and NRC KI, Russian Federation; JINR; MESTD, Serbia; MSSR, Slovakia; ARRS and MIZ\v{S}, Slovenia; DST/NRF, South Africa; MINECO, Spain; SRC and Wallenberg Foundation, Sweden; SERI, SNSF and Cantons of Bern and Geneva, Switzerland; MOST, Taiwan; TAEK, Turkey; STFC, United Kingdom; DOE and NSF, United States of America. In addition, individual groups and members have received support from BCKDF, the Canada Council, CANARIE, CRC, Compute Canada, FQRNT, and the Ontario Innovation Trust, Canada; EPLANET, ERC, FP7, Horizon 2020 and Marie Sk?odowska-Curie Actions, European Union; Investissements d'Avenir Labex and Idex, ANR, Region Auvergne and Fondation Partager le Savoir, France; DFG and AvH Foundation, Germany; Herakleitos, Thales and Aristeia programmes co-financed by EU-ESF and the Greek NSRF; BSF, GIF and Minerva, Israel; BRF, Norway; the Royal Society and Leverhulme Trust, United Kingdom.

The crucial computing support from all WLCG partners is acknowledged
gratefully, in particular from CERN and the ATLAS Tier-1 facilities at
TRIUMF (Canada), NDGF (Denmark, Norway, Sweden), CC-IN2P3 (France),
KIT/GridKA (Germany), INFN-CNAF (Italy), NL-T1 (Netherlands), PIC (Spain),
ASGC (Taiwan), RAL (UK) and BNL (USA) and in the Tier-2 facilities
worldwide.

\printbibliography

\clearpage

\newpage 
\begin{flushleft}
{\Large The ATLAS Collaboration}

\bigskip

G.~Aad$^{\rm 85}$,
B.~Abbott$^{\rm 113}$,
J.~Abdallah$^{\rm 151}$,
O.~Abdinov$^{\rm 11}$,
R.~Aben$^{\rm 107}$,
M.~Abolins$^{\rm 90}$,
O.S.~AbouZeid$^{\rm 158}$,
H.~Abramowicz$^{\rm 153}$,
H.~Abreu$^{\rm 152}$,
R.~Abreu$^{\rm 116}$
% ...
\bigskip
\\
% ...
$^{11}$ Institute of Physics, Azerbaijan Academy of Sciences, Baku, Azerbaijan\\
% ...
$^{85}$ CPPM, Aix-Marseille Universit{\'e} and CNRS/IN2P3, Marseille, France\\
% ...
$^{90}$ Department of Physics and Astronomy, Michigan State University, East Lansing MI, United States of America\\
% ...
$^{107}$ Nikhef National Institute for Subatomic Physics and University of Amsterdam, Amsterdam, Netherlands\\
% ...
$^{113}$ Homer L. Dodge Department of Physics and Astronomy, University of Oklahoma, Norman OK, United States of America\\
% ...
$^{116}$ Center for High Energy Physics, University of Oregon, Eugene OR, United States of America\\
% ...
$^{151}$ Institute of Physics, Academia Sinica, Taipei, Taiwan\\
$^{152}$ Department of Physics, Technion: Israel Institute of Technology, Haifa, Israel\\
$^{153}$ Raymond and Beverly Sackler School of Physics and Astronomy, Tel Aviv University, Tel Aviv, Israel\\
% ...
$^{158}$ Department of Physics, University of Toronto, Toronto ON, Canada\\
% ...
\end{flushleft}


\end{document}